\documentclass[aps,prb,twocolumn,showpacs,superscriptaddress,10pt]{revtex4-1}

\usepackage[T1]{fontenc}
\usepackage[utf8]{inputenc}
\usepackage{amsmath}
\usepackage{amsfonts}
\usepackage{amssymb}
\usepackage{mathrsfs}
\usepackage{graphicx}
\usepackage{color}
\usepackage[usenames,dvipsnames]{xcolor}

\usepackage{tikz}
\usetikzlibrary{positioning,arrows}
\usetikzlibrary{decorations.pathmorphing}
\usetikzlibrary{decorations.markings}
\usetikzlibrary{calc}
\usetikzlibrary{patterns}

\tikzset{
phys/.style={thick, postaction={decorate}, decoration={markings, mark=at position .7 with {\arrow[]{triangle 45}}}},
res/.style={thick, decorate, decoration={snake,segment length=7, amplitude=1.5}},
arrow/.style={thick, draw=white, postaction={decorate}, decoration={markings, mark=at position .6 with {\arrow[black]{triangle 45}}}}
 }

\newcommand{\xx}{\mathbf{x}}
\newcommand{\jj}{\mathbf{j}}

\newcommand{\zz}{\mathbf{z}}

\newcommand{\dd}{\mathrm{d}}
\newcommand{\ee}{\mathrm{e}}
\newcommand{\kk}{\mathbf{k}}
\newcommand{\qq}{\mathbf{q}}
\newcommand{\vecpi}{\boldsymbol\Pi}
\newcommand{\vecphi}{\boldsymbol\phi}

\begin{document}
\title{Short-time universal scaling and light-cone dynamics after a quench in an isolated quantum system in $d$ spatial dimensions.}
\author{Alessio Chiocchetta}
\affiliation{SISSA --- International School for Advanced Studies and INFN, via Bonomea 265, 34136 Trieste, Italy}
\author{Marco Tavora}
\affiliation{Department of Physics, New York University, 4 Washington Place, New York, NY 10003, USA}
\author{Andrea Gambassi}
\affiliation{SISSA --- International School for Advanced Studies and INFN, via Bonomea 265, 34136 Trieste, Italy}
\author{Aditi Mitra}
\affiliation{Department of Physics, New York University, 4 Washington Place, New York, NY 10003, USA}

\begin{abstract}
{We investigate the effects of fluctuations on the dynamics of an}
isolated quantum system represented by a $\phi^4$ field theory with $O(N)$ symmetry
after a quench in $d>2$ spatial dimensions.
A perturbative renormalization-group approach involving a dimensional expansion in $\epsilon=4-d$
is employed {in order to} study the evolution within a prethermal regime controlled by elastic dephasing.
In particular, {we focus on} a quench from a disordered initial state to the critical point,
which introduces an effective temporal boundary
in the evolution. At this boundary, the relevant fields acquire an anomalous scaling dimension, while the
evolution of both the order parameter and its correlation and response functions display universal aging.
Since the relevant excitations propagate ballistically, a light cone in real space emerges.
At longer times, the onset of inelastic scattering appears as secularly growing self-generated dissipation
in the effective Keldysh field theory, with the strength of the dissipative vertices providing an estimate for the time needed to leave the prethermal regime.
\end{abstract}

\pacs{05.70.Ln, 64.60.Ht, 64.70.Tg}

\date{\today}
\maketitle

\section{Introduction}
\label{sec:introduction}

Experimental progress in cold atomic
gases\cite{Lamacraft2012,Yukalov2011,Bloch2008,Greiner2002b} and pump-probe spectroscopy
of strongly correlated materials\cite{Smallwood12,Cavalleri14}, has revived
interest in understanding how an isolated quantum many-body system,
driven far from equilibrium by an initial perturbation such as
{an abrupt change of a global parameter of its Hamiltonian (\emph{quench}) $H$}, evolves
in time\cite{PolkovnikovRMP,Eisert2015,Calabrese2016}.
{After almost ten years of intense theoretical and experimental investigation, the post-quench dynamics of the system, initially prepared in the ground state of the pre-quench Hamiltonian $H_0$, can be schematically described in the following way.
In the generic case of a non-integrable post-quench Hamiltonian, the system relaxes towards a Gibbs thermal state\cite{Deutsch1991,Srednicki1994,Rigol2008,Biroli2010,Canovi2011,Tavora13b,Tavora13,Rigol2016,Yin2016}
in the sense that the expectation value 
of a generic local observable 
can alternatively be determined (in the thermodynamic limit) by an effective thermal ensemble $\rho$.
If, instead, the post-quench Hamiltonian is integrable and therefore it is characterized by an extensive number of mutually commuting (local or quasi-local) integrals of motion which constrain the dynamics, then the eventual asymptotic state is described by the so-called generalized Gibbs ensemble \cite{Rigol2007,Iucci2009,Jaynes1957,Barthel2008,Goldstein2014,Pozsgay2014,Mierzejewski2014,Wouters2014,Essler2016} (GGE).  The intermediate case in which the post-quench Hamiltonian $H$ is obtained by weakly perturbing (in a non-integrable fashion) an integrable one, displays, after an initial  post-quench transient, a quasi-stationary \emph{prethermal} regime\cite{Berges2004a,Langen2013a,Kitagawa2011,Kollar2011,Moeckel2009,Moeckel2010,Marino2012,Marcuzzi2013,Mitra2013,VanDenWorm2013,Bertini2015}
determined by the integrable part of $H$ and approximately described by the corresponding GGE. At longer times, the effect of the perturbation becomes predominant, causing the crossover towards the eventual thermal state. The effective strength of the non-integrable perturbation determines the actual duration of this prethermalization.
However, a prethermal regime may appear because of a variety of possibly different physical phenomena, resulting in what are collectively referred to as
\emph{non-thermal fixed points}\cite{Berges2008,Berges2009,Langen2016}, among which we mention the generation of turbulence\cite{Berges2011,Berges2012,PineiroOrioli2015}, of topological
defects\cite{Nowak2012,Karl2013,Schole2012}, and the presence of conservation laws which slow down the transport of particles and energy~\cite{Lux14}.}

{Remarkably, a novel class of critical-like phenomena may occur before the system thermalizes. These are characterized by a prethermal regime\cite{Eckstein2009,Sciolla2010,Gambassi2011,Schiro2010,Schiro2011,Sciolla2011,Sciolla2013,Chandran2013, Smacchia2015,Chiocchetta2015,Maraga2015}
where the non-equilibrium or dynamical ``phase transitions'' (DPTs) are signalled by a change, upon varying a parameter of the quench, in the qualitative features of the time evolution of physical observables. Qualitatively different temporal behavior 
play the role of different ``phases''. These}
DPTs are also of fundamental interest because they might be characterized by a scale-invariant behavior with
some \emph{universal} features which are independent of the microscopic details of the system, making them resemble
critical phenomena encountered in equilibrium and non-equilibrium statistical physics.
So far, these transitions have been investigated within a mean-field
approximation~\cite{Sciolla2010,Gambassi2011,Schiro2010,Sciolla2011,Chichinadze2016},
while exact results are available for certain solvable
models~\cite{Barankov2006,Yuzbashyan2006b,Yuzbashyan2006b,Sciolla2013,Chandran2013,Smacchia2015,Maraga2015,Maraga2016}, {all characterized by the absence of thermalization and therefore by a neverending prethermal regime.}

{In the present work, instead, we explore the dynamical behavior and the DPT beyond the mean-field approximation, i.e., we account for the effects of fluctuations ---  following the analysis of Ref.~\onlinecite{Chiocchetta2015}, --- which are relevant in spatial dimensionality $d$ smaller than the upper critical dimensionality $d_c$ of the system. In particular, we detail a perturbative renormalization group (RG) study of the dynamics of the system described in Sec.~\ref{sec:model} and perform a dimensional expansion in the deviation $\epsilon \equiv d_c-d$. While}
most of the available analytic methods and efficient algorithms for numerical simulations applies to one-dimensional
systems\cite{PolkovnikovRMP}, the analysis presented here is not tied to any specific dimensionality $d$ and therefore one can explore its role in determining the features of the observed transition.
For the particular model we investigate here, it turns out that the lower critical dimensionality $d_l$ is equal to 2, and therefore our results are expected to be qualitatively correct for $d >d_l = 2$, as discussed further below.

As anticipated in Ref.~\onlinecite{Chiocchetta2015},
we identify a novel temporal scaling
{due to the coarse-grained features of the initial condition affecting the behavior of the system in the early stage of its dynamics, but for times longer than the microscopic scale. Technically this implies that the
fields at the time of the quench
acquire an anomalous scaling dimension
because of the sharp temporal boundary introduced by the quench.} 
Close to the dynamical critical point,
this leads to  the phenomenon of \emph{quantum aging}, analogous to the one observed in classical systems after
a quench to 
equilibrium~\cite{Gambassi2005,Biroli2015} and non-equilibrium~\cite{Baumann2007} critical points.

This aging is characterized by the emergence of the following scaling behavior in the correlation and response functions $G_K$ and $G_R$, respectively,
at two times $t$ and $t'$ after the quench, as a function of the wavevector $k$,
\begin{subequations}
\begin{align}
G_K(k,t,t') & = \frac{1}{k^{2-2\theta_N}}\mathcal{G}_K(kt,kt'), \label{eq:GK-scaling-form}\\
G_R(k,t,t') & = \frac{1}{k} \left(\frac{t'}{t}\right)^{\theta_N} \mathcal{G}_R(kt,kt'), \label{eq:GR-scaling-form}
\end{align}
\end{subequations}
with $\mathcal{G}_K(x,y) \sim (xy)^{1-\theta_N}$ for $x,y \ll 1$ and $\mathcal{G}_K(x,y) \sim 1$ for $x,y \gg 1$, while $\mathcal{G}_R(x,y) \sim x $ for $y\ll x \ll 1$.
The leading algebraic behaviour is described by the new universal exponent $\theta_N$, called \emph{initial-slip} exponent, which we calculate at the leading order in the dimensional expansion. The scaling forms~\eqref{eq:GK-scaling-form} and~\eqref{eq:GR-scaling-form}, which emerge here within a perturbative RG analysis, were also confirmed by the exact solution of the model in the limit $N\to \infty$ of the $O(N)$ model~\cite{Maraga2015}.

The analysis presented here parallel the one for classical dissipative systems~\cite{Janssen1988},
open quantum systems~\cite{Buchold2014,Gagel2014,Gagel2015,Lang2016}, {and imaginary time evolution of quantum systems~\cite{Yin2014}}, but with important differences.
First, while our results describe the approach to a prethermal state,
those of Refs.~\onlinecite{Janssen1988,Gagel2014,Gagel2015} characterize the evolution towards the eventual equilibrium thermal state resulting from coupling the system to an ideal thermal bath. As a consequence of this difference,
an anomalous scaling occurs here in the $k$-dependence of correlation functions, which is discussed
in Sec.~\ref{sec:momentum-distribution}.
Second, the relevant collective excitations of the system investigated here propagate ballistically and therefore a light cone is present in the spatio-temporal correlation~\cite{Calabrese2006,Calabrese2007,Mitra2013}
and response~\cite{Foini2011,Foini2012,Marcuzzi2014} functions, which are analyzed in Sec.~\ref{sec:light-cone} and which
contrast with the diffusive collective excitations of open classical~\cite{Janssen1988}  and quantum~\cite{Buchold2014,Gagel2014,Gagel2015} systems, as determined by the presence of the thermal bath.
{Another relevant consequence of this ballistic character is that the effects of the initial condition, i.e., of the ``temporal boundary'' propagate into the ``temporal bulk'', rendering the distinction between ``bulk'' and ``boundary'' less marked than in the case of dissipative classical systems, where the memory of the initial condition is generally lost exponentially fast in time.}
Third, a major {related} difference with the classical case 
is that the quantum dynamics has to preserve the canonical commutation relations, which imposes a constraint on the scaling dimensions of the relevant fields and therefore affects the scaling forms of $G_{R,K}$.

Although most of the qualitative features discussed here are expected to be quite generic to isolated quantum systems, we focus for concreteness
on the quench of one described by an $N$-component order parameter with Hamiltonian $H(r,u)$ characterized by an
$O(N)$ symmetry. The system is initially prepared in the symmetric ground state of the same Hamiltonian as the one which rules its subsequent time evolution, but with different values of the
parameters $r$ and $u$, which control the distance from the DPT and the strength of the interaction, respectively.
Various properties of this quench protocol have been
the object of study~\cite{Calabrese2006,Calabrese2007,Sotiriadis2009,Sotiriadis2010}, as summarized in Sec.~\ref{sec:model}.
Based on the perturbative renormalization group approach mentioned above and used in Ref.~\onlinecite{Chiocchetta2015}, we demonstrate here
 that a DPT occurs in the pre-thermal regime after the quench and we characterize its universal features.
We argue that this DPT is closely related to the one predicted in the limit $N\to\infty$ of the present
model~\cite{Sciolla2013,Chandran2013,Smacchia2015}.
In particular, by deriving the equivalent Wilson (Sec.~\ref{sec:RG-Wilson}) and Callan-Symanzik{-like} (Sec.~\ref{sec:RG-CS})
flow equations for the relevant couplings and correlation functions, respectively,
we highlight both similarities and differences between the DPT investigated here and the
phase transition occurring within this model in thermodynamic equilibrium.

Remarkably, when the pre-quench value $\Omega_0^2$ of the parameter $r$ is very large on the scale specified further below,
$\Omega_0$ plays the role of an effective temperature for the post-quench system and, accordingly, it
induces a dimensional crossover on the DPT in complete analogy to what happens in
equilibrium quantum phase transitions upon increasing the temperature~\cite{Fisher1988,Sondhi1997,Sachdevbook}.
In particular, due to this finite effective temperature, the system is taken out of the quantum critical regime~\cite{Sachdevbook} and therefore
its upper critical dimensionality $d_c$ is the same as for the classical critical system, i.e., $d_c=4$.

The rest of the presentation is organized as follows:
in Sec.~\ref{sec:model} we introduce the model and discuss the relevant dynamical quantities of the Gaussian theory with Hamiltonian $H(r,u=0)$
after a quench of the parameter $r$. In this context, we highlight in Sec.~\ref{sec:light-cone-G} the emergence of a light cone in the response and
correlation function after a quench at the Gaussian {(dynamical)} critical point $r=0$.
In Sec.~\ref{sec:perturbation-theory} we account perturbatively for the effects of interactions, identifying the
logarithmic corrections associated with the relevant correlation functions. We present the scaling forms
{although the RG resummation required to justify the scaling forms are presented later, in Sec.~\ref{sec:RG-CS}. We also show in
Sec.~\ref{sec:perturbation-theory} that these scaling forms are verified non-perturbatively in the limit $N\to\infty$ of the present model. In particular, the}
resulting expressions of the response and correlation functions in momentum space at
long wavelengths are presented in Sec.~\ref{sec:Greens-momentum-space},
of the momentum distribution in Sec.~\ref{sec:momentum-distribution}, while those
of the same functions in real space are discussed in Sec.~\ref{sec:light-cone}, together with the emerging light cone.
Section~\ref{sec:magnetization} presents results for the time evolution of the average order parameter
when a small symmetry-breaking field is added to the pre-quench Hamiltonian, which are obtained using both perturbation theory
and the self-consistent Hartree-Fock approximation which is exact for $N\to\infty$.

The perturbative analysis mentioned above is then complemented by the construction of appropriate RG equations based on both the
Wilson approach in Sec.~\ref{sec:RG-Wilson} --- which highlights the eventual emergence of dissipative terms ---
and the Callan-Symanzik{-like flow} equation for the relevant correlation functions in Sec.~\ref{sec:RG-CS}.
The summary and conclusions are then presented in Sec.~\ref{sec:conclusions}, while intermediate and technical details
of the analysis are relegated to a the various Appendices.

{As a cautionary note, we emphasize here that the present work (together with Refs.~\onlinecite{Chiocchetta2015,Maraga2015}) provides convincing evidence of the emergence of scaling behavior in the post-quench dynamics of quantum systems. However, a comprehensive picture of the structure of the ultraviolet singularities of the underlying model on the continuum is far from being achieved, especially beyond the perturbative order considered here. In fact, while the emergence of scaling laws in quantum statistical systems in equilibrium both at zero and finite temperature\cite{ZinnJustinbook},
as well as in the static and dynamic behavior of classical statistical systems, both in the bulk and at surfaces\cite{Diehl1981,Diehl1997,Diehl1986}
is based on a detailed understanding of these properties (and of the renormalizability of the associated continuum theory), the present case is sufficiently complicated that this would require a dedicated study, going well beyond the purposes of this investigation. We emphasize that the present study is already a challenging first step in this direction. Despite the peculiarities of the quantum quench, some similarities
(spelled out in Secs.~\ref{sec:Gaussian-theory}, \ref{sec:deep-quench} and in Ref.~\onlinecite{Chiocchetta2015}) between the present case and the well-studied one of quantum systems at finite temperature~\cite{Sachdev1997,Sondhi1997,Sachdevbook},
%
%
as well as with the analytic
continuation\cite{Calabrese2006,Calabrese2007,Marcuzzi2014}
of equilibrium classical systems confined within a film
geometry,\cite{ZinnJustinbook,Cardy1988,Brankov2000}
%
%
suggests that the corresponding theories on the continuum might have a similar structure, though important differences remain due to the boundary conditions imposed by the quench, as discussed in Sec.~\ref{sec:quench-action}.}

\section{The Model}
\label{sec:model}

\subsection{Quench protocol}
\label{sec:protocol}

We consider a system described by a real vector order parameter $\vecphi \equiv \vecphi(\xx)=(\phi_1,\ldots,\phi_N)$ with $N$
components and  the following Hamiltonian in $d$ spatial dimensions, with $O(N)$ symmetry:
\begin{equation}
\label{eq:Hamiltonian}
H(r,u)= \int_\xx \left[ \frac{1}{2}\vecpi^2 + \frac{1}{2}(\nabla \vecphi)^2 + \frac{r}{2}\vecphi^2+\frac{u}{4!N}\left(\vecphi^2\right)^2\right],
\end{equation}
where $\int_\xx \equiv \int\dd^d x$ while  $\vecpi \equiv \vecpi(\xx)$ is the momentum conjugate to $\vecphi$, which satisfies the canonical commutation relations $[\phi_i(\xx),\Pi_j(\xx')] = i \delta^{(d)}(\xx-\xx')\delta_{ij}$. Note that only scalar products such as $\vecphi^2 = \vecphi\cdot\vecphi= \sum_{i=1}^N \phi_i^2$ enter Eq.~\eqref{eq:Hamiltonian}, in order to guarantee its invariance under $O(N)$ transformations.
The coupling $u>0$ parametrizes the strength of interaction, while $r$ controls the distance from the critical point of DPT, as discussed below.
The quench consists of a sudden change of both $r$ and $u$ at time $t=0$ according to
\begin{equation}
H(\Omega_0^2, 0) \equiv H_0 \to H(r,u),
\end{equation}
where we denote the pre-quench value of $r$ as $r(t=0^-) = \Omega_0^2$ for later convenience. We will assume that this value is positive, so that the system is initially in the disordered phase and any correlation function involving an odd number of fields $\vecphi$ vanishes.
This protocol is the same as the one studied in Refs.~\onlinecite{Chiocchetta2015,Maraga2015}, while in Refs.~\onlinecite{Gambassi2011,Sciolla2013,Chandran2013,Smacchia2015} the interaction strength $u$ had the same finite value before and after the quench. 
For both protocols the system is shown to undergo a DPT at some (protocol-dependent)
critical value $r=r_c$. However, one expects the two transitions to be essentially the same. Indeed, as pointed out in
Refs.~\onlinecite{Chiocchetta2015,Maraga2015}, a non-zero interaction in the pre-quench state renormalizes only
the value of $\Omega_0^2$, leaving unchanged the short-range nature of the correlations in the initial state, which actually determines the features of the DPT, as we argue below: accordingly, the two protocols discussed above are essentially equivalent in this respect.
This fact can be regarded as a consequence of the \emph{universality} associated with the DPT: as shown in Sec.~\ref{sec:Callan-Symanzik}, the scaling forms and the value of the critical exponents characterizing the correlation functions are independent of the actual value of $\Omega_0$.

\subsection{Gaussian model:  correlation functions in momentum space}
\label{sec:Gaussian-theory}

In this section we summarize the analysis of the quantum quench in a bosonic free field theory~\cite{Calabrese2006,Calabrese2007,Sotiriadis2009} which corresponds to the special case $u=0$ of the quench we are interested in and which provides the basis for the perturbative calculations discussed in Sec~\ref{sec:perturbation-theory}.
As is clear from the analysis presented below, for the resulting Hamiltonian $H(r,0)$ to have a spectrum bounded from below  it is necessary to assume $r\ge 0$.
Introducing the Fourier transform in momentum space of the fields $\vecphi$ and $\vecpi$ as $\vecphi(\xx) = \int_\kk  \vecphi_{\kk} \,\ee^{i\kk\cdot \xx}$ and $\vecpi(\xx) = \int_\kk  \vecpi_{\kk} \,\ee^{i\kk\cdot \xx}$, respectively, where $\int_\kk \equiv \int \dd^d k /(2\pi)^d$,  Eq.~\eqref{eq:Hamiltonian} can be written as
\begin{equation}
H(r,0) = \frac{1}{2} \int_{\kk} \left( \left| \vecpi_{\kk} \right|^2 + \omega _k^2 \, \left| \vecphi_\kk \right|^2 \right),
\label{eq:H0ms}
\end{equation}
where
\begin{equation}
\label{eq:dispersion-postquench}
\omega_k \equiv \sqrt{k^2 + r}
\end{equation}
is the dispersion relation, while $ \left| \vecpi_{\kk} \right|^2  \equiv \vecpi_\kk\cdot\vecpi_\kk^\dagger = \vecpi_\kk\cdot\vecpi_{-\kk}$ and analogous for $\vecphi$.
The Heisenberg equations of motion for the operators after the quench $\Omega_0^2 \to r$,
derived from  $H(r,0) $, is therefore $\ddot{\vecphi}_{\kk} + \omega_k^2\vecphi_{\kk} = 0$,
with solution
\begin{equation}
\vecphi_{\kk}(t)  = \vecphi_{\kk}(0)\cos(\omega_k t) + \vecpi_{\kk}(0) \frac{\sin(\omega_k t)}{\omega_k}.
\label{eq:field-ev}
\end{equation}
In order to calculate the expectation values of the components $\phi_{j,\kk}(0)$ and $\Pi_{j,\kk}(0)$ (with $j=1,\ldots,N$) on the initial state, it is convenient to introduce
the standard bosonic annihilation and creation operators $b_{j,\kk}$ and $b_{j,\kk}^\dagger$, respectively, defined as
\begin{equation}
\label{eq:bose-ops-def}
\left\{
\begin{aligned}
\phi_{j,\kk}  & = \frac{1}{\sqrt{2\omega_k}} \left( b_{j,\kk} + b_{j,-\kk}^\dagger \right),\\
\Pi_{j,\kk}  & = -i\sqrt{\frac{\omega_k}{2}} \left( b_{j,\kk} - b_{j,-\kk}^\dagger \right),
\end{aligned}
\right.
\end{equation}
{with $[b_{j,\kk},b^\dagger_{j',\kk'}] = \delta_{jj'}\delta_{\kk,\kk'}$ the standard bosonic commutation relations and $\delta_{\kk,-\kk'} \equiv (2\pi)^d\, \delta^{(d)}(\kk+\kk')$.}
Once expressed in terms of these operators, Eq.~\eqref{eq:H0ms} becomes
\begin{equation}
H(r,0) = \sum_{j = 1}^N {\int_{\kk} {{\omega _k}} \,b_{j,\kk}^\dag {b_{j,\kk}}},
\label{eq:Hpq-G}
\end{equation}
up to an inconsequential additive constant.
Assuming the pre-quench state to be in equilibrium at a temperature $T = \beta^{-1}$, with Hamiltonian $H_0 = H(\Omega_0^2,0)$, the density matrix of the system is given by $\rho_0 = \mathcal{Z}^{-1}\ee^{-\beta H_0}$, where $\mathcal{Z} =\text{tr}(\ee^{-\beta H_0})$. Accordingly, one can evaluate the following
statistical averages over $\rho_0$ by introducing the pre-quench operators $b^0_{j,\kk}$ and $b^{0,\dagger}_{j,\kk}$ as in Eq.~\eqref{eq:bose-ops-def}:
\begin{subequations}
\begin{align}
\langle \phi_{i,\kk}(0)\phi_{j,\kk'}(0) \rangle 	& = \delta_{\kk,-\kk'}\delta_{ij} \frac{1}{2\omega_{0k}}\coth(\beta\omega_{0k}/2), \label{eq:field-av-1}\\
\langle \Pi_{i,\kk}(0)\Pi_{j,\kk'}(0) \rangle 		& = \delta_{\kk,-\kk'}\delta_{ij}  \frac{\omega_{0k}}{2}\coth(\beta\omega_{0k}/2),\label{eq:field-av-2} \\
{\langle \{\phi_{i,\kk}(0),\Pi_{j,\kk'}(0)\}\rangle} 		& = 0,
\end{align}
\end{subequations}
and
\begin{equation}
\label{eq:dispersion-prequench}
\omega_{0k}=\sqrt{k^2+\Omega_0^2}
\end{equation}
is the pre-quench dispersion relation.
Since the initial state does not break the $O(N)$ symmetry one has $\langle \phi_{j,\kk}(0)\rangle =\langle \Pi_{j,\kk}(0)\rangle = 0$ and therefore Eq.~\eqref{eq:field-ev} implies that $\langle \phi_{j,\kk}(t) \rangle = 0$ at all times $t$.
The correlation functions of the field $\vecphi$ during the evolution can be easily determined on the basis of Eqs.~\eqref{eq:field-ev}, \eqref{eq:field-av-1}, and \eqref{eq:field-av-2}. Hereafter, we focus on the retarded and Keldysh Green's functions, defined respectively as
\begin{subequations}
\begin{align}
iG_{jl,R}(|\xx-\xx'|,t,t') & =  \vartheta(t-t') \langle \left[\phi_j(\xx,t),\phi_l(\xx',t')\right]\rangle , \label{eq:defGR-gen}\\
iG_{jl,K}(|\xx-\xx'|,t,t') & =  \langle\left\{\phi_j(\xx,t),\phi_l(\xx',t')\right\}\rangle, \label{eq:defGK-gen}
\end{align}
\end{subequations}
where $\vartheta(t<0) = 0$ and $\vartheta(t\ge 0) = 1$.
Note that, as a consequence of the invariance of the Hamiltonian under spatial translations and rotations, $G_{R/K}$  depend only on the distance $|\xx-\xx'|$ between the points $\xx$ and $\xx'$ at which the  fields are evaluated.
Accordingly, it is convenient to consider their Fourier transforms which are related to the Fourier components of the field $\vecphi_\kk$ via
\begin{align}
\delta_{\kk,-\kk'} \, iG_{jl,R}(k,t,t') & = \vartheta(t-t') \langle \left[\phi_{j,\kk}(t),\phi_{l,\kk'}(t')\right]\rangle,\\
\delta_{\kk,-\kk'} \, iG_{jl,K}(k,t,t') & = \langle\left\{\phi_{j,\kk}(t),\phi_{l,\kk'}(t')\right\}\rangle.
\label{eq:defGRGKqspace}
\end{align}
In the absence of symmetry breaking, the $O(N)$ symmetry implies that these functions do not vanish only for $j=l$, i.e., $G_{jl,R/K} = \delta _{jl} G_{R/K}$. Accordingly, in what follows, their dependence on the field components is no longer indicated.
The Gaussian Green's functions (henceforth denoted by a subscript $0$) in momentum space can be immediately determined by using the expression of the time evolution of the field $\phi_{j,\kk}$ in
Eq.~\eqref{eq:field-ev} and the averages over the initial condition in Eqs.~\eqref{eq:field-av-1} and \eqref{eq:field-av-2}, which yield
\begin{subequations}
\label{eq:GRK-gauss}
\begin{align}
G_{0R}(k,t,t') & = - \vartheta(t-t') \frac{\sin (\omega_k(t-t'))}{\omega_k}, \label{eq:GR-gauss}\\
G_{0K}(k,t,t') & = -i\frac{\coth(\beta\omega_{0k}/2)}{\omega_k} [ K_+\cos (\omega_k(t-t')) \nonumber \\
				& \qquad \qquad \qquad  + K_-\cos(\omega_k (t+t'))], \label{eq:GK-gauss}
\end{align}
\end{subequations}
where
\begin{equation}
K_\pm = \frac{1}{2} \left( \frac{\omega_k}{\omega_{0k}} \pm \frac{\omega_{0k}}{\omega_k} \right),
\label{eq:defKpm}
\end{equation}
with $\omega_k$ and $\omega_{0k}$ given in Eqs.~\eqref{eq:dispersion-postquench} and~\eqref{eq:dispersion-prequench}, respectively.
While the retarded Green's function $G_{0R}$, within this Gaussian approximation, does not depend on the initial state and it is therefore time-translation invariant (TTI), the Keldysh Green's function $G_{0K}$ acquires a non-TTI contribution as a consequence of the quantum quench.
As we discuss in Secs.~\ref{sec:perturbation-theory}, \ref{sec:RG-Wilson}, and \ref{sec:Callan-Symanzik}, this contribution plays an important role in the DPT, as it eventually generates an algebraic behavior in time.
Note that, in the absence of a quench, $\omega_k = \omega_{0k}$ and therefore $K_+ = 1$ and $K_- = 0$: correspondingly, the $G_{0K}$ in Eq.~\eqref{eq:GK-gauss} recovers its equilibrium TTI expression.
In addition, if the temperature $T$ of the initial state vanishes $T=0$, i.e., the system is in the ground state of the pre-quench Hamiltonian
$H_0$, the $G_{R/K}$ in Eqs.~\eqref{eq:GR-gauss} and \eqref{eq:GK-gauss} at small wavevectors
$k \ll \Omega_0$ and at $r=0$ read
\begin{subequations}
\begin{align}
G_{0R}(k,t,t') & = - \vartheta(t-t') \frac{\sin (k(t-t'))}{k}, \label{eq:GR-gauss-dq} \\
G_{0K}(k,t,t') & = -i\frac{\Omega_0}{2k^2} [ \cos (k(t-t'))  -\cos (k (t+t'))], \label{eq:GK-gauss-dq}
\end{align}
\end{subequations}
which can be cast in the scaling forms in Eqs.~\eqref{eq:GK-scaling-form} and \eqref{eq:GR-scaling-form}
with $\theta_N=0$. For later convenience, we note that these expressions become, at short times $t$, $t'\ll k^{-1}$,
\begin{subequations}
\begin{align}
G_{0R}(k,t,t') & \simeq - \vartheta(t-t') (t-t'), \label{eq:GR-gauss-dq-smallk} \\
G_{0K}(k,t,t') & \simeq -i \Omega_0 t t'. \label{eq:GK-gauss-dq-smallk}
\end{align}
\end{subequations}
The scaling in Eqs.~\eqref{eq:GR-gauss-dq} and~\eqref{eq:GK-gauss-dq} is related to a Gaussian fixed point. In fact, one recognizes from Eqs.~\eqref{eq:GR-gauss} and~\eqref{eq:GK-gauss} that, for momenta $k \ll \Omega_0$, the correlation length is simply given by $\xi = r^{-1/2}$, due to the dependence of this expression on the combination $k^2+r$. Accordingly, for $r \to r_c = 0$, this correlation length $\xi$ diverges, thus signalling the onset of a transition and, correspondingly, the emergence of scale invariance into the correlation functions. Moreover, recalling the definition of the critical exponent $\nu$, i.e., $\xi \sim |r - r_c|^{-\nu}$, one immediately realizes that in the present case $\nu = 1/2$, which is also the value expected at the Gaussian fixed point for the corresponding equilibrium model~\cite{Mabook,Goldenfeldbook,Sachdevbook}.

A relevant quantity we consider below is the number $n_\kk$ of particles with momentum $\kk$ after the quench, defined as $n_\kk = b^\dagger_\kk b_\kk$ in terms of the operators introduced in Eq.~\eqref{eq:bose-ops-def}, where the index of the field component has been omitted for clarity.
The operator $n_\kk$ can be conveniently expressed in terms of the field $\phi_\kk$ and its conjugate momentum $\Pi_\kk$ as
\begin{equation}
n_\kk + \frac{1}{2}\delta_{\kk,\kk}  =  \frac{1}{4\omega_k}(\{\Pi_\kk,\Pi_{-\kk}\}+ \omega_k^2\{\phi_\kk,\phi_{-\kk}\}).
\end{equation}
Since each term in this expression is proportional to the (infinite) volume $V = \int \dd^d x$ of the system, we consider the associated finite \emph{momentum density} $\mathcal{N}_\kk \equiv \langle n_\kk \rangle/ V$, which can be expressed in terms of the Green's functions in momentum space as
\begin{equation}
\mathcal{N}_\kk  + \frac{1}{2} = \frac{i}{4\omega_k}[G_K^\Pi(k,t,t)+ \omega_k^2 G_K(k,t,t)],
\end{equation}
where we introduced $i \delta_{\kk,-\kk'}\,G_K^\Pi(t,t') =\langle\{\Pi_\kk(t),\Pi_{\kk'}(t')\} \rangle $ in analogy with $G_K$ in Eq.~\eqref{eq:defGK-gen}.
Taking into account the Heisenberg equations of motion $\dot{\phi}_\kk = \Pi_\kk$, the equal-time $G_K^\Pi$ can be expressed as
\begin{equation}
G_K^\Pi(k,t,t) =  \partial_t\partial_{t'} G_K(k,t,t') \bigg|_{t=t'} \equiv \ddot{G}_K(k,t,t),
\end{equation}
from which it follows that the momentum density $\mathcal{N}_\kk$ can be expressed in terms of $G_K$ only:
\begin{equation}
\label{eq:nk}
\mathcal{N}_\kk  + \frac{1}{2} = \frac{i}{4\omega_k}[ \ddot{G}_K(k,t,t)+ \omega_k^2 G_K(k,t,t)].
\end{equation}
The r.h.s. of this equation can be calculated within the Gaussian approximation (hence the subscript $0$) by using Eq.~\eqref{eq:GK-gauss},  which yields
\begin{equation}
\label{eq:nkbare}
\mathcal{N}_{0\kk}  + \frac{1}{2} = \frac{1}{2}K_+\coth(\beta \omega_{0k}/2).
\end{equation}
Note that $\mathcal{N}_{0\kk}$ does not depend on time, because the post-quench Hamiltonian $H(r,0)$ can be written as in Eq.~\eqref{eq:Hpq-G}, i.e., as a linear combination of the momentum densities which are therefore conserved quantities. In addition, the number of excitations after the quench is finite even at $T=0$ as a consequence of the energy injected into the system upon quenching. In the absence of a quench, instead, $K_+ = 1$ and Eq.~\eqref{eq:nkbare} renders the Bose equilibrium distribution $\mathcal{N}_{0\kk} = 1/[\exp(\beta\omega_{0k}) - 1]$.

\subsection{Deep quenches limit and effective temperature}
\label{sec:deep-quench}

In the rest of the discussion, we mostly focus on the limit of \emph{deep quench} $\Omega^2_0 \gg r$, i.e.,
on the case in which there is a significant difference between the pre- and post-quench values of the control parameter $r$.
Since the post-quench value $r_c$ of $r$ at which the DPT takes place scales
as~\cite{Chiocchetta2015} $r_c \sim \epsilon \Lambda^2$ (see also Sec.~\ref{sec:perturbation-theory}) where $\Lambda$ is
some ultra-violet cut-off inherent the microscopic structure of the system, the deep-quench limit can be equivalently expressed as
$\Omega_0 \gg \Lambda$. Interpreting $\Lambda$ as being related to the inverse of the lattice spacing of the underlying microscopic lattice,
the condition $\Omega_0 \gg \Lambda$ implies that the correlation length $\simeq \Omega_0^{-1}$ of fluctuations in the
pre-quench state is smaller than the lattice spacing, i.e., the system is in a highly disordered state.
In turn, as was realized in Refs.~\onlinecite{Calabrese2006,Calabrese2007,Sotiriadis2009,Sotiriadis2010},
this disordered initial state resembles a high-temperature state and, in fact, the momentum density in
Eq.~\eqref{eq:nkbare} takes the form
\begin{equation}
\mathcal{N}_{0\kk} \simeq  \frac{T_\text{eff}}{\omega_k}
\end{equation}
of a thermal one in the deep-quench limit, with an effective temperature given by
\begin{equation}
T_\text{eff} = \Omega_0/4.
\label{eq:Teff-G}
\end{equation}
This similarity is made even more apparent by considering the fluctuation-dissipation theorem (FDT)~\cite{Kubo1966} which relates in frequency-momentum space the Keldysh and
retarded Green's functions of a system in thermal equilibrium at temperature
$\beta^{-1}$ as
\begin{equation}
G_K(\omega,k) = \coth(\beta \omega/2)[G_R(k,\omega) - G_R(k,-\omega)].
\label{eq:FDT}
\end{equation}
Out of equilibrium, one can \emph{define} an effective temperature $T_\text{eff} = \beta_\text{eff}^{-1}$ such that $G_{K/R}$ satisfy
the FDT~\cite{Cugliandolo2011,Foini2011,Mitra2012,Foini2012}, which generically depends on both frequency $\omega$ and momentum $k$ as a consequence
of the lack of thermalization.
In the present case, considering only the stationary part of Eq.~\eqref{eq:GK-gauss}, the Fourier transform of $G_{0K}$ is related to the one of
$G_{0R}$ in Eq.~\eqref{eq:GR-gauss} via
\begin{equation}
G_{0K}(k,\omega) = \frac{\Omega_0}{2\omega}[G_{0R}(k,\omega) - G_{0R}(k,-\omega)],
\end{equation}
which takes the form of the FDT in Eq.~\eqref{eq:FDT}  with the same (high) temperature
$T = T_\text{eff}$ as defined above in Eq.~\eqref{eq:Teff-G} from the behavior of $\mathcal{N}_{0\kk}$.
As shown in Sec.~\ref{sec:RG-Wilson}, the fact that the system appears to be ``thermal'' in the deep-quench limit has important consequences on its
critical properties. Indeed, it behaves effectively as a $d$-dimensional classical system rather than a $d+1$-dimensional one,
the latter expected for a closed quantum system at zero temperature. Accordingly, the effect of a deep quench on a DPT is heuristically the same as
that of a non-vanishing temperature on a quantum phase transition, where the temperature is so high that the system falls out of the so-called
quantum-critical regime\cite{Sachdevbook,Sondhi1997,Fisher1988}. However, as anticipated in Sec.~\ref{sec:introduction} and discussed below,
 this DPT is characterized by novel universal non-equilibrium properties, absent in the transition at equilibrium.

\subsection{Gaussian model: correlation functions in real space and light-cone dynamics for a critical quench}
\label{sec:light-cone-G}

In this section we discuss the properties of the Green's functions
in real space $G_{R/K}(x,t,t')$, with $x=|\xx_1 - \xx_2|$,
highlighting the emergence of a light cone in the dynamics in both correlation~\cite{Calabrese2006} and response~\cite{Foini2012,Marcuzzi2014} functions,
which has been observed experimentally~\cite{Cheneau2012,Langen2013b}
in the correlation function of a one-dimensional quantum gas.
The emergence of a light cone is due~\cite{Calabrese2006,Foini2012,Marcuzzi2014} to the fact that the entangled quasi-particle pairs
generated by the quench propagate ballistically with a velocity $v$, causing a qualitative crossover
in the behavior of the Green's functions
from short times, at which they behave as in the initial state,
to long times at which the effect of
the quench dominate; this is accompanied by an enhancement of these functions right on the light cone.

Since $G_{R/K}(x,t,t')$ depend separately on the two times $t$ and $t'$, there are two kind of light cones
emerging in their structure: one for $x = t+t'$ and one for $x = t-t'$.
For the Keldysh Green's function $G_K(x,t,t)$ at equal times , the enhancement at the light cone can be physically
interpreted as due to the simultaneous arrival at positions $\xx_1$ and $\xx_2$ of highly entangled
excitations generated by the quench~\cite{Calabrese2006}.
For the specific Hamiltonian in Eq.~\eqref{eq:Hamiltonian} with $u=0$, the present normalization set the velocity of propagation to $v=1$.
While in principle the value of $v$ is affected by the presence of the interaction,
this is not the case up to one loop in perturbation theory, as shown in the rest of this section.
For the retarded Green's function $G_R(x,t,t')$ the enhancement at the light cone can also be again understood from the
ballistic propagation of excitations with velocity $v=1$: a perturbation created at $x=0$ at time $t'$
cannot be felt at position $x$ until the condition $x=|t-t'|$ is obeyed. After this time, the effect of the initial perturbation
decreases upon increasing $t$ for a fixed value of $x$, and so the response function approaches zero.

Exploiting the spatial isotropy and translational invariance, $G_{R/K}(x,t,t')$ in $d$ spatial dimensions
can be calculated from their Fourier transforms $G_{R/K}(k,t,t')$ in Eqs.~\eqref{eq:GR-gauss} and
\eqref{eq:GK-gauss}
as
\begin{equation}
\label{Gft}
\begin{split}
G_{R,K}(x,t,t') &=\frac{1}{x^{d/2-1}\left(2\pi\right)^{d/2}} \!\!\int_0^\Lambda\dd k\, k^{d/2} \\
					&\quad\quad\times J_{d/2-1}(kx) G_{R,K}(k,t,t'),
\end{split}
\end{equation}
where we included a sharp ultra-violet cut-off $\Lambda$ in the integral over $\kk$ and $J_{\alpha}$ indicates the Bessel function of the first kind. This expression is obtained by exploiting the fact the functions $G_{R,K}$ depend only on the modulus $k$ of the wavevector $\kk$: one can then perform a change of variables using hyperspherical coordinates and then integrate over the angular variables\cite{Stein1971}.

In this section we focus on the case of a critical quench, corresponding to a vanishing post-quench value of the parameter $r=0$. Correspondingly, the correlation length $\xi = r^{-1/2}$ diverges, causing the emergence of universal scaling forms and algebraic decays also in the light-cone structure of the Green's functions.
Let us first consider spatial dimension $d=4$. The Gaussian retarded Green's function $G_{0R}^{d=4}(x,t,t')$ follows from Eqs.~\eqref{Gft} and \eqref{eq:GR-gauss}:
\begin{equation}
G_{0R}^{d=4}(x,t,t')= -\frac{1}{4\pi^2 x^3} \int_0^{\Lambda x}\!\! \dd y \, y J_1(y) \sin(y(t-t')/x),\label{grrealspaceG}
\end{equation}
where we assume, for simplicity, $t>t'$. Denoting by $\tau = t-t'$ the difference between the two times, for $\Lambda x\gg 1$ and $\Lambda \tau \gg1$ we find that $G_{0R}$ exhibits a light cone, similarly to what was observed in other models~\cite{Foini2011,Foini2012,Marcuzzi2014}:
\begin{subequations}
\begin{align}
&G_{0R}^{d=4}(x \gg \tau ) = \frac{\Lambda^3}{4\pi^{5/2} \left( \Lambda x \right)^{5/2}} \sin(\Lambda \tau )
\nonumber\\
& \qquad\qquad \qquad\qquad \times \left[ {\sin (\Lambda x) + \cos (\Lambda x)} \right]\simeq 0, \\
&G_{0R}^{d=4}(x = \tau )  = - \frac{ \Lambda^3 }{ 12\pi ^{5/2} \left( \Lambda \tau  \right)^{3/2} }, \label{eq:GR-gauss-lc-2}\\
&G_{0R}^{d=4}(x \ll \tau )  =  - \frac{ \Lambda^2 }{ 4\pi^{5/2} \tau \left( {\Lambda x} \right)^{3/2}}\cos (\Lambda \tau )
\simeq 0.
\end{align}
\end{subequations}
Inside ($x\ll \tau$) and outside ($x\gg \tau$) the light cone, ${G_{0R}}$ vanishes on average due to the rapidly oscillating terms,
while exactly on the light cone $x=\tau=t-t'$ it does not, and actually  is characterized by an algebraic temporal decay $\propto \tau^{-3/2}$.
Accordingly $G_{0R}(x,\tau)$ is peaked at $x=\tau$ and vanishes away from this point in a manner which depends on the ultraviolet (UV) physics.

In Sec.~\ref{sec:light-cone} we show that this basic behavior is preserved even in presence of interactions, provided that the dynamics is in
the collisionless prethermal regime. However, the algebraic decay at $x=|t-t'|$ will be modified in two ways: first, it will depend on whether the initial
perturbation occurs at a short time or at a long time relative to a microscopic time scale $\Lambda^{-1}$, which is not captured by
the Gaussian quench discussed here. Second, the algebraic decay will
acquire corrections described by the anomalous exponent $\theta_N$ for $t'$ at short times.

The Gaussian Keldysh Green's function $G_{0K}^{d=4}(x,t,t)$ at equal times, in real space, and in the deep-quench limit follows from Eqs.~\eqref{Gft} and \eqref{eq:GK-gauss-dq}:
\begin{equation}
\label{eq:GK-real-space-d4}
iG_{0K}^{d=4}(x,t,t) = \frac{\Omega _0}{8\pi^2 x} \int_0^\Lambda  \dd k\,  [ 1 - \cos(2kt)] J_1(kx).
\end{equation}
Evaluating the integral, the light cone is observed to emerge for $\Lambda x\gg 1$ and $\Lambda t\gg 1$, upon crossing the line $x=2t$,
\begin{subequations}
\label{eq:GK-rs-lc}
\begin{align}
&iG_{0K}^{d=4}(x\gg 2t) \simeq  {\cal O}\left(\frac{J_0(\Lambda x)}{x^2}\right) \simeq 0, \label{eq:GK-rs-lc1} \\
&iG_{0K}^{d=4}(x=2t)  \simeq \frac{\Omega_0}{8\pi ^{5/2}}\frac{\Lambda ^2}{(\Lambda x)^{3/2}},
\label{eq:GK-rs-lc2}\\
&iG_{0K}^{d=4}(x\ll 2t) \simeq \frac{\Omega_0}{8\pi^2}\frac{\Lambda^2}{(\Lambda x)^2}.\label{eq:GK-rs-lc3}
\end{align}
\end{subequations}
Outside the light cone, i.e., for $x\gg 2t$, the behavior of $G_{0K}$ is primarily determined by the initial state
as the effect of the quench has not yet set in.
Since the initial state is gapped, with two-point correlations decaying rapidly upon increasing their distance,
$G_{0K}$ vanishes outside the light cone for $\Lambda x \gg 1$.
Inside the light cone ($x \ll 2t$), instead, a time-independent value is obtained, which is characterized by an algebraic spatial decay $\propto x^{-2}$. 
Finally, right on the light cone $x=2t$, the correlator $G_{0K}$ is enhanced, showing a slower algebraic
decay $\propto x^{-3/2}$.

While above we focused on the case $d=4$, it is interesting to study the behavior of $G_{0K}$ in generic spatial dimensionality $d$,
which we compare further below in Sec.~\ref{sec:light-cone} with the results of the perturbative dimensional expansion in the presence of interactions.
Note that $G_{0K}$ outside the light con  vanishes for the reason indicated above; thus we discuss here its behavior
at and inside the light cone.
Instead of regularizing the momentum integrals via a sharp cut-off as we did in Eq.~\eqref{Gft},
we consider below a generic cut-off function $f(k/\Lambda)$ such that $f(x\ll 1) =1$,
with an exponential decay as $x \gg 1$.
In view of the asymptotic form of Bessel functions~\cite{Abramowitzbook}, and noting that
they oscillate in phase with $G_{0K}(k,t,t)$ on the light cone, we find (see Eq.~\eqref{Gft})
\begin{align}
\label{eq:GK-lc-computation}
&iG_{0K}(x=2t)\simeq \frac{\Omega_0}{x^{d/2-1}\left(2\pi\right)^{d/2}}\int_0^{\infty}\!\!\!\dd k\, k^{d/2}
\frac{f(k/\Lambda)}{k^2\sqrt{k x}} \nonumber\\
&\qquad \propto \frac{1}{x^{d-2}}\int_0^{\infty}\!\!\!
\dd y\, y^{(d-5)/2} f\left(\frac{y}{\Lambda x}\right)\nonumber\\
&\qquad \simeq  \frac{1}{x^{d-2}}\int_0^{\Lambda x}\!\!\dd y \, y^{(d-5)/2}\propto \frac{1}{x^{(d-1)/2}}.
\end{align}
Inside the light cone $x\ll 2t$ we find, instead,
\begin{equation}
\begin{split}
&iG_{0K}(x\ll 2t)\simeq\frac{\Omega_0}{x^{d/2-1}\left(2\pi\right)^{d/2}}\\
&\qquad\qquad \times
\int_0^{\infty}\!\!\dd k\, k^{d/2} f(k/\Lambda)J_{d/2-1}(kx)
\frac{1}{k^2}\\
&\propto \frac{1}{x^{d-2}}\int_0^{\infty}\!\!\dd y \, y^{-2+d/2}J_{d/2-1}(y)f\left(\frac{y}{\Lambda x}\right)\propto \frac{1}{x^{d-2}},
\end{split}
\end{equation}
where we replaced  $\sin^2(kt)$ appearing in $G_{0K}(k,t,t)$ (see Eq.~\eqref{eq:GK-gauss-dq}) with its temporal mean value $1/2$ and in the last line the cut-off function $f$ has been replaced by $1$ because the rapidly oscillating Bessel function suppresses the integral at large arguments.
In summary, for a deep quench to the critical point of the Gaussian theory in $d$
spatial dimensions one finds
\begin{subequations}
\label{gk0}
\begin{align}
&iG_{0K}(x\gg 2t) \simeq 0,\\
&iG_{0K}(x=2t)\propto \frac{1}{x^{(d-1)/2}},\label{gk0b}\\
&iG_{0K}(x\ll 2t)\propto \frac{1}{x^{d-2}}. \label{gk0a}
\end{align}
\end{subequations}
The response function on the other hand can be simply derived by using Eqs.~\eqref{eq:GR-gauss-dq} and~\eqref{Gft},
\begin{equation}
G_{0R}(x=t-t')\propto  \frac{1}{x^{(d-1)/2}}\label{gRd}
\end{equation}
and vanishes away from $x=t-t'$ as discussed above.

As anticipated above and shown in Sec.~\ref{sec:light-cone}, these expressions acquire sizable corrections when a
finite value of the post-quench interaction $u$ is included, with corrections taking the form of an anomalous scaling
which modifies the exponents appearing in Eqs.~\eqref{gk0} and~\eqref{gRd}.

\subsection{Keldysh action for a quench}
\label{sec:quench-action}

In the sections which follow we show how the interaction modifies the dynamics after a quench compared to those of the Gaussian model discussed above. To this end, we use perturbation theory and renormalization-group techniques, the formulation of which is particularly simple within the Keldysh functional formalism~\cite{Altlandbook2010,Kamenevbook}. For later convenience, in this section we briefly recall how a quench can be describe within it.

{\subsubsection{Construction of the formalism}}

For simplicity, we assume that the system is prepared in a thermal state described by the density matrix $\rho_0$ introduced after Eq.~\eqref{eq:Hpq-G}, where now $H_0$ is a generic pre-quench Hamiltonian, not necessarily quadratic.
After the quench, the dynamics of the system is ruled by the post-quench Hamiltonian $H$ obtained from $H_0$ by changing some parameters at time $t=0$. Accordingly, the expectation value of any operator $\mathcal O$ can be expressed in the following functional form~\cite{Kamenevbook}:
\begin{equation}
\label{eq:KeldyshAction}
\langle {\mathcal O}(t)\rangle = \int \mathcal{D}\vecphi \;\mathcal{O}[\vecphi_f(t)]\,\ee^{iS_K},
\end{equation}
where $S_K = S_K[\vecphi_I,\vecphi_f,\vecphi_b]$ is the Keldysh action, which is a functional of the fields $\vecphi_f$, $\vecphi_b$ and $\vecphi_I$, referred to, respectively, as \emph{forward}, \emph{backward} and \emph{initial} fields; $\mathcal{D}\vecphi \equiv D[\vecphi_f, \vecphi_b, \vecphi_I]$ is the functional measure.
Note that the r.h.s. of Eq.~\eqref{eq:KeldyshAction} involves $N$-component vector fields $\vecphi_{f,b,I}$, with $\vecphi_f$ replacing the original operator field $\vecphi$ in the formal expression $\mathcal{O}[\vecphi]$ of $\mathcal O$.
The action $S_K= S_s + S_b$ consists of the two parts $S_s$ and $S_b$ corresponding, respectively, to the initial state and the post-quench dynamics, where
\begin{subequations}
\begin{align}
S_s &  = i \int_0^\beta \dd \tau \; \mathcal{L}^E_0(\vecphi_I), \label{eq:SsKeld} \\
S_b & =  \int_0^{\infty}\dd \tau \left[\mathcal{L}(\vecphi_f)-\mathcal{L}(\vecphi_b)\right],\label{eq:SbKeld}
\end{align}
\end{subequations}
where $\mathcal{L}(\vecphi)$ is the Lagrangian associated with the Hamiltonian~\eqref{eq:Hamiltonian}, i.e.,
\begin{equation}
\label{eq:Lagrangian}
\mathcal{L}(\vecphi) = \int_\xx \left[ \frac{1}{2}\dot{\vecphi}^2 - \frac{1}{2}(\nabla \vecphi)^2 - \frac{r}{2}\vecphi^2  - \frac{u}{4!N}\left(\vecphi^2\right)^2\right],
\end{equation}
while $\mathcal{L}^E_0(\vecphi)$ is the Euclidean Lagrangian associated with the pre-quench Hamiltonian, namely
\begin{equation}
\mathcal{L}^E_0(\vecphi) = \int_\xx \left[ \frac{1}{2}\dot{\vecphi}^2 + \frac{1}{2}(\nabla \vecphi)^2 + \frac{\Omega_0^2}{2}\vecphi^2 \right].
\end{equation}

In order to simplify the notation, hereafter the spatial and temporal dependence of the fields is not explicitly indicated.
The fields $\vecphi_f$, $\vecphi_b$ and $\vecphi_I$ in Eq.~\eqref{eq:KeldyshAction} are not actually independent but are related by the boundary conditions
\begin{equation}
\begin{cases}
\vecphi_f(t) = \vecphi_b(t), \\
\vecphi_b(0)  = \vecphi_I(0), \\
\vecphi_f(0) = \vecphi_I(\beta).
\end{cases}
\label{eq:K-bc}
\end{equation}
Note that if the observable $\mathcal O$ in Eq.~\eqref{eq:KeldyshAction} is calculated at $t = 0^-$,
then these boundary conditions reduce to the periodic one $\vecphi_I(0) = \vecphi_I(\beta)$ usually encountered in a Matsubara path-integral\cite{Altlandbook2010}.
The functional representation in Eqs.~\eqref{eq:KeldyshAction}, \eqref{eq:SsKeld}, and \eqref{eq:SbKeld} correspond to the Schwinger-Keldysh contour\cite{Kamenevbook}:
the usual forward-backward branches of the contour of integration are supplemented by an initial branch in imaginary time which appears as a consequence of the initial density matrix $\rho_0$ represented by $S_s$ in Eq.~\eqref{eq:SsKeld}.
The actions $S_s$ and $S_b$ are referred to as \emph{surface} and \emph{bulk} action, respectively, for a reason which will become clear further below.
It is convenient to express the forward and backward fields $\vecphi_{f}$ and $\vecphi_{b}$, respectively, in the so-called retarded-advanced-Keldysh basis\cite{Kamenevbook} via
$\vecphi_f = (\vecphi_c + \vecphi_q)/\sqrt{2}$ and $\vecphi_b =(\vecphi_c - \vecphi_q)/\sqrt{2}$,
where the fields $\vecphi_c$ and $\vecphi_q$ are referred to as \emph{classical} and \emph{quantum} fields, respectively.
In these terms, the retarded and Keldysh Green's functions in Eqs.~\eqref{eq:defGR-gen} and \eqref{eq:defGK-gen} read\cite{Kamenevbook}
\begin{align}
iG_R(|\xx-\xx'|,t,t') & =  \langle \phi_c(\xx,t)\phi_q(\xx',t') \rangle, \label{eq:GR-phic-phiq}\\
iG_K(|\xx-\xx'|,t,t') & =  \langle \phi_c(\xx,t)\phi_c(\xx',t') \rangle,\label{eq:GK-phic-phiq}
\end{align}
where, as before, we do not indicate the indices of the field components of $\vecphi_{c,q}$ because  the $O(N)$ symmetry forces them to be equal for the fields inside the expectation values.
The expression of $G_{R/K}$ given in Eqs.~\eqref{eq:GK-gauss} and \eqref{eq:GR-gauss} can alternatively be calculated  within the functional formalism introduced above, as we discuss in App.~\ref{app:functional-Green}.
For the quench protocol described in Sec.~\ref{sec:protocol}, $S_s$ and $S_b$ can be written as
\begin{align}
\label{eq:SK-s}
S_s 	& = i\int_\xx \,\int_0^{\beta }\dd \tau\, \left[ \frac{1}{2} (\dot{\vecphi}_I)^2 + \frac{1}{2} (\mathbf \nabla \vecphi_I)^2 + \frac{\Omega^2_0}{2}\vecphi_I^2\right], \\
S_b 	& =  \int_\xx \,\int_0^{+\infty}\dd t\,\left[\dot{\vecphi}_q \cdot\dot{\vecphi}_c -  (\mathbf \nabla \vecphi_q)\cdot(\mathbf \nabla \vecphi_c) - r\,\vecphi_q \cdot \vecphi_c \right. \nonumber\\
  	   	& \qquad\qquad \left. - \frac{2u_c}{4!N}(\vecphi_q \cdot \vecphi_c)\vecphi_c^2- \frac{2u_q}{4!N}
(\vecphi_c \cdot \vecphi_q)\vecphi_q^2\right].
\label{eq:SK-b}
\end{align}
Note that the coupling constant $u$ in Eq.~\eqref{eq:Hamiltonian} is indicated in Eq.~\eqref{eq:SK-b} as $u_{c,q}$: although, in principle, $u_c=u_q=u$, the couplings of the terms $(\vecphi_q \cdot \vecphi_c)\vecphi_c^2$ and $(\vecphi_c \cdot \vecphi_q)\vecphi_q^2$ deriving from $\left(\vecphi^2\right)^2$ may behave differently under RG and therefore we denote them with different symbols.
Given that $S_s$ is Gaussian, the functional integral in Eq.~\eqref{eq:KeldyshAction} with an observable $\mathcal O(t)$ at time $t\neq 0$ can be simplified\cite{Calzetta1988} by calculating the integral over $\vecphi_I$.
For convenience, we first rewrite Eq.~\eqref{eq:SK-s} in momentum space, where it reads
\begin{equation}
S_s = \frac{i}{2}\int_\kk  \,\int_0^{\beta }\dd \tau\, \left( \dot{\vecphi}_I^2 + \omega_{0k}^2 \vecphi_I^2\right),
\end{equation}
with $\omega_{0k}$ given in Eq.~\eqref{eq:dispersion-prequench}.
In order to integrate out $\vecphi_I$, we solve the saddle-point  equation with the boundary conditions $\vecphi_I(\beta) = \vecphi_f(0)$ and $\vecphi_I(0) = \vecphi_b(0)$, we insert the solution back in $S_s$ and then we perform the integral in $\tau$.
The resulting action is
\begin{equation}
\label{eq:initial-action}
S_s  = i\int_\kk \, \frac{\omega_{0k}}{2} \left[\vecphi_{0c}^2 \tanh(\beta\omega_{0k}/2)+ \vecphi_{0q}^2 \coth(\beta\omega_{0k}/2)\right],
\end{equation}
where $\vecphi_{0c}\equiv\vecphi_c(t=0)$ and $\vecphi_{0q}\equiv\vecphi_q(t=0)$.
Accordingly, the initial action $S_s$ can be regarded as a functional which weights the value of $\vecphi_{c,q}$ at $t = 0$, and therefore it determines the initial conditions for the subsequent evolution.
In order to show this, e.g., for  $T=0$, one can consider the saddle-point equations for the action $S_K = S_b + S_s$, with $S_b$ and $S_s$ given in Eqs.~\eqref{eq:SK-b} and ~\eqref{eq:initial-action}, respectively, finding
\begin{equation}
\label{eq:BC}
i\,\omega_{0k}\vecphi_c(0)  = 
\dot\vecphi_q(0)
\quad \text{and}\quad
i\,\omega_{0k}\vecphi_q(0) 
= \dot\vecphi_c(0),
\end{equation}
where the time derivatives on the r.h.s. of these equations come from the integration by parts in the bulk action $S_b$.
Note that in the generalized boundary conditions \eqref{eq:BC}, classical and quantum fields are coupled. In addition, for $\omega_{0k}\to +\infty$, Dirichlet boundary conditions are effectively recovered.
Since the quench induces a breaking of time-translational invariance, the Keldysh action in Eq.~\eqref{eq:KeldyshAction} can be formally regarded as the action of a semi-infinite
system\cite{Lubensky1975,Bray1977a,Bray1977b,Diehl1981,Diehl1986,Diehl1997}
in $d+1$ dimensions in which the bulk is described by $S_b$, the $d$-dimensional surface by $S_s$, while $t$ measures the distance from that boundary.
{Note that this is just a suggestive analogy as there is a fundamental difference between these two cases, with far reaching consequences: while in semi-infinite systems the influence of the boundary generically decays exponentially upon increasing the distance from the surface, this does not generically happen in the present non-equilibrium dynamics, due to the oscillatory dependence of the propagators $G_{0R/0K}$ in Eq.~\eqref{eq:GRK-gauss} on time, which is not qualitatively altered by the one-loop corrections considered further below in Sec.~\ref{sec:perturbation-theory}. However, though somehow artificial, such a separation between bulk and boundary, is anyhow useful.}

{\subsubsection{Structure of the resulting action}}

%
%
{Before proceeding with the analysis of the perturbation theory and of the RG flow of the coupling associated with Eqs.~\eqref{eq:SK-s} and \eqref{eq:SK-b}, we would like to comment briefly below on the structure of the corresponding field theory, comparing it with those of theories emerging in similar contexts.}

{In the absence of a quench (and therefore with the appropriate $S_s$, i.e., with the Euclidean Lagrangian density associated with the Hamiltonian in Eq.~\eqref{eq:Hamiltonian}),
%
%
time-translational invariance is not broken and it eventually reduces to an equilibrium thermal field theory\cite{Kamenevbook}  (characterized by a certain symmetry related to the FDT in Eq.~\eqref{eq:FDT}, see, e.g., Ref.~\onlinecite{Sieberer2015})
%
%
at the temperature $\beta^{-1}$ set by the integral along the imaginary time in $S_s$. The resulting field theory is effectively a $d+1$-dimensional one ($z=1$ the dynamical exponent), but confined within a film of thickness $\beta$ and periodic boundary conditions in the direction across the film. Because of these boundary conditions, no breaking of translational invariance occurs and the only complication compared to a bulk theory arises because the zero mode of the transverse fluctuations needs to be accounted for non-perturbatively\cite{ZinnJustinbook,Cardy1988};
%
%
as a result, some quantities display a non-analytic dependence\cite{Sachdev1997,Diehl2006,Diehl2011}
%
%
on the interaction strength $u=u_c=u_q$, while the theory has an upper critical dimensionality $d_c=4$ for $\beta^{-1}\neq 0$, with a subtle dimensional crossover as $\beta^{-1}\to 0$.}

{In the presence of a quench, the situation becomes more subtle due to the breaking of time-translational invariance, and the consequent loss of a number of relationships between the scaling dimensions of the fields $\vecphi_{c,q,I}$ which, as discussed further below in
Sec.~\ref{sec:engineering-dimensions}, may result in different effective theories depending also on the initial state of the quench. Note that this is never the case both in semi-infinite equilibrium systems\cite{Diehl1986,Diehl1997}
%
%
or non-equilibrium bulk systems after a temperature
quench\cite{Janssen1988,Gambassi2005},
as the boundary condition does not affect the effective theory in the bulk.
In the limit of the deep quench we focus on here, power counting indicates, for example, that the resulting effective theory has $u_q=0$, with an upper critical dimensionality $d_c=4$, somehow in analogy to the equilibrium case at finite temperature mentioned above. Note, however, that in the present case, the formal mapping onto a film of a certain thickness does not apply in the form which is valid for equilibrium systems and thus the problem with the zero mode might not arise, as the perturbative calculations in Sec.~\ref{sec:perturbation-theory} suggest.}

{However, a similar mapping was introduced in Refs.~\onlinecite{Calabrese2006,Calabrese2007} for determining the post-quench evolution of observables such as a correlation function and later for a response function in Ref.~\onlinecite{Marcuzzi2014}, where a detailed comparison with the Keldysh formalism discussed above was also presented. Within this formalism, the evolution of correlation functions of a certain observable is obtained from the  analytic continuation of the corresponding correlation function in the system augmented of one bounded spatial dimension, i.e., in a film, with boundary conditions imposed by the initial state of the quench. The analytic continuation of the coordinate across the film to imaginary values then plays the role of the real time of the evolution. However, this mapping is seemingly limited to the case of dynamics long after the quench and therefore is unable to describe the phenomena described here, though an investigation of the relationship between these two approaches would be certainly fruitful, and is well beyond the scope of the present work.}

\subsection{Keldysh Green's function as a propagation of the initial state}
\label{sec:Keldysh-propagation}

In the absence of an initial condition (e.g., assuming the quench to occur at $t=-\infty$),
the bulk action $S_b$ in Eq.~\eqref{eq:SK-b} is characterized by having only the retarded Green's function $G_{0R}$ as the Gaussian propagator. This can be readily seen from the absence of a term $\vecphi_q^2$ in the bulk action \eqref{eq:SK-b} when $u_{q,c}=0$.
(In passing we note that this is the reason why an infinitesimally small term $\propto \vecphi_q^2$ which satisfies the fluctuation-dissipation theorem~\cite{Kamenevbook} has to be added to this action in order to recover the equilibrium Keldysh Green's function $G_K$.)
In the presence of the quench at $t=0$, instead, correlations appear already  in the Gaussian theory  because of the ``forward propagation'' of those which are present in the initial state. In fact, the Keldysh Green's function can be generically written as
\begin{equation}
iG_K(1,2) = \langle \phi_c(1)\phi_c(2) \,\ee^{ - \frac{1}{2}\int_\kk\left( \omega_{0k} \phi^2_{0q} - \dot{\phi}^2_{0q}/\omega_{0k} \right) }\rangle_b,
\label{eq:GK-in-state}
\end{equation}
where $n\equiv (\kk_n,t_n)$ and $\langle\dots\rangle_b$ denotes the average taken on the bulk action only, i.e., without the quench. This expression follows from Eq.~\eqref{eq:KeldyshAction}, while the argument of the exponential from Eqs.~\eqref{eq:initial-action}  and \eqref{eq:BC}, where, for the purpose of illustration, we assumed an initial state with $\beta^{-1}=0$.
Within the Gaussian approximation $u_{c,q}=0$ and in the deep-quench limit the second term in the exponential of Eq.~\eqref{eq:GK-in-state} is negligible compared to the first one, such that this relation takes the simpler form
\begin{equation}
\label{eq:GKvsGR}
iG_K(k,t,t') = \Omega_0 G_R(k,t,0) G_R(k,t',0),
\end{equation}
which follows from applying Wick's theorem and which suggests that, indeed, $G_K$ can be simply regarded as the forward propagation in time via $G_R$ of the correlations in the initial state.
Actually, one can show from the Dyson equations~\cite{Kamenevbook} of the theory (see App.~\ref{app:Dyson}) that Eq.~\eqref{eq:GKvsGR} holds beyond the Gaussian approximation, provided that the Keldysh component of the self-energy vanishes. In the present case, however, such a component is generated in the perturbative expansion at two loops via the sunset diagram and therefore Eq.~\eqref{eq:GKvsGR}  applies only up to one loop.
The physical interpretation of this fact is that, from the RG point of view, the non-vanishing diagrams contributing to the Keldysh Green's function $G_K$ also generate an interaction vertex $\propto \vecphi_q^2$ in the bulk action $S_b$, which are then responsible for the emergence of a $G_K$ within the bulk. More generally, such a vertex in the bulk is also expected to destabilize the pre-thermal state and eventually induce
thermalization within the system~\cite{Mitra2011,Mitra2012,Mitra12b}: in this sense, the validity of
Eq.~\eqref{eq:GKvsGR} can be regarded as the hallmark of the pre-thermal regime.
Note that in the limit $N\to\infty$ of the present model, the Keldysh component of the self-energy vanishes to all orders in perturbation theory, which renders the previous relation an exact identity~\cite{Maraga2015}.

\section{Perturbation theory}
\label{sec:perturbation-theory}

In this Section we consider the perturbative effect of the quenched interaction on the Gaussian theory described in Sec.~\ref{sec:Gaussian-theory}. The interaction $\propto (\vecphi^2)^2$ in the post-quench Hamiltonian \eqref{eq:Hamiltonian} generates the  two vertices $u_c(\vecphi_q \cdot \vecphi_c)\vecphi_c^2$ and $u_q(\vecphi_c\cdot \vecphi_q)\vecphi_q^2$  in the action \eqref{eq:SK-b} of the Keldysh formalism, which we refer to as \emph{classical} and \emph{quantum vertices}, respectively. Though, in principle, $u_{c,q}=u$ we allow $u_c \neq u_q$ in view of the possibility that these two couplings flow differently under RG,
{as it actually happens, being $u_q$ irrelevant for $d>2$,}
see Sec.~\ref{sec:RG-Wilson}.
At one loop, these two vertices cause: (a) a renormalization of the post-quench parameter $r$, which, compared to the Gaussian approximation, shifts its critical value to $r_c \neq 0$; (b) a logarithmic dependence of $G_{R/K}(k,t,t')$ on times $t$, $t'$ and momenta $k$, suggesting the emergence of a corresponding algebraic behavior once the perturbative series is resummed.
These effects are found in the deep-quench limit for $d=4$,
which is recognized to be the upper critical dimensionality of the system in Sec.~\ref{sec:RG-Wilson}.
Although Eq.~\eqref{eq:GKvsGR} applies to the one-loop calculation considered here, it is anyhow instructive to determine separately the correction to both $G_R$ and $G_K$, which is done in Sec.~\ref{sec:Greens-momentum-space}. Based on these results, we analyze in Sec.~\ref{sec:momentum-distribution} the momentum distribution, while in Sec.~\ref{sec:light-cone} we investigate the Green's functions in real space, highlighting the effects on the light cone which we discussed in Sec.~\ref{sec:light-cone-G} within the Gaussian approximation.

\subsection{Green's function in momentum space}
\label{sec:Greens-momentum-space}

The one-loop corrections to $iG_{R/K}$ are due to the tadpole diagrams associated with the two interaction vertices. However,  the absence of a Gaussian propagator of the form $\langle \phi_q\phi_q\rangle$ prevents the quantum vertex  {(which is anyhow irrelevant, see Sec.~\ref{sec:RG-Wilson})} from contributing and therefore the only contribution is due to the classical vertex $2 u_c (\vecphi_q \cdot \vecphi_c)\vecphi_c^2/(4!N)$ via the tadpole:
\begin{equation}
\label{eq:tadpole-diagram}
{\cal T}(t) =
\begin{tikzpicture}[baseline={([yshift=-2ex]current bounding box.center)}]
\coordinate[] (o) at (0,0);
\coordinate[] (l) at (-1,0);
\coordinate[] (r) at (1,0);
\draw[res] (l)  -- (o) node[circle,fill,inner sep=1pt]{};
\draw[thick] (o) -- (r);
\draw[thick] (0,0.35) circle (0.35);
\end{tikzpicture} .
\end{equation}
Here and in the diagrams which follow, a wiggly line is associated with the field $\phi_q$, while a normal one to the field $\phi_c$. Accordingly, a Gaussian retarded Green's function corresponds to a line which starts normal and and becomes wiggly, and viceversa, while the Gaussian Keldysh Green's function is represented by a normal line.
This tadpole involves the integral of $iG_{0K}(k,t,t)$ over the momentum $k$ (corresponding to the closed normal line in the diagram), which has to be calculated with a suitable large-$k$ cutoff $\Lambda$.
In the case of a deep quench, the time-dependent tadpole ${\cal T}(t)$ can be written as
\begin{align}
{\cal T}(t) & = - i  \frac{u_c (N+2)}{12N} \int\!\!\frac{\dd^4 k}{(2\pi)^4}\, f(\omega_k/\Lambda) iG_{0K}(k,t,t) \nonumber\\
  			    &\equiv -i [B_0 + B(t)],
\label{eq:Tpole}
\end{align}
where, by using Eq.~\eqref{eq:GK-gauss-dq},
\begin{equation}
\label{eq:ball-TTI}
B_0 \equiv  \frac{{u_c (N+2)}\Omega_0}{24N} \int\!\!\frac{\dd^4 k}{(2\pi)^4} \frac{1}{\omega_k^2} f(\omega_k/\Lambda)
\end{equation}
and
\begin{equation}
\label{eq:ball-NTTI}
B(t) \equiv - \frac{{u_c (N+2)}\Omega_0}{24N} \int\!\!\frac{\dd^4 k}{(2\pi)^4}\frac{\cos(2 \omega_k t)}{\omega_k^2}f(\omega_k/\Lambda),
\end{equation}
while the function $f(x)$ implements the ultra-violet cut-off as discussed in Sec.~\ref{sec:light-cone-G}. In the following, we assume the convenient smooth function $f(x) = \ee^{-x}$, which is preferable to a {(yet widely used in the literature)} sharp cut-off $f(x) = \vartheta(1-x)$
{as the latter, contrary to the former, is known to produce spurious sub-leading oscillatory factors in time\cite{Maraga2015} and sub-leading long-range correlations in space (see, e.g., Ref.~\onlinecite{Dantchev2003}),
%
%
which can mask the truly universal behaviour. We also verified that the eventual leading scaling behavior is not affected by a different choice of the (smooth) function $f$, providing a further, simple check of its universality.}
For simplicity, we focus below on the case in which the post-quench parameter $r$ is tuned at the value $r=r_c$, such that any macroscopic
length-scale is removed from the correlation functions, which therefore becomes self-similar. In this sense, $r_c$ corresponds to a critical point where the system becomes scale invariant.
As we expect $r_c$ to be perturbatively small in the coupling constant $u_c$ (at least within the dimensional expansion), one can actually set $r=0$ in the perturbative expression for the one-loop correction. Accordingly, from Eqs.~\eqref{eq:ball-TTI} and~\eqref{eq:ball-NTTI}, one finds
\begin{equation}
B_0 = 4\theta_N\Lambda^2, \qquad B(t) =  \theta_N (2\Lambda)^2 \frac{(2\Lambda t)^2-1}{[1+(2\Lambda t)^2]^2},
\label{eq:B0-crit}
\end{equation}
where we introduced, for later convenience, the constant
\begin{equation}
{\theta_N} = \frac{1}{8\pi^2} \frac{N + 2}{96N}\Omega _0 u_c.
\label{thetapert}
\end{equation}
While $B_0$ in Eq.~\eqref{eq:B0-crit} has no finite limit for $\Lambda \to \infty$ and therefore its specific value depends on the specific form of the cut-off function $f$ in Eq.~\eqref{eq:ball-TTI}, $B(t)$ becomes independent of it for $\Lambda t \gg 1$, with
\begin{equation}
B(t\gg \Lambda^{-1}) \simeq \frac{\theta_N}{t^2} .
\label{eq:B0-crit3}
\end{equation}
In order to determine the one-loop correction $\delta G_R$ to $G_R$, the tadpole ${\cal T}(t)$ in Eq.~\eqref{eq:Tpole} has to be integrated with two Gaussian retarded functions $G_{0R}$, i.e.,
\begin{align}
\delta G_R(q,t,t') & =
\begin{tikzpicture}[baseline={([yshift=-2ex]current bounding box.center)}]
\coordinate[] (o) at (0,0);
\coordinate[] (ol) at (-0.8,0);
\coordinate[label={[label distance=2]above:$t$}] (l) at (-1.6,0);
\coordinate[] (or) at (0.8,0);
\coordinate[label={[label distance=2]above:$t'$}] (r) at (1.6,0);
\draw[thick] (l) node[circle,fill,inner sep=1pt]{} -- (ol);
\draw[res] (ol)  -- (o) node[circle,fill,inner sep=1pt]{};
\draw[thick] (o) -- (or);
\draw[res] (or) -- (r) node[circle,fill,inner sep=1pt]{};
\draw[thick] (0,0.35) circle (0.35);
\end{tikzpicture} \nonumber \\
& = \int_0^\infty\!\!\dd \tau\, G_{0R}(q,t,\tau) i{\cal T}(\tau) G_{0R}(q,\tau,t').
\label{eq:dGR-1}
\end{align}
Let us first discuss the contribution to $\delta G_R$ due to the time-independent part $B_0$ of the tadpole $\cal T$, which generates an effective shift of the parameter $r$. In fact, this can be seen by reverting the argument, i.e., by considering the effect that a perturbatively small shift $\delta r$ of the parameter $r$ has on $G_R(q,t,t')$ for $t>t'$.
Since $\delta r$ couples to $-i\vecphi_c \cdot \vecphi_q$ in the exponential factor appearing in Eq.~\eqref{eq:KeldyshAction} (see also Eq.~\eqref{eq:SK-b}) one has, up to first order in $\delta r$,
\begin{equation}
\label{eq:GR-dr}
\begin{split}
G_R(q,t,t')\big|_{r+\delta r} 	
& = G_R(q,t,t')\big|_r  \\
& + \, \delta r \int_0^\infty\!\! \dd \tau  \; G_R(q,t,\tau)\big|_r \,G_R(q,\tau,t')\big|_r,
\end{split}
\end{equation}
where $G_R$ has been expressed as the expectation value in Eq.~\eqref{eq:GR-phic-phiq}.
Comparing Eq.~\eqref{eq:dGR-1} with Eq.~\eqref{eq:GR-dr} one can see that the contribution of $B_0$ is the same as that one of a shift $\delta r$ of the parameter $r$, i.e.,
\begin{equation}
r \mapsto r + B_0 + {\cal O}(u^2).
\label{eq:shift-r}
\end{equation}
Accordingly, the resulting critical value $r_c$ of $r$ becomes $r_c =  - B_0 + {\cal O}(u^2)$.

We consider next the term containing the time-dependent part $B(\tau)$ of the tadpole ${\cal T}(\tau)$ in
Eq.~\eqref{eq:dGR-1}. Because of the causality of the retarded Green's functions $G_{0R}$ in the integrand of this equation, the integration domain in $\tau$ runs from $t'$ to $t$, where henceforth we assume $t>t'$.
Within this domain, $B(\tau)$ is well approximated by Eq.~\eqref{eq:B0-crit3} as soon as  $t$, $t' \gg \Lambda^{-1}$ and, correspondingly, the integral can be easily calculated.
Focussing on a quench to the critical point $r=r_c$,
the correction  $\delta G_R$ to the retarded Green's function $G_R$ beyond its Gaussian expression can be written as
\begin{equation}
\label{eq:dGR-2}
\delta G_R(q,t,t') = -G_{0R}(q,t,t')\theta_N  F_R(qt,qt'),
\end{equation}
with the scaling function
\begin{equation}
\begin{split}
F_R(x,y) & = \frac{\sin(x+y)}{\sin(x-y)}[\text{Ci}(2x) - \text{Ci}(2y) ]  \\
		 & \quad - \frac{\cos(x+y)}{\sin(x-y)}[\text{Si}(2x) - \text{Si}(2y) ],
\end{split}
\end{equation}
where the sine integral $\text{Si}(x)$ and cosine integral $\text{Ci}(x)$ are defined, respectively, as
\begin{equation}
\text{Si}(x) = \int_0^x \dd t\,\frac{\sin t}{t}, \quad \text{Ci}(x) = -\int_x^\infty \dd t\,\frac{\cos t}{t}.
\label{eq:CiSi-def}
\end{equation}
The scaling function $F_R(x,y)$ is a consequence of the absence of length- and time-scales at the critical
point $r=r_c$.
At short times $t$, $t' \ll q^{-1}$, the expansion of the cosine integral $\text{Ci}(x) \simeq \ln x$ renders a logarithmic term and, correspondingly, the leading order of Eq.~\eqref{eq:dGR-2} reads:
\begin{equation}
\delta G_R(q,t>t') =  \theta_N \, (t+t') \ln(t/t').
\end{equation}
When these two times are well separated, i.e., $t'\ll t \ll q^{-1}$, the total retarded Green's function $G_R = G_{0R} + \delta G_R $ reads, to leading order,
\begin{equation}
\label{eq:GR-log}
G_R(q,t,t') = - t \left[1-\theta_N\,\ln(t/t')  + {\cal O}(u^2)\right],
\end{equation}
where we used the fact that, at criticality, $G_{0R}(q,t,t') \simeq (t-t')$ for $t'<t \ll q^{-1}$, see
Eq.~\eqref{eq:GR-gauss-dq-smallk}.
Equation~\eqref{eq:GR-log} suggests that the perturbative series can be resummed such that these logarithms result into an algebraic time dependence
\begin{equation}
G_R(q=0,t\gg t')\simeq - t\left(t'/t\right)^{\theta_N},
\end{equation}
as is actually proven in Sec.~\ref{sec:Callan-Symanzik} with the aid of the RG approach,
{which, inter alia, fixes the value of the still arbitrary coupling constant $u_c$ in Eq.~\eqref{thetapert}, providing a value of $\theta_N$ only as a function of the dimensionality $d$.}
In the analysis above we have assumed that both times $t$ and $t'$ involved in $G_R$ are much longer than the microscopic time scale $\simeq \Lambda^{-1}$. In the opposite case in which the shorter time $t'$ is at the ``temporal boundary'', i.e., $t' \ll \Lambda^{-1} \ll t$ or, equivalently, $t'=0$, it turns out that $G_R$ up to one loop (see App.~\ref{app:singularities})
\begin{equation}
\label{eq:GR-log-div}
G_R(q,t,0) \propto  -t\left[1-\theta_N\,\ln(\Lambda t) \right] ,
\end{equation}
has a logarithmic dependence on $\Lambda t$ and therefore shows a formal divergence as $\Lambda$ grows. This divergence can be regularized only by means of a renormalization, i.e., of a redefinition of the fields appearing in the surface action (see Sec.~\ref{sec:RG-CS}) which is responsible for the emergence of the algebraic scaling at small times suggested by Eq.~\eqref{eq:GR-log}.

Consider now the Keldysh Green's function $G_K$: the one-loop contributions to this function can be expressed in terms of the same
tadpole ${\cal T}(t)$ as the one in Eq.~\eqref{eq:tadpole-diagram} contributing to the retarded Green's function $G_R$, the only difference being in the final integrations.
In particular, $iG_K$ receives two contributions
\begin{align}
\label{eq:dGK-1}
\delta& iG_K(q,t,t')  = \nonumber \\
& =\begin{tikzpicture}[baseline={([yshift=-2ex]current bounding box.center)}]
\coordinate[] (o) at (0,0);
\coordinate[] (ol) at (-0.8,0);
\coordinate[label={[label distance=2]above:$t$}] (l) at (-1.6,0);
\coordinate[] (or) at (0.8,0);
\coordinate[label={[label distance=2]above:$t'$}] (r) at (1.6,0);
\draw[thick] (l) node[circle,fill,inner sep=1pt]{} -- (ol);
\draw[res] (ol)  -- (o) node[circle,fill,inner sep=1pt]{};
\draw[thick] (o) -- (or);
\draw[thick] (or) -- (r) node[circle,fill,inner sep=1pt]{};
\draw[thick] (0,0.35) circle (0.35);
\end{tikzpicture}
 +
\begin{tikzpicture}[baseline={([yshift=-2ex]current bounding box.center)}]
\coordinate[] (o) at (0,0);
\coordinate[] (ol) at (-0.8,0);
\coordinate[label={[label distance=2]above:$t$}] (l) at (-1.6,0);
\coordinate[] (or) at (0.8,0);
\coordinate[label={[label distance=2]above:$t'$}] (r) at (1.6,0);
\draw[thick] (l) node[circle,fill,inner sep=1pt]{} -- (ol);
\draw[thick] (ol)  -- (o) node[circle,fill,inner sep=1pt]{};
\draw[res] (o) -- (or);
\draw[thick] (or) -- (r) node[circle,fill,inner sep=1pt]{};
\draw[thick] (0,0.35) circle (0.35);
\end{tikzpicture} \nonumber \\
& = \int_0^{\infty}\!\!\!\dd \tau \, iG_{0R}(q,t,\tau) {\cal T}(\tau) iG_{0K}(q,\tau,t')  + (t \leftrightarrow  t'),
\end{align}
accordingly to the notation explained after Eq.~\eqref{eq:tadpole-diagram}.
Taking into account the decomposition of ${\cal T}$ in Eq.~\eqref{eq:Tpole} one recognizes,
as in the case of $G_R$, that the term in $\delta iG_K$ which is proportional to $B_0$ is similar to the one which would be generated by a shift $\delta r$ of the parameter $r$ in $G_{0K}$ and therefore it can be absorbed
in a redefinition of this parameter as in Eq.~\eqref{eq:shift-r}.

The explicit calculation (reported in App.~\ref{app:Keldysh-correction}) of the most singular correction $\delta iG_K$ in Eq.~\eqref{eq:dGK-1} renders, to leading order in $q/\Lambda \ll 1$ and for $t$, $t' \gg \Lambda^{-1}$,
\begin{equation}
\label{eq:dGK-2}
\delta iG_K(q,t,t')  = iG_{0K}(q,t,t')\,  \theta_N \left[ 2\ln(q/\Lambda) -  F_K(qt,qt') \right],
\end{equation}
where $\theta_N$ is given in Eq.~\eqref{thetapert}, while $F_K(x,y)$ is a scaling function defined as
\begin{equation}
F_K(x,y)   =  \text{Ci}(2x) + \text{Ci}(2y).
\label{eq:FK}
\end{equation}
Note that the first contribution on the r.h.s.~of Eq.~\eqref{eq:dGK-2} contains a term which grows logarithmically as $\Lambda$ increases.
Differently from the case of the retarded Green's function [see Eq.~\eqref{eq:GR-log-div}], the divergence for $\Lambda \to\infty$ occurs even for finite values of $t$ and $t'$.
This is not surprising, because $G_K$ is proportional to a product of two retarded functions $G_R(t,0)$ with one vanishing time argument [see Eq.~\eqref{eq:GKvsGR}], each of them carrying a logarithmic divergence, see Eq.~\eqref{eq:GR-log-div}.
At short times $t$, $t' \ll q^{-1}$, the scaling function $F_K$ in Eq.~\eqref{eq:dGK-2} reads $F_K(x\ll1,y\ll1) \simeq \ln (xy)$,
with the logarithm coming from the series expansion of the
cosine integral $\mbox{Ci}$ for small arguments [see Eq.~\eqref{eq:CiSi-def}]. Accordingly, the leading behaviour of the total Keldysh Green's function $G_K = G_{0K} + \delta G_K$ at short times is given by
\begin{equation}
iG_K(q,t,t') = iG_{0K}(q,t,t') \left[1 - \theta_N  \ln(\Lambda^2tt')\right].
\end{equation}
As for the retarded Green's function in Eq.~\eqref{eq:GR-log-div}, this expression
suggests that the dependence on time beyond perturbation theory is actually algebraic, of the form
\begin{equation}
iG_K(q=0,t,t') \sim \left(tt'\right)^{1-\theta_N},
\end{equation}
where $G_{0K}(q,t,t')$ at criticality and for $t$, $t'\ll q^{-1}$ is given by
Eq.~\eqref{eq:GK-gauss-dq-smallk}.
The re-summation of the logarithms emerging in perturbation theory will be justified in Secs.~\ref{sec:RG-Wilson} and~\ref{sec:RG-CS}  on the basis of the renormalization-group approach.
In the opposite regime of long times $t \gg q^{-1}$ (or, alternatively, large wavevectors $t^{-1} \ll q\ll \Lambda$), the equal-time Keldysh Green's function $G_K(q,t,t)$ (and, consequently, the momentum density in Eq.~\eqref{eq:nk}) acquires an anomalous dependence on the momentum $q$, as can be seen from the re-summation of the logarithm in Eq.~\eqref{eq:dGK-2}:
\begin{equation}
\label{eq:GK-stationary}
iG_K(q\gg t^{-1}, t, t) \propto q^{-2+2\theta_N},
\end{equation}
where we took into account that $F_K(x\gg 1,y\gg 1) \simeq \sin(2x)/(2x) + \sin(2y)/(2y) $.
Accordingly, at long times, the stationary part of the Keldysh Green's function acquires the anomalous scaling \eqref{eq:GK-stationary}: however, this anomalous scaling is not related to the anomalous dimension of the field $\phi_c(t)$, as it happens in equilibrium systems where the renormalization of $\phi_c(t)$ induces the anomalous dimension $\eta$~\cite{TauberBook2014}.
Instead, the reason for the scaling observed here turns out to be the renormalization of the initial quantum field $\phi_{0q}$: this becomes clear by comparing Eq.~\eqref{eq:GKvsGR} with Eq.~\eqref{eq:GR-log-div}. The initial-slip exponent $\theta_N$ thus characterizes not only the short-time critical behaviour of the Green's functions, but also their long-time form, in contrast to what happens in classical diffusive systems~\cite{Gambassi2005}. This fact can be regarded as a peculiarity of the prethermal state, which retains memory of the initial state as a consequence of the fact that the dynamics induced by the post-quench Hamiltonian is analytically solvable.

\subsection{Momentum distribution}
\label{sec:momentum-distribution}

The one-loop correction to the Gaussian momentum density $\mathcal{N}_{0\kk}$ in Eq.~\eqref{eq:nkbare}
can be straightforwardly calculated on the basis of the definition \eqref{eq:nk}, taking into account the perturbative correction $\delta G_K$
to the Keldysh Green's function $G_K$ in Eq.~\eqref{eq:dGK-2}.
Accordingly, the total momentum density $\mathcal{N}_\kk = \mathcal{N}_{0\kk} + \delta \mathcal{N}_\kk$ at small momenta $k \ll \Lambda $ and microscopically
long times $t \gg \Lambda^{-1}$ reads
\begin{equation}
\mathcal{N}_\kk(t) + \frac{1}{2} = \frac{\Omega_0}{4k}\left\{1 +  2\theta_N [\ln(k/\Lambda) - F_\mathcal{N}(kt)] \right\},
\end{equation}
where the scaling function $F_\mathcal{N}(x)$ is defined as
\begin{equation}
F_\mathcal{N}(x)  = \text{Ci}(2x) - \frac{\sin (2x)}{2x},
\label{eq:FN-mdist}
\end{equation}
with $F_\mathcal{N}(x\ll 1)  \simeq  \ln x$, while $F_\mathcal{N}(x\gg 1)  \simeq  - \cos (2x)/(2x)^2$.
As discussed above in Sec.~\ref{sec:Greens-momentum-space}, the presence of logarithmic corrections suggests that the exact momentum
density exhibits an algebraic dependence on momentum and/or time.
In fact, after a resummation of the leading logarithms --- motivated by the RG approach presented in
Secs.~\ref{sec:RG-Wilson} and \ref{sec:RG-CS} --- the momentum density reads
\begin{equation}
\label{eq:momentum-scaling}
\mathcal{N}_\kk(t)  + \frac{1}{2} \simeq \frac{\Omega_0}{4} \frac{1}{\Lambda}\left(\frac{\Lambda}{k}\right)^{1-2\theta_N} \mathcal{F}(kt),
\end{equation}
where, consistently with the perturbative expansion,
$\mathcal{F}(x) \equiv \exp[-2\theta_N F_\mathcal{N}(x)]$ is a scaling function such that
(see Eq.~\eqref{eq:FN-mdist})
\begin{equation}
\mathcal{F}(x) =
\begin{cases}
x^{-2\theta_N} &\text{for} \quad x \ll 1, \\
1 &  \text{for} \quad  x \gg 1.
\end{cases}
\end{equation}
Accordingly, at a certain time $t$, the momentum distribution $\mathcal{N}_\kk(t)+1/2$ displays
different algebraic behaviors as a function of $k$, i.e.,
the unperturbed one $\sim k^{-1}$ for $k \ll t^{-1}$ and the anomalous one $\sim k^{-1+2\theta_N}$ for $k \gg t^{-1}$,
with the crossover occurring at $k\sim t^{-1}$.

\subsection{Green's functions in real space: light-cone dynamics}
\label{sec:light-cone}

Based on the perturbative expressions reported in the previous section, we focus here on the Green's functions $G_{R,K}$ in real space and determine the corrections to their Gaussian expressions $G^0_{R,K}$ reported in Sec.~\ref{sec:light-cone-G}.
Up to the first order in the perturbative and dimensional expansion they can be written as,
\begin{equation}
G_{R,K} = G_{0R,K} + \delta G_{R,K}^{\delta r} + \delta G_{R,K}^{u} +\delta G_{R,K}^{\epsilon} + {\mathcal O}(u_c^2,\epsilon u_c,\epsilon^2)
\label{eq:exp-Grs}
\end{equation}
where $\delta G^{\delta r}_{R,K}$ is the correction from the renormalization of the parameter $r$, i.e., from the constant part $B_0$ of the tadpole in Eq.~\eqref{eq:Tpole}, $\delta G^u_{R,K}$
contains the universal part coming from the time-dependent part $B(t)$ of the tadpole, while
$\delta G_{R,K}^{\epsilon}$ comes from expanding $G_{0R,K}$ in Eq.~\eqref{Gft} up to the first order in the dimensional expansion, with $\epsilon = 4-d$. Some details of the calculation are given in App.~\ref{app:lc-A}.
In what follows we assume that the systems is poised at the critical value $r=r_c$ discussed after Eq.~\eqref{eq:shift-r},  which cancels exactly the term $\delta G_{R,K}^{\delta r}$ in Eq.~\eqref{eq:exp-Grs}, therefore neglected below.

For the retarded Green's function we find, from Eq.~\eqref{Gft}, that corrections arise only on the light cone, namely (see App.~\ref{app:lc-A})
\begin{equation}
\delta G_{R}^{\epsilon}(x=t-t',t,t') = G_{0R}(x=t-t',t,t') \frac{\epsilon}{2}\ln x
\label{dGeps}
\end{equation}
while, from the Fourier transform of Eq.~\eqref{eq:dGR-2} in hyperspherical coordinates, one finds
\begin{align}
\delta G_R^u & (x,t,t') =  \frac{\theta_N}{4\pi^2x^3} \int_0^{\Lambda x} \!\!\dd y\, y J_1(y) \nonumber\\
& \times \biggl\{ \sin\left(\frac{y(t+t')}{x}\right) \biggl[ {\rm Ci} \left( \frac{2 ty}{x} \right)  -  {\rm Ci}\left(\frac{2 t'y}{x} \right)\biggr] \nonumber\\
& \quad \quad- \cos\left(\frac{y(t+t')}{x} \right) \biggl[  {\rm Si}\left(\frac{2 ty}{x}\right) - {\rm Si} \left( \frac{2 t'y}{x} \right) \biggr] \biggr\},
\label{dGru}
\end{align}
where $\theta_N$ is given by Eq.~\eqref{thetapert}.
For $t' \to 0$, a logarithmic singularity emerges in the previous expression due to ${\rm Ci}(2t'y/x)$ (while $\text{Si}(2 t'y/x)$ vanishes), which may be rewritten as:
\begin{equation}
G_R(x,t\gg t') \simeq G_{0R}(x,t\gg t')\left[ 1 + \theta _N\ln \left( \Lambda t'\right) \right].\label{dgru}
\end{equation}
Combined with Eq.~\eqref{dGeps}, Eq.~\eqref{dgru} and the expression of $G_{0R}$ given in Eq.~\eqref{eq:GR-gauss-lc-2}, this implies
\begin{subequations}
\label{eq:GR-lc}
\begin{eqnarray}
&&G_R(x=t-t',  \Lambda t'\gg 1)\propto t^{- 3/2 + \epsilon/2} , \label{eq:GR-lc1} \\
&&G_R(x=t-t',  \Lambda t'\ll 1) \propto t^{- 3/2 + \epsilon/2}t'^{\theta_N}. \label{eq:GR-lc3}
\end{eqnarray}
\end{subequations}
Accordingly, the presence of interaction affects only the behaviour of $G_R$ on the light cone $x = t-t'$,
provided the initial perturbation was applied at a very short time $t'$,
modifying the Gaussian scaling~\eqref{eq:GR-gauss-lc-2} through an anomalous scaling with respect to $t'$.
{Note, however, that the scaling law $\propto t'^{\theta_N}$ is visible only for $t' \ll \Lambda^{-1}$, i.e., only before a non-universal microscopic time controlled by lattice effects, and therefore its presence is expected to be masked by non-universal contributions which dominate at such short time scales.
}

Turning to $G_K(x,t,t')$, for simplicity we focus here only on its behavior at equal times $t'=t$: this allows us to study the structure of $G_K$ as far as the light cone at $x = t+t' = 2t$ is concerned. Outside this light cone, i.e., for $2t\ll x$, $iG_K$ vanishes as it is essentially determined by
the pre-quench state which is characterized by very short-range correlation. Accordingly, we consider the expression of $iG_K$ only on and inside the
light cone.  In particular, on the light cone we find (see App.~\ref{app:lc-A} for details)
\begin{equation}
\delta G_K^{\epsilon}(x=2t,t,t)= G_{0K}(x=2t,t,t) \frac{\epsilon}{2}\ln x ,
\label{dGKe2}
\end{equation}
while inside it (see App.~\ref{app:lc-A}),
\begin{eqnarray}
\delta G_K^{\epsilon}(x\ll 2t,t,t)=  G_{0K}(x\ll 2t,t,t)\, \epsilon\ln x.
\label{dGKe1}
\end{eqnarray}

The loop correction $\delta G_K^u$ to $G_K(x,t,t)$ follows from the Fourier transform in hyperspherical coordinates of Eqs.~\eqref{eq:dGK-2} and \eqref{eq:FK}, and takes the form
\begin{eqnarray}
&&i\delta G_K^u(x,t,t)=2\theta_N\frac{\Omega_0}{8\pi^2 x} \int_0^{\Lambda} \!\!\dd q\, J_1(q x) \nonumber\\
&& \times \biggl\{ \left[1-\cos(2qt)\right]\left[-{\rm Ci}(2qt)+ \ln (q/\Lambda)\right] \nonumber\\
&&\quad\quad +\sin(2qt){\rm Si}(2qt)\biggr\}.
\label{dGKU}
\end{eqnarray}
From this expression it is straightforward to see that at long times, inside the light cone (see App.~\ref{app:lc-A})
\begin{eqnarray}
i\delta G_K^u(x\ll 2t,t,t)= -2\theta_N G_{0K}(x\ll 2t,t,t).
\label{dGKu1}
\end{eqnarray}
On the light cone $x=2t$, instead, there are no logarithmic corrections to $\delta G_K^u$,
as shown in App.~\ref{app:lc-A}.
Combining Eqs.~\eqref{dGKe1} and Eq.~\eqref{dGKu1} with $G_{0K}$ in Eqs.~\eqref{eq:GK-rs-lc3} we obtain the expression of the scaling behavior of $G_K$ inside the
light cone, while combining Eq.~\eqref{dGKe2} with Eq.~\eqref{eq:GK-rs-lc2}, we obtain that one for the scaling on the light cone. The corresponding final expressions are (see also Eq.~\eqref{eq:GK-rs-lc1}):
\begin{subequations}
\label{eq:GK-lc}
\begin{eqnarray}
&&iG_K(x\gg 2t)\simeq 0, \label{eq:GK-lc1}\\
&&iG_K(x=2t)\simeq t^{-3/2 + \epsilon/2}, \label{eq:GK-lc2}\\
&&iG_K(x\ll 2t)\simeq x^{-2+\epsilon-2\theta_N}. \label{eq:GK-lc3}
\end{eqnarray}
\end{subequations}
In summary, interactions modify the correlation function $G_K(x,t,t)$
as follows: on the light cone $x=2t$, $G_K$ still decays $\propto 1/x^{(d-1)/2}$ upon increasing $x$, as in the Gaussian case (see Eq.~\eqref{gk0b}), while inside the light cone, i.e., for $x\ll 2t$, this decay changes qualitatively compared to that case (see Eq.~\eqref{gk0a}) and becomes faster $\propto 1/x^{d-2+2\theta_N}$.
The latter is consistent with the slower decay~\cite{Chiocchetta2015} of the momentum distribution
$iG_K(q\gg t^{-1})$ upon increasing $q$, compared to its Gaussian expression, found in the previous section, see Eq.~\eqref{eq:GK-stationary}.
{It is also interesting to note that while in quenches of one-dimensional systems ($d=1$), which belong to the Luttinger liquid universality class, interactions
do modify the scaling of $G_K$ on the light cone\cite{Mitra2013}, this is not the case for the present system, at least up to the first order in perturbation theory.}
%

%
\begin{figure}
\centerline{
\setlength{\tabcolsep}{0.05cm}
\begin{tabular}{lc}
\includegraphics[width=5cm]{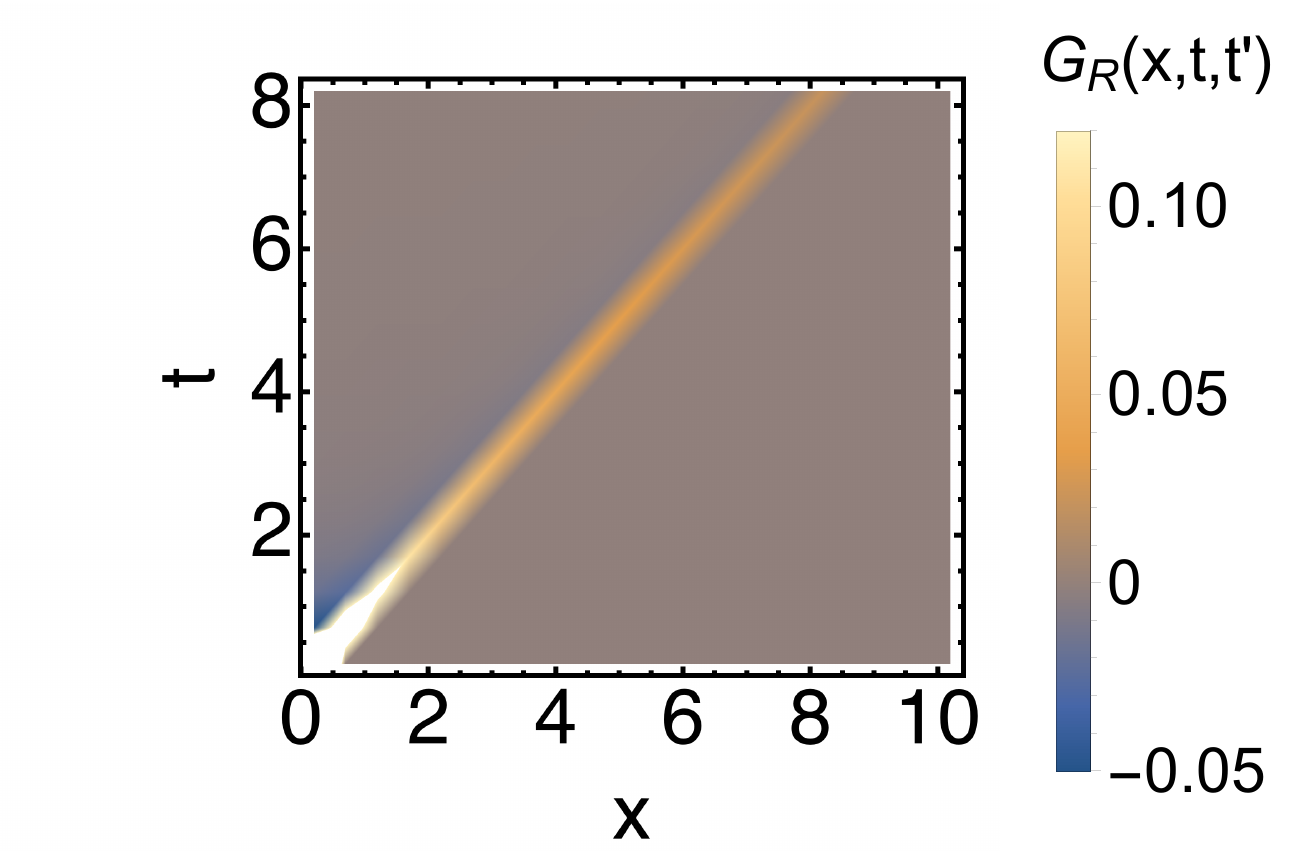} &
\includegraphics[width=4cm]{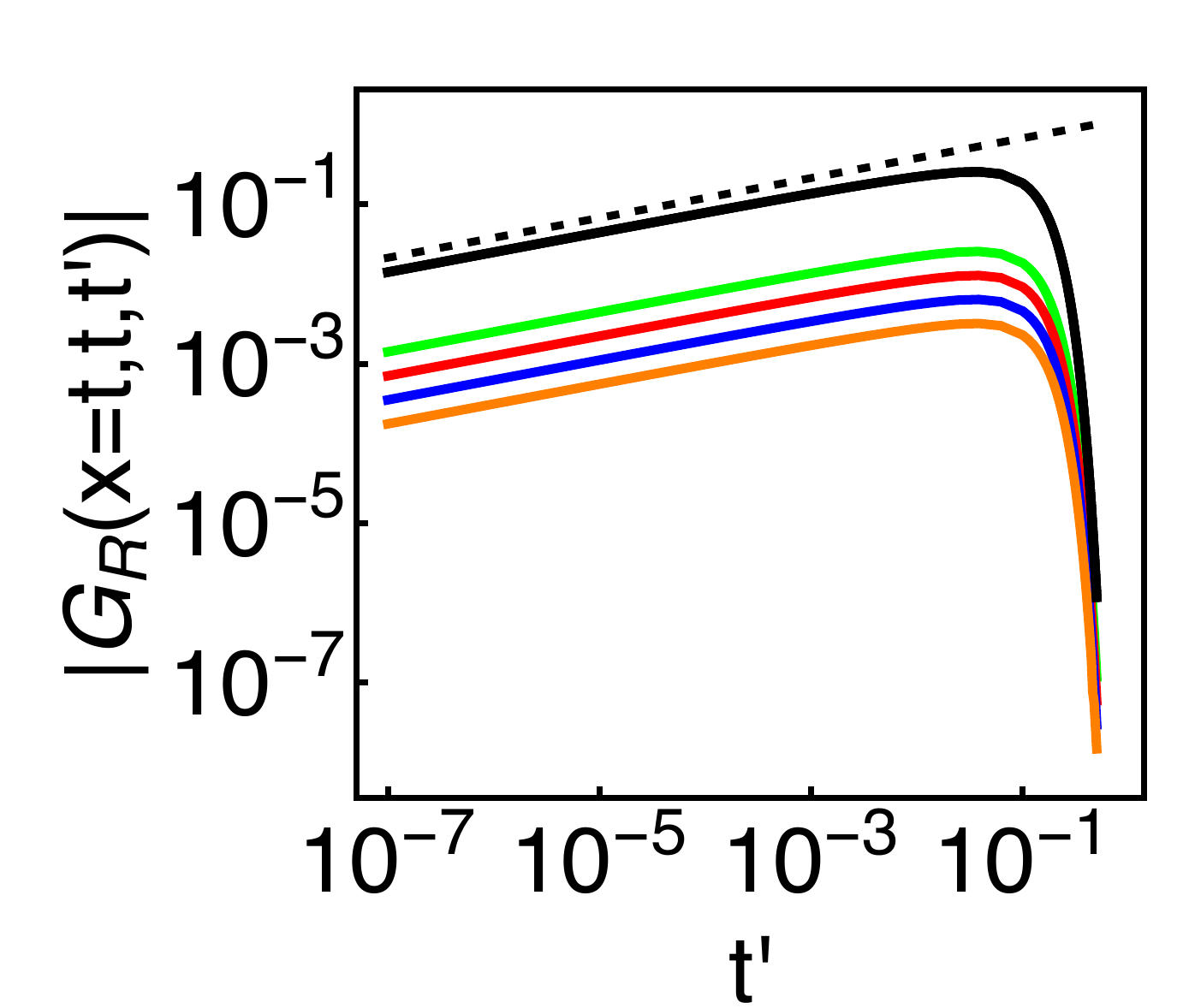}
\end{tabular}
}
\caption{(Color online). $G_R(x,t,t')$ after a quench to the dynamical critical point for the $O(N)$ model with $N\to \infty$ and spatial dimension $d=3$, with an UV cut-off $\Lambda = \pi/2$.
Left panel: $G_R(x,t,t')$ as a function of $x$ and $t$ for $t' = 0.1$. The numerical value of $G_R$ is expressed as indicated by the color code in the legend. Right panel: plot of $G_R(x,t,t')$ on a double logarithmic scale as a function of $t'$ for various values of $x=t$. The coloured solid lines correspond to sections with $t = 80$ (lower, orange line), $t = 40$ (lower-middle, blue line), $t=20$ (upper-middle, red line), and $t=10$ (upper, green line), respectively. The solid black line corresponds to a rescaling of the solid coloured line with $t$, i.e., to $tG_R(x,t,t')$, over which the coloured solid lines collapse.
For small values of $t'$, $G_R(x,t,t')$ displays an algebraic growth $ \propto {t'}^{0.25}$, as indicated by the corresponding dashed line, which is in agreement with Eq.~\eqref{eq:GR-lc3}.}
\label{fig:lightcone-GR}
\end{figure}

\begin{figure}
\centerline{
\setlength{\tabcolsep}{0.05cm}
\begin{tabular}{lc}
\includegraphics[width=4.7cm]{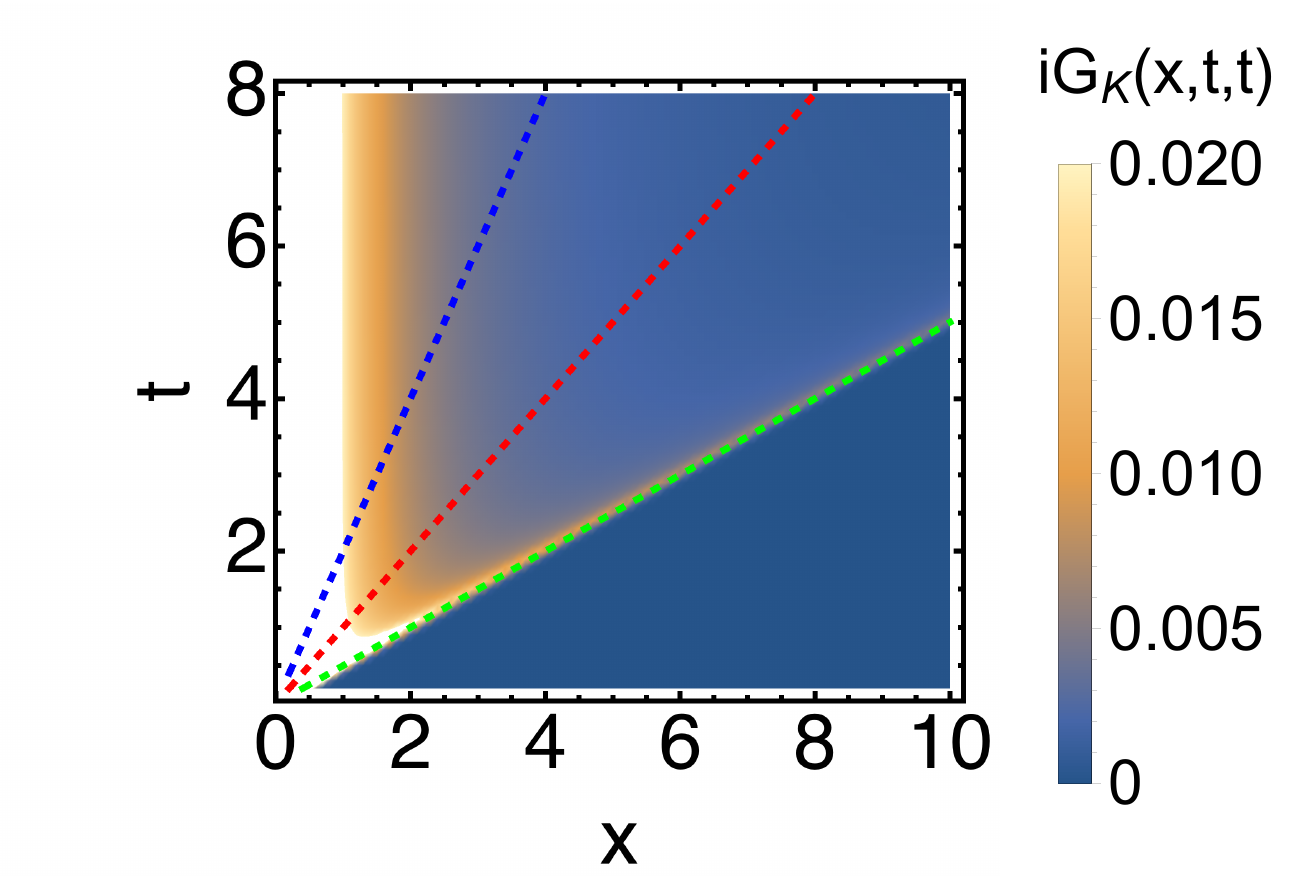} &
\includegraphics[width=3.8cm]{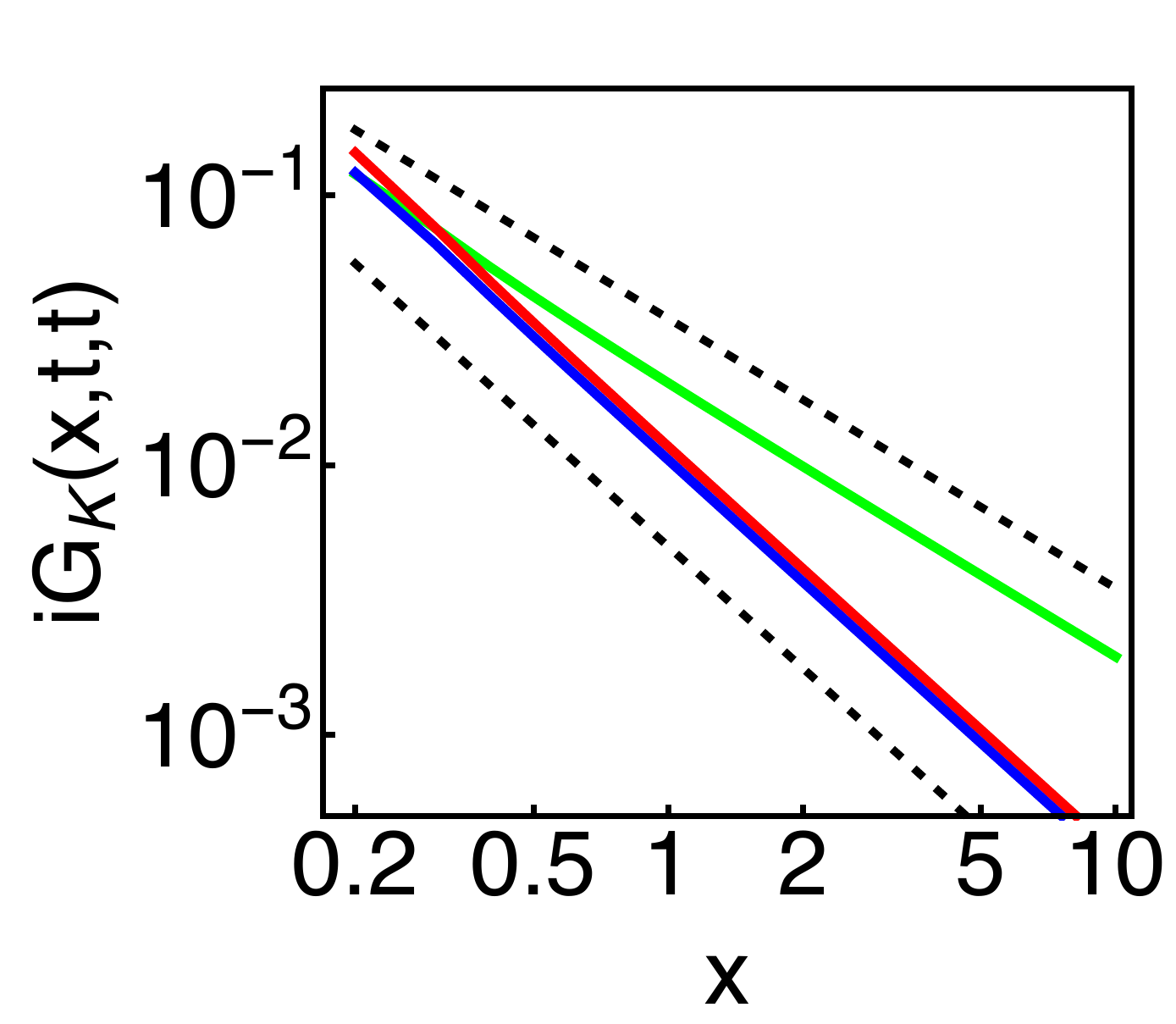}
\end{tabular}
}
\caption{(Color online). $G_K(x,t,t)$ after a quench to the dynamical critical critical point for the $O(N)$ model with $N\to \infty$ and spatial dimension $d=3$, with an UV cut-off $\Lambda = \pi/2$.
Left panel: $G_K(x,t,t)$ as a function of $x$ and $t$. The numerical value of $G_K$ is expressed as indicated by the color code in the legend.  The dashed lines correspond to the sections $x = t/2$ (upper, blue line), $x= t$ (middle, red line) and $x = 2t$ (lower, green line), respectively, which are highlighted in the right panel. Right panel:  plot of $G_K(x,t,t)$ on a double logarithmic scale as a function of $x$ along the sections of the $(x,t)$-plane highlighted in the left panel, i.e., the dashed lines correspond to the sections $x = t/2$ (lower, blue line), $x= t$ (middle, red line) and $x = 2t$ (upper, green line), respectively. Along the first two and the last sections, $G_K(x,t,t)$ displays an algebraic decay $\propto x^{-3/2}$ and $\propto x^{-1}$, respectively, as indicated by the corresponding lower and upper dashed lines. This behaviour is in agreement with Eqs.~\eqref{eq:GK-lc2} and~\eqref{eq:GK-lc3}.}
\label{fig:lightcone-GK}
\end{figure}

In order to test the validity of the scaling behavior summarized above (see Eqs.~\eqref{eq:GR-lc} and~\eqref{eq:GK-lc}) beyond perturbation theory,
we consider the $O(N)$ model in the exactly solvable limit $N\to\infty$ which has been considered, e.g., in Ref.~\onlinecite{Maraga2015}. In particular, after a deep quench to the critical point, $G_R(x,t,t')$ and
$i G_K(x,t,t)$ can be calculated, respectively, via a Fourier transform of $2\text{Im}\left[f_\kk(t)f^*_\kk(t')\right]$ and $2 |f_\kk(t)|^2$,
where the evolution of the complex coefficients $f_{\mathbf k}$ is determined numerically, according to Eqs.~(13), (7) and (11) of
Ref.~\onlinecite{Maraga2015}.

In Fig.~\ref{fig:lightcone-GR} we report the resulting $G_R(x,t,t')$ in spatial dimension $d=3$, as a function of $x$ and $t$
(with fixed $t' \ll t$) on the left panel, and as a function of $t'$ on the {lightcone $x=t-t' \simeq t$} on the right one.
The light cone of $G_R(x,t,t')$ is clearly visible in the left panel: in fact, upon varying $x$ for fixed $t,t' \ll t$, $G_R(x,t,t')$ vanishes for $x \ll t$ and $x \gg t$, while it grows as it
approaches the light cone. The colored solid curves on the right panel, instead, show that $G_R(x=t,t,t')$ grows algebraically as
$G_R(x=t,t,t') \propto t'^{\theta_N}$ for small values of $t'$ and various values of $t$, in agreement with Eq.~\eqref{eq:GR-lc3},
with the proper value $\epsilon = 1$ and $\theta_N \to \theta_\infty = 1/4$ (see, c.f., Eq.~\eqref{eq:thNinf}) in $d=3$. Moreover,
by rescaling the colored solid curves by $t$, i.e., by plotting $tG_R(x=t,t,t')$ as a function of $t'$ with fixed $t$, all the curves collapse on the master curve indicated by the solid black line,
in agreement with the algebraic dependence on $t$ predicted by Eq.~\eqref{eq:GR-lc3}. Analogous agreement with the scaling behaviors in Eq.~\eqref{eq:GR-lc} is found for $d\neq 3$, showing that these relations hold beyond perturbation theory.

In Fig.~\ref{fig:lightcone-GK} we report $G_K(x,t,t)$ for the same model in spatial dimensionality $d=3$, as a function of $x$ and $t$ on the left panel, while on the right panel $iG_K$
is shown as a function of $x$ along the cuts in the $(x,t)$-plane indicated by the corresponding dashed lines on the left panel.
The light cone of $iG_K(x,t,t)$ is clearly visible in the left panel: in fact, $iG_K(x,t,t)$ vanishes for $x>2t$, while it grows as it approaches either this light cone or the line $x=0$ (where the white color indicates large values compared to those displayed in the rest of the plot and indicated by the color code) from within the region with $x<2t$, and it vanishes upon increasing $x$ and $t$.  The curves on the right panel, instead, show that $G_K(x,t,t) \propto x^{-3/2}$ (lowermost and intermediate solid lines) as $x$ grows within the light cone, i.e., with $x = t/2$ and $x=t$,  while  $G_K(x,t,t) \propto x^{-1}$ (uppermost line) exactly on the light cone $x=2t$, in agreement with Eqs.~\eqref{eq:GK-lc2} and \eqref{eq:GK-lc3}, respectively,  with the proper value $\epsilon = 1$ and $\theta_N \to \theta_\infty = 1/4$ (see, c.f., Eq.~\eqref{eq:thNinf}) in $d=3$.
Analogous agreement with the scaling behaviors in Eq.~\eqref{eq:GK-lc} is found for $d\neq 3$, showing that these relations hold beyond perturbation theory.

\section{Magnetization dynamics}
\label{sec:magnetization}

In the previous sections we focused on the temporal evolution of two-time quantities after a quench, assuming that the $O(N)$ symmetry of the initial state is not broken by the dynamics. However, as it happens in classical systems (see, e.g., Ref.~\onlinecite{Gambassi2005}), universal features are expected to emerge also in the evolution of one-time quantities such as the order parameter, when the symmetry of the initial state is broken by a suitable small external field ${\mathbf h}(\xx)$, which is then switched off at $t>0$.
As we show below, the short-time behavior of the mean order parameter ${\mathbf M}(\xx,t)$ (henceforth referred to as the magnetization) actually provides one additional direct measure of the short-time exponent $\theta_N$ encountered in the previous sections.
In Secs.~\ref{sec:magp} and \ref{maghf} below, we study the evolution of ${\mathbf M}$ in perturbation theory and by solving the Hartree-Fock equations (which  are exact in the limit $N\to\infty$), respectively.

\subsection{Magnetization dynamics from perturbation theory}
\label{sec:magp}

The inclusion of a symmetry-breaking field in the pre-quench Hamiltonian $H_0$ induces a non-vanishing value of the order parameter ${\mathbf M}(\xx,t) \equiv \langle \vecphi(\xx,t)\rangle$, which evolves after the quench.
If the post-quench Hamiltonian $H$ is tuned to its critical point,
one expects this evolution to be characterized by universal exponents.
More precisely, in this section we investigate the effect of modifying $H_0 =H(\Omega_0^2,0)$ in Eq.~\eqref{eq:Hamiltonian} as:
\begin{equation}
\label{eq:Hamiltonian-SB}
H_0 = \int_\xx \, \left[ \frac{1}{2}\vecpi^2 + \frac{1}{2}(\mathbf \nabla \vecphi)^2 + \frac{\Omega_0^2}{2}\vecphi^2 - {\mathbf h}\cdot \vecphi  \right],
\end{equation}
where ${\mathbf h} \equiv {\mathbf h}(\xx)$ is the external field which breaks explicitly the $O(N)$ symmetry of $H_0$ and therefore of the initial state.
Within the Keldysh formalism, this corresponds to a change in the initial action \eqref{eq:initial-action}, which now reads, in momentum space:
\begin{align}
\label{eq:initial-action-M}
S_s
& = i\int_\kk \, \frac{\omega_{0k}}{2} \left[\left(\vecphi_{0c,\kk}-\sqrt{2}\frac{{\mathbf h}_\kk}{\omega^2_{0k}}\right)^2 \tanh(\beta\omega_{0k}/2)\right. \nonumber \\
& \qquad \qquad \left.+ \vecphi_{0q,\kk}^2 \coth(\beta\omega_{0k}/2) - 2	\beta \frac{{{|\mathbf h}_\kk|}^2}{\omega^3_{0k}}\right],
\end{align}
where $\mathbf{h}_\kk$ indicates the Fourier transform of $\mathbf{h}(\xx)$ and $\omega_{0k}$ is given in Eq.~\eqref{eq:dispersion-prequench}.
Recalling that the integration by parts of $\dot{\vecphi}_q\cdot\dot{\vecphi}_c$ in the bulk action \eqref{eq:SK-b} generates a term proportional to $\dot{\vecphi}_{0q}\cdot\vecphi_{0c}$ in the initial action $S_s$, one can easily see from the change of variables $\vecphi_{0c,\kk} \to \vecphi_{0c,\kk} + \sqrt{2}{\mathbf h}_\kk/\omega^2_{0k}$ that
taking the functional average $\langle \dots \rangle_{\mathbf h}$ with ${\mathbf h} \neq {\mathbf 0}$ is equivalent to calculating it with  ${\mathbf h} = {\mathbf 0}$, but with a modified weight, i.e.,
\begin{equation}
\label{eq:magnetization-insertion}
\langle \dots \rangle_{\mathbf h} = \langle \dots \ee^{-i\sqrt{2}\int_\kk {\mathbf h}_\kk  \cdot \dot{\vecphi}_{0q,\kk}/\omega^2_{0k}}\rangle_{\mathbf h=0}.
\end{equation}
This means that any average in the presence of the field ${\mathbf h}$ can be calculated as in its absence by considering the insertion of the operator shown in the r.h.s. of  Eq.~\eqref{eq:magnetization-insertion}. Let us first consider the Gaussian approximation, i.e., set $u=0$.
Using either Eq.~\eqref{eq:magnetization-insertion} or by repeating the calculations of
Sec.~\ref{sec:Gaussian-theory} with the pre-quench Hamiltonian \eqref{eq:Hamiltonian-SB}, one finds that
\begin{equation}
\label{eq:M-gauss}
\langle \vecphi_\kk(t) \rangle =  \frac{{\mathbf h}_\kk}{\omega_{0k}^2}\cos (\omega_k t).
\end{equation}
Moreover, while $G_R$ is not affected by ${\mathbf h}\neq {\mathbf 0}$, the Keldysh Green's function is modified because of the presence of a non-vanishing $\langle \vecphi_\kk(t) \rangle$ as:
\begin{equation}
\begin{split}
\langle \{\phi_{i,\kk}(t),\phi_{j,\kk'}(t')\}\rangle & = \delta_{i,j}\,\delta_{\kk,-\kk'} iG_K(k,t,t')  \\
&	\qquad 	+ 2\langle \phi_{i,\kk}(t) \rangle \langle \phi_{j,\kk'}(t') \rangle,
\end{split}
\end{equation}
where $G_K(k,t,t')$ is the same as in Eq.~\eqref{eq:GK-gauss}, i.e., the Keldysh Green's function corresponding to $\langle \vecphi_\kk(t) \rangle = {\mathbf 0}$. In the following, we assume the initial external field ${\mathbf h}$ to be spatially homogeneous and aligned with the first component of the field, i.e.,
\begin{equation}
\label{eq:hi-init}
h_i(\xx) = \delta_{i,1}M_0 \Omega_0^2.
\end{equation}
Note that the residual  symmetry of the initial state under $O(N-1)$ transformations involving the components $i=2,\ldots,N$ of the field which are transverse to $\mathbf{h}$ (collectively indicated by $\bot$) implies that the only non-vanishing post-quench component of the magnetization is the one along $i=1$ (referred to as longitudinal and indicated by $\|$), i.e.,
\begin{equation}
\label{eq:sb-defM}
M_i(\xx,t) = \delta_{i,1} M(t).
\end{equation}
Accordingly, from Eq.~\eqref{eq:M-gauss}, one finds
\begin{equation}
M(t) = M_0\cos (\sqrt{r}t)\label{eq:Mgaussian},
\end{equation}
which shows that the order parameter $M$ oscillates indefinitely with angular frequency $\sqrt{r}$ and that at the Gaussian critical point $r=0$ it does not evolve.

Beyond the Gaussian approximation $u=0$ considered above, one has to take into account the presence of interaction in the post-quench Hamiltonian, which affects the temporal evolution of $M$ in Eq.~\eqref{eq:sb-defM}.
Focussing, for concreteness, on the critical point $r=r_c$ (with $r_c\neq 0$ due to the presence of interaction, see Sec.~\ref{sec:Greens-momentum-space}, in particular Eq.~\eqref{eq:shift-r})
and on a deep quench, the same perturbative expansion illustrated in Sec.~\ref{sec:perturbation-theory} renders, at one loop, a correction $M(t) = M_0 + \delta M(t)$ with
\begin{equation}
\label{eq:M1loop}
\begin{split}
\delta M (t) & = \int_0^{t} \dd t' \,G_{0R}(k=0,t,t') \\
&\qquad \qquad\times \left[B(t')M_0 + \frac{2u_c}{4!N} M_0^3 \right],
\end{split}
\end{equation}
where only the time-dependent part $B(t')$ of the tadpole $i\mathcal{T}(t')$ in Eq.~\eqref{eq:Tpole} contributes to Eq.~\eqref{eq:M1loop} because the time-independent contribution $B_0$ is cancelled by assuming $r=r_c$, as explained for the correlation functions in Sec.~\ref{sec:Greens-momentum-space}. 
Taking into account the explicit expression of $B(t)$ in Eq.~\eqref{eq:B0-crit} and of $G_{0R}$ in Eq.~\eqref{eq:GR-gauss-dq}
it is easy to find that $\delta M$ exhibits a logarithmic dependence on $t$ for $\Lambda t \gg 1$, and a quadratic growth in time:
\begin{equation}
\label{eq:dM}
\delta M(t) \simeq M_0 \theta_N \ln (\Lambda t) - M_0 \frac{u_c}{4!N}(M_0 t)^2,
\end{equation}
where $\theta_N$ is given in Eq.~\eqref{thetapert}.
This expression indicates that the dynamics of the magnetization $M(t)$ is characterized by two regimes: for $t \ll t_i \simeq 1/(M_0\sqrt{u_c})$ the term proportional to $t^2$ is negligible, and therefore $\delta M$ is dominated by the logarithmic term. This suggest that, after a proper resummation of the logarithms {(which is justified further below in Sec.~\ref{sec:RG-CS})}, the magnetization grows algebraically as a function of time as
\begin{equation}
{M(t) \propto M_0 (\Lambda t)^{\theta_N}.}
\label{eq:scal-M-pt}
\end{equation}
On the other hand, when $t \gtrsim t_i$ the second term in the r.h.s. of Eq.~\eqref{eq:dM} becomes dominant, signalling a crossover to a different dynamical regime. Notice that, for these two regimes to be distinguishable, one must require that $\Lambda t_i \gg 1$, i.e., $\Lambda \gg M_0\sqrt{u_c}$.
In Sec.~\ref{sec:magnetization-scaling}, we assess the existence of these two dynamical regimes by showing that a general scaling form can be derived for the magnetization $M$ after a quench to the critical point, which involves the scaling variable associated with $M_0$ as in Eq.~\eqref{eq:scaling-initslip}.

\subsection{Magnetization dynamics from a self-consistent Hartree-Fock approximation}
\label{maghf}

In the previous subsection, the post-quench dynamics of the model in Eq.~\eqref{eq:initial-action-M} has been studied via a perturbative dimensional expansion around the upper critical dimensionality
{$d=d_c$. Within this approach, the temporal evolution of ${\mathbf M}(\xx,t)$ turns out to be compatible with the algebraic law in Eq.~\eqref{eq:scal-M-pt}, which is actually derived in Sec.~\ref{sec:magnetization-scaling} on the basis of suitable RG equations. However, in order to check the emergence of such an algebraic behavior beyond perturbation theory, we consider here the case $N\to\infty$ of the present model which, as discussed in detail in Ref.~\onlinecite{Maraga2015} for ${\mathbf h} = {\mathbf 0}$, is exactly solvable beyond perturbation theory, i.e., for a generic value of the spatial dimensionality $d$. }
%
In fact, in this limit, the Hartree-Fock or self-consistent approximation where the ${(\vecphi^2)}^2$ interaction is approximated by a quadratic ``mean-field'' term $\propto \vecphi^2 \langle \vecphi^2\rangle$ becomes exact.
The latter corresponds to an effective, time-dependent square ``mass'' $r(t)$ which is calculated self-consistently, see, e.g.,
Refs.~\onlinecite{Sciolla2013,Chandran2013,Smacchia2015,Maraga2015}.
{As in Sec.~\ref{sec:magp}, the initial magnetic field $\bf h$ is taken along the direction of the first component of the field, see Eq.~\eqref{eq:hi-init}, and therefore the post-quench order parameter is given by Eq.~\eqref{eq:sb-defM} with $M(0) = \sqrt{N} M_0$, where the multiplicative factor $\sqrt{N}$ is needed in order to have a well-defined limit $N\to \infty$, as discussed below.}

The one-loop equations governing the time evolution of the system are~\cite{Sciolla2013}:
\begin{subequations}
\label{eq:HFNfin}
\begin{align}
\label{eq:HFNfin-1}
& \left[ {\partial _t^2 + r(t) - \frac{u}{3}M^2(t)} \right]M(t)= 0,\\
& \left[ {\partial _t^2 + {q^2} + r(t)} \right.\nonumber\\
& \qquad\left. { - \frac{u}{{6N}}iG_K^\parallel (x = 0,t,t)} \right]G_K^\parallel (q,t,t') = 0,\label{eq:HFNfin-2}\\
& \left[ {\partial _t^2 + {q^2} + r(t) - \frac{u}{3}M^2(t)} \right]G_K^ \bot (q,t,t') = 0,\label{eq:HFNfin-3}\\
& r(t)= r + \frac{u}{2}\left[ M^2(t) + \frac{1}{2N}iG_K^\parallel (x = 0,t,t) \right.\nonumber\\
& \left. \qquad\qquad + \frac{N - 1}{6N}iG_K^ \bot (x = 0,t,t) \right],\label{eq:HFNfin-4}
\end{align}
\end{subequations}
where $G_K^{\|}$ and $G_K^{\bot}$ are the Keldysh Green's functions of the longitudinal and of a generic transverse component of the field, respectively, while $r(t)$ is the time-dependent effective parameter self-consistently determined from Eq.~\eqref{eq:HFNfin-4}

\begin{figure}
\includegraphics[totalheight=6cm]{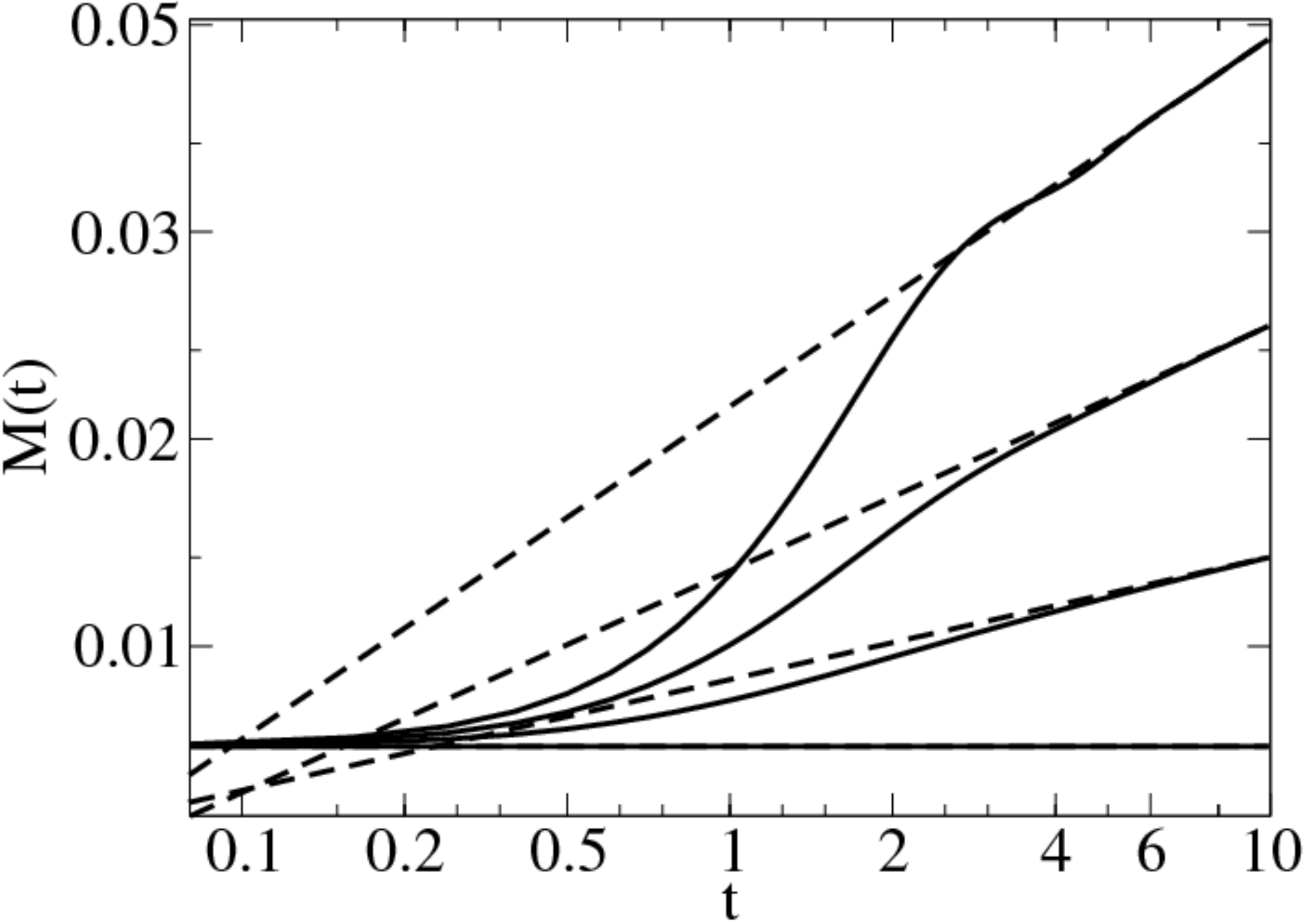}
\caption{Magnetization $M(t)$ as a function of time $t$ (on a double logarithmic scale) for various values of the spatial dimensionality $d = 2.5$, 3, 3.5, and 4, from top to bottom. The solid lines indicate the numerical solution of the self-consistent equations \eqref{eq:HFNinf} with $r=r_c$ and $M_0 = 0.005$.
The dashed lines, instead, indicate the corresponding algebraic increase $\propto t^{\theta _\infty }$
(see Eq.~\eqref{magnetizationanalytic}) with $\theta_\infty$  reported in Eq.~\eqref{eq:thNinf}.
These numerical solutions correspond to the case $\Omega_0=25$, $\Lambda=1$ and to an exponential cutoff function $\ee^{-q}$ introduced for the regularization of the integrals as in Eq.~\eqref{eq:smacchia-ansatz} (see also the discussion in Ref.~\onlinecite{Maraga2015}).
}
\label{phiplot}
\end{figure}

{
To solve this system of equations we need to specify initial conditions for the magnetization and the Green's functions.
Since, from the Heisenberg equations of motion $\dot{\phi} = \Pi$, with $\phi$ and $\Pi$ the components of the fields and its conjugate momentum along $\mathbf{h}$, and since in the initial state $\langle \Pi(t=0) \rangle = 0$, one concludes that $\dot{M}(t)|_{t=0} = 0$.
The initial conditions for the $G_K^{\parallel , \bot }(q,t,t')$, ${\partial _t}G_K^{\parallel , \bot }(q,t,t')$ and  ${\partial _t}{\partial _{t'}}G_K^ \bot (q,t,t')$ are~\cite{Sciolla2013}:
\begin{subequations}
\begin{align}
&G_K^{\parallel , \bot }(q,t = 0,t' = 0) = \frac{1}{\sqrt{q^2 + r_{\parallel , \bot }(0)}},\\
&{\partial _t}G_K^{\parallel , \bot }(q,t = 0,t' = 0) = 0,\\
&{\partial _t}{\partial _{t'}}G_K^{\parallel , \bot }(q,t = 0,t' = 0) = \sqrt {q^2 + r_{\parallel,\bot}(0)},
\end{align}
\end{subequations}
where we introduced
\begin{subequations}
\begin{align}
&r_\parallel(0) = r(0),\\
&r_ \bot(0) = r(0) - \frac{u}{3}M_0^2.
\end{align}
\end{subequations}
The set of Eqs.~\eqref{eq:HFNfin} is not analytically solvable and in general one has to resort to approximations or to a numerical evaluation. However, in the limit $N\to\infty$, one can see from Eqs.~\eqref{eq:HFNfin} that the quadratic term $\propto M^2(t)$ can be neglected for times $t \ll t_i$, with $t_i$ defined by the condition
\begin{equation}
r + \frac{u}{12}iG_K^ \bot (x = 0,t_i,t_i) \simeq \frac{u}{3}M^2(t_i);
\label{deftiv2}
\end{equation}
with these assumptions, Eqs.~\eqref{eq:HFNfin} simplify to:
}
\begin{subequations}
\label{eq:HFNinf}
\begin{align}
& \left[ {\partial _t^2 + r(t)} \right]M(t)= 0, \label{eq:HFNinf-1}\\
& \left[ {\partial _t^2 + {q^2} + r(t)} \right]G_K^{\parallel , \bot }(q,t,t') = 0,\label{eq:HFNinf-2}\\
& r(t) = r + \frac{u}{{12}}iG_K^ \bot (x = 0,t,t). \label{eq:HFNinf-3}
\end{align}
\end{subequations}
The critical value $r_c$ of the parameter $r$ can be obtained via the ansatz suggested in Ref.~\onlinecite{Smacchia2015}, namely:
\begin{equation}
\label{eq:smacchia-ansatz}
r_c =  - \frac{u}{4!}  \int_q   \ee^{ - q/\Lambda }  \frac{ 2 q^2 + \Omega _0}{q^2\sqrt{q^2 + \Omega _0} },
\end{equation}
where the integral is regularized through the smooth cut-off function $\exp(-q/\Lambda)$. This regularization is used to evaluate all the integrals over momentum in the self-consistent Eqs.~\eqref{eq:HFNinf}: the choice of a smooth cutoff is crucial in order to avoid spurious non-universal oscillations which would mask the real universal behaviour of the dynamical quantities~\cite{Maraga2015}.
For $r=r_c$, the effective parameter $r(t)$ at long times $\Lambda t\gg 1$ has (up to rapidly oscillatory terms) the following universal behavior\cite{Maraga2015}
\begin{equation}
r(t) = \frac{a}{t^2},
\label{effectivemass}
\end{equation}
in which the value of $a$ can be obtained analytically and for $2<d<4$ is given by\cite{Maraga2015}
\begin{equation}
a = \frac{d}{4}\left( {1 - \frac{d}{4}} \right).
\end{equation}
By inserting this expression into Eq.~\eqref{eq:HFNinf-1}, we then obtain
\begin{equation}
M(t\gg\Lambda^{-1}) \propto (\Lambda t)^{\theta _\infty },
\label{magnetizationanalytic}
\end{equation}
where~\cite{Maraga2015}
\begin{equation}
\label{eq:thNinf}
\theta _\infty = 1-\frac{d}{4},
\end{equation}
which agrees with Eq.~\eqref{thetapert} in the limit
$N\to\infty$ and up to the first order in $\epsilon = 4-d$ after the fixed-point value of the coupling constant $u_c$ is introduced in that expression, according to what we discuss further below in Sec.~\ref{sec:one-loop-corrections} (see, in particular, Eq.~\eqref{thetaNexpression}).
Accordingly, for times smaller than $t_i$ estimated below (but much longer than the microscopic scale $\Lambda^{-1}$), the order parameter increase is controlled by the universal exponent ${\theta_\infty }$.
The time dependence of $M(t)$ obtained from the numerical solution of Eqs.~\eqref{eq:HFNinf-1}, \eqref{eq:HFNinf-2}, and \eqref{eq:HFNinf-3} with $r=r_c$ and various values of the spatial dimensionality $d$ are shown in Fig.~\ref{phiplot}, where they are also compared with the analytical prediction in
Eq.~\eqref{magnetizationanalytic}.

{
In order to estimate the timescale $t_i$ beyond which Eqs.~\eqref{eq:HFNinf} are no longer valid, we analyse Eq.~\eqref{deftiv2}.
First, note that, as $d \to 4$,  $M(t) \simeq {\cal  A}(\Lambda t)^{\theta_\infty}$ becomes almost independent of $t$ and equal to $M_0$ (see Fig.~\ref{phiplot}) and therefore ${\cal  A} \simeq M_0$.
Hence, for $d$ sufficiently close to 4 we may safely approximate $M(t) \simeq M_0(\Lambda t)^{\theta_\infty} $. Using this approximation in Eq.~\eqref{deftiv2} and solving for $t_i$ with the aid of Eq.~\eqref{effectivemass}, we obtain, to leading order in $\theta_\infty$, the following expression for $t_i$:
\begin{equation}
\label{eq:ti-estimate}
\Lambda t_i \simeq \frac{\sqrt{3} \Lambda }{M_0}\sqrt{\frac{a}{u}}  \propto \frac{1}{M_0}.
\end{equation}
This estimate of $t_i$ is compatible for $d \to 4$ with the scaling derived from the RG analysis in Sec.~\ref{sec:magnetization-scaling}, see Eq.~\eqref{eq:scaling-initslip} .
}

\section{Renormalization group: Wilson approach}
\label{sec:RG-Wilson}

In the previous sections we discussed the emergence of scaling behaviors on the basis of the results of perturbative calculations at the lowest order in the interaction strength, with the sole exception of the exact results presented in Sec.~\ref{maghf}. Here and in Sec.~\ref{sec:RG-CS}, instead, we demonstrate that such scaling behaviors follow from the existence of a fixed point in the renormalization-group (RG) flow which we determine
within the Wilson and Callan-Symanzik approach, respectively. In particular, this allows us to determine the fixed-point value of $u_c$ on which $\theta_N$ in Eq.~\eqref{thetapert} depends and which is not fixed by the perturbative calculations above.

\subsection{Canonical dimensions}
\label{sec:engineering-dimensions}

In order to determine the canonical  dimensions of the relevant coupling constants and parameters in the Keldysh actions \eqref{eq:SK-s} and \eqref{eq:SK-b} which describe the non-equilibrium dynamics of the model, we consider the scaling of the corresponding Gaussian theory in the case of a deep quench $\Omega_0 \gg \Lambda$, with a vanishing temperature $T=0$ of the initial state. Accordingly, the only operators appearing in the initial action $S_s$  in Eq.~\eqref{eq:initial-action} are:
\begin{equation}
\omega_{0k}\,\vecphi_{c,q}^2 = \left[\Omega_0 + \frac{k^2}{2\Omega_0} + O(k^4/\Omega^3_0) \right]\vecphi_{c,q}^2.
\end{equation}
Since we consider $\Omega_0 \gg \Lambda$ and the leading term in the expansion of $\omega_{0k}$ (see Eq.~\eqref{eq:dispersion-prequench}) is the most relevant in the RG sense, we can approximate $\omega_{0k} \simeq \Omega_0$. However, as the classical and quantum fields $\vecphi_{c,q}$ can in principle have different flows under RG, we allow the coefficients of the corresponding terms $\vecphi_c^2$ and $\vecphi_q^2$ in the initial action to be different, denoting them by $\Omega_{0c}$ and $\Omega_{0q}$, respectively.

In order to study the canonical scaling of the model, we introduce an arbitrary momentum scale $\mu$, such that space and time coordinates scale as $\xx \sim \mu^{-1}$ and $t\sim \mu^{-z}$, where $z$ is the dynamical critical exponent with $z=1$ within the Gaussian approximation.
While in thermal equilibrium, the canonical dimensions of the fields and of the couplings follow immediately from requiring the action to be dimensionless,  this requirement is no longer sufficient in the present case, given the large number of fields involved in the total Keldysh action $S_s+S_b$.
Accordingly, in order to fix these dimensions we impose the following two conditions: (i) the canonical dimensions of the fields at the surface $t=0$ are the same as in the bulk $t>0$ and (ii) $\Omega_{0q}$ is dimensionless.
Condition (i) --- which do not necessarily hold beyond the Gaussian approximation --- basically amounts at requiring the continuity of the fields during the quench and, in fact, it is the functional translation of the operatorial conditions $\vecphi(0^-)=\vecphi(0^+)$ and
$\vecpi(0^-)=\vecpi(0^+)$~\cite{Calabrese2006,Calabrese2007,Maraga2015}.
Condition (ii), instead, is based on the fact that $\Omega_{0q}$ plays the role of an effective temperature --- see the discussion in Sec.~\ref{sec:deep-quench} --- and temperatures are known to be scale-invariant in equilibrium quantum systems outside the quantum critical regime\cite{Fisher1988}.
Using these conditions, from Eqs.~\eqref{eq:SK-b} and \eqref{eq:SK-s}
one finds the scaling of the fields
\begin{equation}
\label{eq:GaussianScalingFields}
\vecphi_c(x) \sim \mu^{(d-2)/2}, \qquad \vecphi_q(x) \sim \mu^{d/2},
\end{equation}
and of the couplings
\begin{equation}
\label{eq:GaussianScalingCouplings}
\Omega_{0c} \sim \mu^2, \quad r \sim \mu^2, \quad u_c \sim \mu^{4-d}, \quad u_q \sim \mu^{2-d},
\end{equation}
from which it follows that the upper critical dimensionality is $d=d_c  = 4$. Notice that the quantum vertex $u_q$ is irrelevant for $d > 2$. This conclusion is consistent with the dimensional crossover presented and discussed in Sec.~\ref{sec:deep-quench}: the upper critical dimensionality, which is 3 for a quantum system in equilibrium at $T=0$, is increased to 4 by the effect of a deep quench.
Note, in addition, that  the positive canonical dimension of the classical initial ``mass'' $\Omega_{0c}$ implies that it grows indefinitely under RG and therefore its stable fixed-point value is $\Omega_{0c} = +\infty$.

\subsection{One-loop corrections}
\label{sec:one-loop-corrections}

In order to account for the effects of the interaction, in this subsection we derive the RG equations in perturbation theory up to one loop, based on a momentum-shell integration\cite{Wilson1974}.
It turns out that this is sufficient to highlight the emergence of an initial-slip exponent, while additional effects, such as the emergence of dissipative terms and thermalization,
can be captured only by including contributions at two or more loops, which account for inelastic processes\cite{Giraud2010,Mitra2011,Mitra2012,Mitra12b}.

We assume that the actions in Eqs.~\eqref{eq:SK-s} and \eqref{eq:SK-b} are defined in momentum space with momenta of modulus $k < \Lambda$, where $\Lambda$ is a cutoff set by the inverse of some microscopic length scale, e.g., a possible lattice spacing.
In order to implement the RG transformation in momentum space~\cite{Mabook,Goldenfeldbook}
we decompose each field $\phi(k) \in \{ \vecphi_c, \vecphi_q\}$ as a sum of a ``slow'' and a ``fast'' component $\phi_<(k)$ and $\phi_>(k)$, respectively, i.e., as $\phi = \phi_< + \phi_>$, where $\phi_<(k)$ involves only modes with $0\leq k < \Lambda -\dd\Lambda$, while $\phi_>(k)$ involves  the remaining ones within the  momentum shell $\Lambda -\dd\Lambda \leq k \leq \Lambda$ of small thickness $\dd\Lambda = \delta\ell\, \Lambda$, with $0<\delta\ell \ll 1$.
Then, the fast fields are integrated out from the action: this generates new terms which renormalize the bare couplings of the action of the remaining slow fields. Here we do this integration perturbatively in the coupling constant and therefore the exponential factor in Eq.~\eqref{eq:KeldyshAction} is first split into the Gaussian and interaction parts $S_\mathrm{G}+S_\mathrm{int}$, then expanded up to second order in the interaction terms, with the remaining expectation values of the fast fields calculated with respect to their Gaussian action $S_\mathrm{G}$. Finally, the result is re-exponentiated using a cumulant expansion:
\begin{equation}
\label{eq:RGSK}
\begin{split}
\int \mathcal{D} &[\phi_>,\phi_<] \;  \ee^{i( S_\mathrm{G} +  S_\mathrm{int})} \\
& \simeq \int \mathcal{D} [\phi_>,\phi_<]\; \ee^{i S_\mathrm{G}} \left[ 1+ i\, S_\mathrm{int} -\frac{1}{2} S_\mathrm{int}^2 \right]\\
& \simeq \int \mathcal{D}\phi_<\; \exp\left\{iS^<_\mathrm{G} + i\, \langle S_{\mathrm{int}}\rangle_>  -\frac{1}{2}\langle S_{\mathrm{int}}^2\rangle_>^c  \right\},
\end{split}
\end{equation}
where $S^<_\mathrm{G} = S_\mathrm{G}[\phi_<]$ is the Gaussian action calculated on the slow fields, $\langle\dots\rangle_>$ denotes the functional integration with  respect to $S_\mathrm{G}[\phi_>]$, while the superscript $c$ indicate the connected average. For the case we are interested in,
\begin{equation}
\label{eq:Sint}
\begin{split}
{S_{{\rm{int}}}} = &  - \frac{{2{u_c}}}{{4!N}}\int_{\textbf{x},t>0} ({\vecphi _q} \cdot {\vecphi _c})|{\vecphi _c}{|^2} \\
& -  \frac{{2{u_q}}}{{4!N}}\int_{\textbf{x},t>0} {({\vecphi _q} \cdot {\vecphi _c})|{\vecphi _q}{|^2}}.
\end{split}
\end{equation}
After the integration, coordinates and fields are rescaled in order to restore the initial value of the cutoff:
\begin{equation}
x \to b x, \quad t \to b^z t, \quad \phi_i \to b^{\zeta_i}\phi_i, \quad  \phi_{0,i} \to b^{\zeta_{0,i}}\phi_{0,i},
\end{equation}
where $i = c, q$ and $b = \Lambda/(\Lambda - \dd\Lambda) \simeq 1 + \delta\ell$. The field scaling dimensions $\zeta_i$ and $\zeta_{0,i}$ --- the Gaussian values of which can be read directly from Eq.~\eqref{eq:GaussianScalingFields} --- generically acquire an anomalous contribution due to the interactions. The resulting action for the slow fields (which, after rescaling are renamed as the original fields $\phi$) then acquires contributions from both integration and rescaling and the expression of the new effective parameters (denoted below by $'$) in the action in terms of the original ones constitute the so-called RG recursion relations.

In passing, we mention that, strictly speaking, the functional integral on the l.h.s.~of  Eq.~\eqref{eq:RGSK} is actually equal to one and therefore the renormalization procedure outlined above --- which in equilibrium is typically implemented on the partition function of the statistical system under study --- would look unnecessary. However, this is no longer the case if one either adds externals sources for the field (as done, e.g., in App.~\ref{app:functional-Green}) or, equivalently, implements integration and rescaling directly on correlation functions.

Assuming, for the sake of clarity, a vanishing temperature $T=0$ in the initial state, we find the following RG recursion relations (see App.~\ref{app:Wilson_one_loop} for details) at the lowest order in perturbation theory
\begin{subequations}
\label{eq:RGrecursion}
\begin{align}
r' &= b^2\,\left(r + u_c I_1\right), \\
u_c' &= b^{2-2\zeta_c} \,u_c\, \left(1- u_c I_2\right), \\
u_q' &= b^{2-2\zeta_q} \,u_q\, \left(1- u_q I_2\right),
\end{align}
\end{subequations}
where the integrals $I_{1,2}$ can be decomposed in time-independent and time-dependent parts as $I_{1,2}(t) = I^\text{b}_{1,2} + I_{1,2}^\text{s}(t)$, with the time-independent terms
\begin{subequations}
\label{eq:integral_TI}
\begin{align}
& I_1^{\rm{b}} = \delta\ell\,c_{1,N} \frac{{{\Lambda ^d}}}{{{\omega _\Lambda }}}{K_ + },\\
&I_2^{\rm{b}} = \delta\ell\, c_{2,N} \frac{\Lambda^d }{{\omega _\Lambda ^4}}{\omega _{0\Lambda }},
\end{align}
\end{subequations}
while the time-dependent ones are
\begin{subequations}
\label{eq:integral_TD}
\begin{align}
&I_1^{\rm{s}}(t) = \delta\ell\, c_{1,N} \frac{{{\Lambda ^d}}}{{{\omega _\Lambda }}}{K_ - }\cos (2{\omega _\Lambda }t),\label{eq:integral1_TD}\\
&I_2^{\rm{s}}(t) = \delta\ell\,c_{2,N} \frac{{{\Lambda ^d}}}{{\omega _\Lambda ^3}}\,\Big[ {{K_ - }\, 2{\omega _\Lambda }\,t\,\sin (2{\omega _\Lambda }t)} \nonumber\\
&\qquad \qquad \qquad\qquad \qquad\qquad  { - \frac{{{\omega _{0\Lambda }}}}{{{\omega _\Lambda }}}\cos (2{\omega _\Lambda }t)} \Big],
\label{eq:integral2_TD}
\end{align}
\end{subequations}
where $K_\pm$, $\omega_\Lambda = \omega_{k=\Lambda}$ and $\omega_{0\Lambda} = \omega_{0,k=\Lambda}$, with $\omega_k$ and $\omega_{0k}$ given in Eqs.~\eqref{eq:dispersion-postquench} and~\eqref{eq:dispersion-prequench}, respectively. Notice that, in terms of $\Omega_{0q}$ and $\Omega_{0c}$, the functions $K_\pm$ now reads (see also App.~\ref{app:functional-Green})
\begin{equation}
\label{eq:Kpm-definition}
K_\pm(k) = \frac{1}{2}\left(\sqrt{\frac{k^2+r}{k^2+\Omega^2_{0c}}} \pm \sqrt{\frac{k^2+\Omega^2_{0q}}{k^2+r}}\right).
\end{equation}
In the expressions above we introduced the coefficients
\begin{equation}
\label{eq:c-coefficients}
c_{1,N} = \frac{{N + 2}}{{12N}}{a_d} \quad\mbox{and}\quad c_{2,N} = \frac{{N + 8}}{{24N}}{a_d},
\end{equation}
where $a_d = \Omega_d/(2\pi)^d$, with $\Omega_d = 2\pi^{d/2}/\Gamma(d/2)$ is the $d$-dimensional solid angle, $\Gamma(x)$ being the gamma function.
Note that $I_{1,2}$ are time-dependent quantities as a consequence of the non-TTI term of $G_{0K}$ in Eq.~\eqref{eq:GK-gauss}.

In order to cast the RG equations in a convenient differential form, we take into account that $b \simeq 1+ \delta \ell$, with $\delta\ell \ll 1$ and we keep from Eq.~\eqref{eq:RGrecursion} only the first order in $\delta \ell$.
Neglecting for the time being the time-dependent terms $I^\text{s}_1$ and $I^\text{s}_2$ (see further below for a discussion), we find the differential equations:
\begin{subequations}
\label{eq:RG-unified}
\begin{align}
&\frac{\dd r}{\dd\ell} = 2r + \,u_c\frac{c_{1,N}}{2} \frac{\Lambda^d}{\omega_\Lambda } \left[ \sqrt{\frac{\Lambda ^2 + r}{\Lambda ^2 + \Omega _{0c}^2}}  + \sqrt {\frac{\Lambda ^2 + \Omega _{0q}^2}{\Lambda ^2 + r}}  \right], \label{eq:RG_r}\\
&\frac{\dd u_c}{\dd\ell} = u_c\left[ (d_c - d) - c_{2,N} \, u_c\frac{\Lambda ^d}{\omega _\Lambda^4}\sqrt{{\Lambda ^2} + \Omega _{0q}^2}  \right],\label{eq:RG_uc}\\
&\frac{\dd u_q}{\dd\ell} = u_q\left[ (d_q - d) - c_{2,N} \,u_c\frac{\Lambda ^d}{\omega _\Lambda ^4}\sqrt{\Lambda ^2 + \Omega _{0q}^2}\right]. \label{eq:RG_uq}
\end{align}
\end{subequations}
For later convenience, we defined here the classical and quantum ``upper critical dimensions'' $d_c$ and $d_q$, respectively, as $d_{c,q} = 2 + d - 2\zeta_{c,q}$.

While Eqs.~\eqref{eq:RG-unified} are valid in the generic case, henceforth we focus on a deep quench. As noted in Sec.~\ref{sec:engineering-dimensions}, $\Omega_{0c}$ has a stable fixed point at $\Omega^*_{0,c} = +\infty$ for a deep quench and, correspondingly, the first term in brackets in Eq.~\eqref{eq:RG_r} vanishes.
Moreover, from the Gaussian scaling we find $d_c = 4$, $d_q = 2$, and therefore the quantum vertex is irrelevant for $d>2$ and can be neglected. Finally, 
using the fact that $\Omega_{0q} \gg \Lambda $, Eqs.~\eqref{eq:RG_r} and \eqref{eq:RG_uc} become
\begin{subequations}
\begin{align}
&&\frac{\dd r}{\dd\ell } = 2r + u_c\Omega _{0q}\frac{c_{1,N}}{2}\frac{\Lambda^{4 - \epsilon }}{\Lambda ^2 + r}, \label{eq:RGE1}\\
&&\frac{\dd u_c}{\dd\ell } = u_c\left[ \epsilon  - u_c\Omega _{0q}c_{2,N}\Lambda^{- \epsilon } \right],\label{eq:RGE2}
\end{align}
\end{subequations}
where $\epsilon = 4 - d$, while $c_{1,N}$ and $c_{2,N}$ can be calculated here for $d=4$.
These flow equations admit a nontrivial fixed point $Q_{\rm dy}({\Omega _{0q}})$ in the $(r,u_c)$-plane, which, at the lowest order in the dimensional expansion, is given by
\begin{align}
Q_{\rm dy}({\Omega _0})
& = \left( -\epsilon \Lambda^2 \frac{c_{1,N}}{4 c_{2,N}}, \frac{\epsilon}{\Omega_{0q} c_{2,N}}\right)  \nonumber\\
& =  \left(  - \frac{\epsilon \Lambda^2(N + 2)}{2(N + 8)},\epsilon \frac{192\pi^2N}{\Omega_{0q}(N + 8)} \right),
\label{eq:dynamicalfixedpoint}
\end{align}
which corresponds to the dynamical transition discussed in Sec.~\ref{sec:introduction}. This point turns out to be stable for $\epsilon >0$, while it becomes unstable for $\epsilon < 0$, i.e, $d>d_c$ such that, as it happens in equilibrium, the resulting theory is effectively Gaussian.
Right before Eqs.~\eqref{eq:RG-unified} we mentioned that these equations have been determined by neglecting the time-dependent parts $I_{1,2}^\text{s}$ of the integrals $I_{1,2}$ in Eqs.~\eqref{eq:RGrecursion}. These parts are responsible for rapid oscillations on the temporal scale $\sim \omega_\Lambda^{-1} \simeq \Lambda^{-1}$ set by the microscopic structure of the model, which however do not contribute to time averages of the slow fields on much longer time scales. We emphasize that these terms are not universal as they strongly depend on the regularization. For instance, if a soft cut-off was used instead of a sharp one, these fast-oscillating terms would be replaced by functions which decay smoothly to zero on a temporal scale $\simeq \Lambda^{-1}$ (cf. Ref~\onlinecite{Maraga2015}), and therefore they would clearly not contribute to Eqs.~\eqref{eq:RG-unified}.

Equations~\eqref{eq:RG_r} and~\eqref{eq:RG_uc} are remarkably similar to those for the same model \emph{in equilibrium}
at a finite temperature $T=\beta^{-1}$ (see, e.g., Ref.~\onlinecite{Fisher1988}) which, in turn, admit a non-trivial fixed point $Q_\text{eq}(T)$~\cite{Sondhi1997,Sachdevbook}. In fact, both of them can be cast in the form
\begin{subequations}
\label{eq:RG-unified-general}
\begin{align}
\label{eq:RG-unified1}
\frac{\dd r}{\dd \ell} &= 2r + \frac{c_{1,N}}{2}\,u\,F(r), \\
\frac{\dd u}{\dd \ell} &=  u\left[ (d_c-d) + c_{2,N}\,u\,F'(0)\right],\label{eq:RG-unified2}
\end{align}
\end{subequations}
where $u$ indicates, in equilibrium, the coupling constant of the model in Eq.~\eqref{eq:Hamiltonian}, while out of equilibrium, it stands for $u_c$. For a quench, $F(r)$ reads
\begin{equation}
F^\text{neq}(r) = \frac{\Lambda^d}{\sqrt{\Lambda^2 + r}} \left( \sqrt{ \frac{\Lambda^2 +r}{\Lambda^2+\Omega^2_{0c}} } + \sqrt{ \frac{\Lambda^2 +\Omega^2_{0q}}{\Lambda^2+r}} \right),
\end{equation}
while in equilibrium,
\begin{equation}
F^\text{eq}(r) = \frac{\Lambda^d}{\sqrt{\Lambda^2 + r}} 2 \coth\left(\beta\sqrt{\Lambda^2+r}/2\right).
\end{equation}
The relationship between the thermal and the dynamical fixed points $Q_\text{eq}$ and $Q_\text{dy}$ has been extensively discussed in Ref.~\onlinecite{Chiocchetta2015}.
Note that the upper critical dimension is $d_c = 3$ both in the case of a vanishing initial mass (i.e., $\Omega_{0q}^2 = \Omega_{0c}^2 = r$) and of the equilibrium theory at temperature $T=0$;
instead, the upper critical dimension is $d_c = 4$ for both the case of a deep quench and of equilibrium at a finite temperature.
These similarities provide further support to the observations made in Sec.~\ref{sec:deep-quench} that the stationary state after the quench resembles an equilibrium state, with the initial value $\Omega_{0q}$ of the parameter $r$ in the former case playing the role of an effective temperature in the latter. Accordingly, the deep quench is responsible for a change of the upper critical dimensionality of the dynamical transition, in the very same way as a finite temperature modifies the upper critical dimensionality for an equilibrium quantum phase transition.

Although the analysis here reported only applies in dimensions above the lower critical one $d_l$, this ``dimensional crossover'' induced by the quench appears to be a generic feature of non-equilibrium isolated quantum systems.
In fact, the same effect was shown to appear for a quench in the integrable one-dimensional Ising chain in transverse field~\cite{Fagotti2011,Fagotti2012a,Fagotti2012b}, where the eventual (non-thermal) stationary state does not display any critical behaviour.
Correspondingly, the equilibrium Ising chain in a transverse field~\cite{Sachdevbook} displays a quantum phase transition at $T=0$ (as it can be mapped on a classical Ising model in two spatial dimensions), while no phase transition occurs for $T>0$ (as the corresponding classical Ising model would be in one spatial dimension). In addition, a recent experiment~\cite{Nicklas2015} on a one-dimensional Bose gas with two components suggested the absence of scaling behaviour in the long-time dynamics of the system. In this case, the dynamics was shown to be captured by the effective Hamiltonian~\eqref{eq:Hamiltonian} with $N=1$, which, at equilibrium, is characterized by a quantum phase transition at $T=0$ which however disappears for $T>0$.  Accordingly, we do not expect any DPT in this case.

A remarkable consequence of the structure of RG equations \eqref{eq:RG-unified-general} is that the value of the critical exponent $\nu$ --- which can be obtained as the eigenvalue of the linearized RG flow around the fixed point~\cite{Mabook,Goldenfeldbook}
$u=u^* = -\epsilon/[c_{2,N} F'(0)] + \mathcal{O}(\epsilon^2)$ --- is independent of the specific form of $F(r)$ and it is given by
$\nu^{-1} = 2 - c_{1,N} \epsilon/(2 c_{2,N}) + \mathcal{O}(\epsilon^2)$, i.e., by
\begin{equation}
\nu = \frac{1}{2} + \frac{N+2}{N+8}\frac{\epsilon}{4} + \mathcal{O}(\epsilon^2),
\end{equation}
with $\epsilon = d_c-d$. Accordingly, both the dynamical and the equilibrium transitions are characterized by the same exponent $\nu$, a fact which was already noticed in Ref.~\onlinecite{Smacchia2015} for the $O(N)$ model in the limit $N\to \infty$.
However, one may wonder if also the other independent critical exponent of the equilibrium theory, i.e., the anomalous dimension $\eta$, is the same as the one of the dynamical phase transition. The answer to this question is far from being trivial, since the computation of $\eta$ requires a two-loop calculation which would also generate dissipative terms, responsible also for the destabilization of the prethermal state.

The analysis presented above reveals the existence of the fixed point $Q_\text{dy}$ (see Eq.~\eqref{eq:dynamicalfixedpoint}) of the RG flow, which is approached upon considering increasingly long-distance properties whenever the value of the parameter $r$ in Eq.~\eqref{eq:Hamiltonian} equals a certain critical value $r_c$, which depends, inter alia, on $u_c$ and which corresponds to the stable manifold of the fixed point $Q_{\rm dy}$. Accordingly, the coupling constant $u_c$ which appears in the perturbative calculations discussed in  Sec.~\ref{sec:perturbation-theory} with an unknown value can be substituted with its fixed-point value $u_c^*$ corresponding to $Q_{\rm dy}$. In particular, inserting Eq.~\eqref{eq:dynamicalfixedpoint} into Eq.~\eqref{thetapert}, and noting that $\Omega_{0q}=\Omega_0$, the initial-slip exponent eventually reads:
\begin{equation}
\theta_N = \frac{\epsilon}{4} \frac{N+2}{N+8}  + \mathcal{O}(\epsilon^2).
\label{thetaNexpression}
\end{equation}
The value of $\theta_N$ in the limit $N \to \infty$ agrees with the exact results presented in Ref.~\onlinecite{Maraga2015} and reported here in Eq.~\eqref{eq:thNinf}.

\subsection{Dissipative and secular terms}
\label{sec:dissipative-terms}

The integration procedure outlined in Sec.~\ref{sec:one-loop-corrections} generates a number of terms in addition to those which are present in the action before integration.
Physically, as the system evolves in time, inelastic collisions induce a thermalization process, and the system starts acting as a bath for itself.
It is then natural to wonder whether dissipative terms might emerge~\cite{Mitra2011}. In particular, they would be encoded of the form $ig_1\vecphi_q^2$, $ig^A_2(\vecphi_q \cdot \vecphi_c)^2$ and
$ig^B_2 \vecphi_q^2 \vecphi^2_c$ in the bulk action, which indeed correspond to having, respectively, additive ($g_1$) and multiplicative ($g^{A,B}_2$) Gaussian noise acting on the
system~\cite{Kamenevbook}.
Based on the canonical power counting associated with the deep quench (see Sec.~\ref{sec:one-loop-corrections}, with $d_c=4$), these couplings turn out to scale as
\begin{equation}
g_1 \sim \mu, \qquad g_2^A \sim g_2^B \sim \mu^{3-d};
\end{equation}
accordingly $g_1$ is expected to be relevant in all spatial dimensions while $g^{A,B}_2$ are relevant only for $d<3$. If these terms are generated under RG, their effect is to change significantly the properties of the resulting action and to induce a crossover towards a different fixed point, with a different canonical scaling. However, in order for $g_1$ to be generated one needs to consider at least a two-loop contribution and therefore one is allowed to neglect it in our one-loop analysis.
On the other hand, corrections to $g^{A,B}_2$ are generated at one loop (see App.~\ref{app:Wilson_one_loop}) as
\begin{equation}
\label{eq:secular-term-1}
\delta S_{({\text{diss}})}^{(2)} =  i \int_{\xx,t} \left[ \delta g_2^A(t) \left( \vecphi_c^< \cdot \vecphi_q^< \right)^2 + \delta g^B_2(t)  (\vecphi_c^<)^2 (\vecphi_q^<)^2 \right],
\end{equation}
with
\begin{subequations}
\label{eq:secular-term-2}
\begin{align}
\delta g_2^A(t) & = \delta \ell \, \frac{a_d}{144} \frac{N+6}{N^2} \frac{\Lambda^d}{\omega_\Lambda^2} \left[u_c^2\left(K_+^2 +K_-^2\right)- u_c u_q\right] \, t, \label{eq:secular-term-2a}\\
\delta g_2^B(t) & = \delta \ell \, \frac{a_d}{72} \frac{1}{N^2} \frac{\Lambda^d}{\omega_\Lambda^2} \left[u_c^2\left(K_+^2 +K_-^2\right)- u_c u_q\right] \, t, \label{eq:secular-term-2b}
\end{align}
\end{subequations}
with $K_\pm$ given in Eq.~\eqref{eq:Kpm-definition} and $a_d$ after Eq.~\eqref{eq:c-coefficients}.
In the absence of a quench of the parameter $r$, i.e., with $r=\Omega_0^2$ one has $K_+=1$, $K_-=0$ and $u_c=u_q$; accordingly, the dissipative vertices $\delta g_2^{A,B}(t)$ vanish at this order in perturbation theory, because the quench of the coupling constant $u$ (occurring from a vanishing to a non-vanishing value) does not affect the Gaussian propagators and has consequences only at higher-order terms in perturbation theory.  The dissipation is also absent in the limit $N\to\infty$, as it can be seen from the scaling $\propto N^{-1}$ of the terms in Eq.~\eqref{eq:secular-term-2}: this implies that the prethermal state is actually the true steady-state of the model after the quench, in agreement with the fact that the Hamiltonian $H$ becomes solvable in this limit~\cite{Maraga2015}.
In all the other cases, the dissipative correction increases upon increasing the time $t$ and accordingly, though formally irrelevant for $d>3$, $\delta g_2^{A,B}(t)$ grows indefinitely, spoiling the perturbative expansion. This kind of divergence is due to the presence of the so-called secular terms in the simple perturbative expansion which has been done here. Although several techniques exist to avoid these terms\cite{Berges2004b,Berges2015,Moeckel2008,Tavora13b},
we emphasize that they affect significantly only the long-time dynamics, while being essentially inconsequential as far as the short-time behaviour we are considering here is concerned.
An estimate of the time at which the contributions of Eq.~\eqref{eq:secular-term-2} become important can be obtained by requiring them to be of the same order as the other couplings present in the action. Considering, for simplicity, the values of $r$, $u_c$ and $u_q$ at their fixed points in the deep-quench limit \eqref{eq:dynamicalfixedpoint}, one can readily derive the RG equations for $g_2^{A,B}$ from Eq.~\eqref{eq:secular-term-2}.
Then, by identifying the dimensionless RG flow parameter $\ell$ with the time $t$ elapsed since the quench, i.e., by setting
$\ell = \ln(\Lambda t)$, one finds
\begin{align}
\Lambda^{d-3}\frac{\dd g_2^A}{\dd(\Lambda t)} = \epsilon^2 \frac{N+6}{(N+8)^2} \frac{2}{a_d} , \\
\Lambda^{d-3}\frac{\dd g_2^B}{\dd(\Lambda t)} = \epsilon^2 \frac{1}{(N+8)^2} \frac{1}{a_d} .
\end{align}
Accordingly, for large $N$, the dissipative couplings $g_2^{A,B}(t)$  become of order one at time scales $\Lambda t_A^* \sim N/\epsilon^2$
and $\Lambda t_B^* \sim N^2/\epsilon^2$, respectively.
{
Note, however, that corrections to $g_2^{A,B}$ are generated also for $d>4$ (i.e., $\epsilon<0$), when the coupling $u_c$ is irrelevant: in fact, $u_c$ still generates anyhow terms which are responsible for thermalization, providing thus an instance of a sort of \emph{dangerously irrelevant} operator~\cite{Mitra2011}. In this case, the time scales $t_A^*$ and $t_B^*$ are modified to $\Lambda t_A^* \sim N/(u_c^0)^2$ and $\Lambda t_B^* \sim N^2/(u_c^0)^2$, where $u_c^0$ indicates the microscopic value of the coupling constant $u_c$.
Accordingly, for large values of $N$ or small values of $\max\{\epsilon,u_c^0\}$, the dissipative vertices can be disregarded for quite long times after the quench.
}

\section{Renormalization group: Callan-Symanzik approach}
\label{sec:RG-CS}

In spite of its transparent physical interpretation, the Wilson RG discussed in the previous section is not very practical for carrying out actual calculations. In fact, as we showed in Section~\ref{sec:RG-Wilson}, it still leaves a certain degree of arbitrariness in fixing the dimensions of all the fields and therefore one has to supplement the analysis with additional physical arguments, as we did in order to fix the upper critical dimensionality to $d_c=4$. More severely, Wilson RG is limited in providing predictions of some critical exponents beyond one loop\cite{Wegner1973} and therefore different techniques have been developed for extending the RG  analysis.
In view of these difficulties, in this section we discuss the issue of the emergence of the short-time universal scaling within an alternative renormalization scheme, inspired by the well-established field-theoretical approach (see, e.g., Ref.~\onlinecite{TauberBook2014}).
The idea is to regularize the correlation functions of the relevant fields by redefining the couplings and the fields so as to remove the terms in perturbation theory which would be divergent upon increasing the ultraviolet cutoff $\Lambda$, suitably introduced into the original (bare) action.  As a result, the effective action as well as all the correlation functions generated by it, once expressed in terms of these renormalized couplings and fields, is free of divergences. This renormalized action depends on the arbitrary scale $\mu$ which defines the momentum scale of the effective theory. Using the fact that the original correlation functions --- determined by the bare action --- are actually independent of $\mu$, one can derive a  Callan-Symanzik flow equation for the correlation functions which involves also the RG equations for the relevant couplings of the theory.
Note that, because of the the breaking of TTI induced by the quench, it is no longer viable to define a Fourier transform of the correlation functions and consequently the Callan-Symanzik equations discussed below will be derived for the correlation functions rather than the vertex functions which are usually introduced when TTI is not broken.

\subsection{Renormalization of the initial fields}
\label{sec:RG-Green}

As anticipated above, the divergences upon increasing the cutoff $\Lambda$ which emerge in perturbation theory, can be cancelled by a suitable renormalization of coupling constants and fields: in practice, this is achieved by calculating the relevant correlation functions at particular values of the times and momenta --- the so-called normalization point (NP) --- of the involved fields and by absorbing the resulting divergences in a redefinition of coupling and fields. In particular, we define the NP as  times $t = \mu^{-1}$ and vanishing momenta $q =0$ for all fields involved in the correlation function.
For simplicity, we focus below on the critical point $r=r_c$, which, as shown in Sec.~\ref{sec:perturbation-theory}, is shifted away from its Gaussian value $r_c=0$ because of the interaction.

We consider first the retarded Green's function $G_R(k,t,t') = \langle \phi_c(t)\phi_q(t')\rangle$, where, for the sake of simplicity, the indices of the field components have been suppressed and the fields inside the correlators are consequently assumed to refer to the same component
For simplicity, the dependence on $k$ is understood on the l.h.s. of the previous equality. As shown in App.~\ref{app:singularities}, $G_R$ is finite in the formal limit $\Lambda\to\infty$ as long as both times $t>t'$ do not vanish, while it  grows $\propto \ln (\Lambda t)$ when the earlier time $t'$ vanishes, i.e., when the involved quantum field is the one at the boundary $\vecphi_{0q}$.
In fact, at the normalization point,
\begin{equation}
\label{eq:GR1R}
\langle \phi_c(t)\phi_{0q}\rangle\big|_\text{NP}  = \left[ 1 - \theta_N\,\ln (\Lambda/\mu) \right]\times\text{finite  terms},
\end{equation}
where by ``finite terms'' we mean an expression which is finite as $\Lambda\to\infty$.
In order to reabsorb the logarithmic divergence in Eq.~\eqref{eq:GR1R}, we redefine the initial quantum field as $\vecphi_{q0} = Z_0^{1/2}\vecphi^R_{0q}$, where $\vecphi_{q0}^R$ is the corresponding renormalized field and the renormalization constant $Z_0$ is determined such that the renormalized retarded Green's function $G_R^R(t,0) \equiv \langle \phi_c(t)\phi^R_{0q} \rangle = Z_0^{-1/2} G_R(t,0)$, remains finite as $\Lambda\to \infty$.
This requires
\begin{equation}
\label{eq:Z0}
Z^{1/2}_0 = 1 - \theta_N \ln (\Lambda/\mu),
\end{equation}
up to the lowest order in perturbation theory considered in this work.
Since Eq.~\eqref{eq:GKvsGR} holds up to the same order, the renormalized Keldysh Green's function $G^R_K$ can be simply defined as:
\begin{equation}
\label{eq:GK-ren}
G^R_K(k,t,t') = \Omega_{0q} G_R^R(k,t,0)G_R^R(k,t',0) = Z_0^{-1} G_K(k,t,t').
\end{equation}
As shown in Sec.~\ref{sec:Greens-momentum-space},
the perturbative correction to $G_K(k,t,t')$ generates a term $\propto \ln (\Lambda/\mu)$ for generic values of the times $t$ and $t'$. After the introduction of the renormalization constant as in Eq.~\eqref{eq:GK-ren}, this logarithmic dependence is actually removed from $G^R_K$.

\subsection{Renormalization of the coupling constant}
\label{sec:RG-coupling}
In addition to the renormalizations introduced above, which render finite $G_{K,R}^R$, one has also to consider that the coupling constant $u_c$ is renormalized by the interactions and that this renormalization is essential for determining the RG flow of the whole system.
In order to determine it, we study the perturbative corrections to the four-point function $\langle \phi _{cj} (1)\phi_{ql} (2)\phi_{qm} (3)\phi_{qn} (4)\rangle $ to second order in $u_{c}$, where, with a more convenient notation, $\phi_{cj,qj}(n)$ denotes the $j$-component of the $N$-component field $\vecphi_{c,q}$ calculated at the space-time point $n \equiv (\xx_n,t_n)$.
Taking into account the possible diagrams, one finds:
\begin{align}
\label{eq:four-point-function}
&\langle \phi _{cj} (1)\phi_{ql} (2)\phi_{qm} (3)\phi_{qn} (4)\rangle \nonumber\\
& =
\begin{tikzpicture}[baseline={([yshift=-.5ex]current bounding box.center)}]
\coordinate[] (o) at (0,0);
\coordinate[] (no1) at (-0.5,0.25);
\coordinate[label={[label distance=2]left:$1$}] (no2) at (-1,0.5);
\coordinate[] (ne1) at (0.5, 0.25);
\coordinate[label={[label distance=2]right:$3$}] (ne2) at (1, 0.5);
\coordinate[] (so1) at (-0.5,-0.25);
\coordinate[label={[label distance=2]left:$2$}] (so2) at (-1,-0.5);
\coordinate[] (se1) at (0.5,-0.25);
\coordinate[label={[label distance=2]right:$4$}] (se2) at (1,-0.5);
\draw[res] (o) node[circle,fill,inner sep=1pt]{} -- (no1);
\draw[thick] (no2) node[circle,fill,inner sep=1pt]{} -- (no1);
\draw[thick] (so1) -- (o);
\draw[res] (so1) -- (so2) node[circle,fill,inner sep=1pt]{};
\draw[thick] (ne1) -- (o);
\draw[res] (ne1) -- (ne2) node[circle,fill,inner sep=1pt]{};
\draw[thick] (o) node[circle,fill,inner sep=1pt]{} -- (se1);
\draw[res] (se1) -- (se2) node[circle,fill,inner sep=1pt]{};
\end{tikzpicture}
+
\begin{tikzpicture}[baseline={([yshift=-.5ex]current bounding box.center)}]
\coordinate[] (o) at (0,0);
\coordinate[] (ol) at (-0.5,0);
\coordinate[] (or) at (0.5,0);
\coordinate[] (no1) at (-1,0.25);
\coordinate[label={[label distance=2]left:$1$}] (no2) at (-1.5,0.5);
\coordinate[] (ne1) at (1, 0.25);
\coordinate[label={[label distance=2]right:$3$}] (ne2) at (1.5, 0.5);
\coordinate[] (so1) at (-1,-0.25);
\coordinate[label={[label distance=2]left:$2$}] (so2) at (-1.5,-0.5);
\coordinate[] (se1) at (1,-0.25);
\coordinate[label={[label distance=2]right:$4$}] (se2) at (1.5,-0.5);
\draw[res] (or) arc (0:-90:0.5);
\draw[thick] (or) arc (0:270:0.5);
\draw[res] (ol) node[circle,fill,inner sep=1pt]{} -- (no1);
\draw[thick] (no2) node[circle,fill,inner sep=1pt]{} -- (no1);
\draw[thick] (so1) -- (ol);
\draw[res] (so1) -- (so2) node[circle,fill,inner sep=1pt]{};
\draw[thick] (ne1) -- (or);
\draw[res] (ne1) -- (ne2) node[circle,fill,inner sep=1pt]{};
\draw[thick] (or) node[circle,fill,inner sep=1pt]{} -- (se1);
\draw[res] (se1) -- (se2) node[circle,fill,inner sep=1pt]{};
\end{tikzpicture}  \nonumber \\
& = \langle \phi _{cj} (1)\phi_{ql} (2)\phi_{qm} (3)\phi_{qn} (4) \left[ 1 - i\frac{2u_c}{4!N}\int_{1'}  \, (\vecphi_q \cdot \vecphi_c)\vecphi_c^2 \right. \nonumber \\
&\,\,\left.  - \frac{1}{2}\left( \frac{2u_c}{4!N} \right)^2\int_{1'}  \, (\vecphi_q \cdot \vecphi_c)\vecphi_c^2\int_{2'} \, (\vecphi_q \cdot \vecphi_c)\vecphi_c^2 \right]\rangle _0\nonumber\\
& \equiv I_0 + I_1 + I_2,
\end{align}
where the integrations run on the points indicated as $1'$ and $2'$ at which the arguments of the corresponding integrands are calculated, while $I_m$ indicates the contribution of order $u_c^m$ to that expansion.
Note that $I_0$ vanishes because, as a consequence of Wick's theorem, it is proportional to a two-point correlation of quantum fields $\langle \phi_q\phi_q\rangle_0 = 0$. The first order term $I_1$, instead, can be written as
\begin{equation}
\label{eq:integral-I1-CS}
\begin{split}
I_1 &= - i\frac{2 u_c}{4!N}\int_{1'}  \! {\langle \phi _{cj} (1)\phi_{ql} (2)\phi_{qm} (3)\phi_{qn} (4) ({\vecphi _q} \cdot {\vecphi _c})\vecphi_c^2}(1')\rangle _0\cr
& =  - i\frac{u_c}{6N} F_{jlmn}\\
& \qquad \times \int_{1'}  \, G_{0R}(1,1' )G_{0R}(1' ,2)G_{0R}(1' ,3)G_{0R}(1' ,4),
\end{split}
\end{equation}
where $F_{jlmn} = \delta_{jl}\delta _{mn} + \delta_{jm}\delta_{ln} + \delta_{jn}\delta_{lm}$.
The calculation of $I_2$ is slightly more involved and it is reported in App.~\ref{app:second_order_RG} (see Eq.~\eqref{eq:integral2}).
Among the many terms generated by the repeated use of Wick's theorem, we need to retain only those which renormalize the local vertex. After discarding the terms which do not fulfil this requirement, one finds
(see Eq.~\eqref{eq:integral-I2})
\begin{equation}
\begin{split}
I_2  & =  \frac{N + 8}{36N}F_{jlmn}\,u_c^2\\
&\qquad \,\,\,\,\, \times\int_{1'}  \, G_{0R}(1,1')G_{0R}(1' ,2)G_{0R}(1' ,3)G_{0R}(1' ,4) \\
&\qquad \,\,\,\,\, \times \int_{2'}  \, G_{0K}(1' ,2' )G_{0R}(1' ,2').
\end{split}
\end{equation}
Accordingly, adding $I_1$ to $I_2$, the four-point correlation function reads:
\begin{equation}
\begin{split}
&\langle \phi _{cj} (1)\phi_{ql} (2)\phi_{qm} (3)\phi_{qn} (4) \rangle  =  - i\frac{u_c}{6N} F_{jlmn} \\
 &\times \int_{1'}  \, G_{0R}(1,1' )G_{0R}(1' ,2)G_{0R}(1' ,3)G_{0R}(1' ,4)(1 + J),
\end{split}
\end{equation}
where the integral $J$, for $d = 4$ and $ \Lambda t \gg 1$ (see App.~\ref{app:second_order_RG}), is given by (see Eqs.~\eqref{eq:J-1} and~\eqref{eq:J-2})
\begin{align}
\label{eq:log}
&J = \frac{N + 8}{6N} u_c \int_\xx \int_0^t \dd t' \, iG_{0K}(\xx,t,t' )G_{0R}(\xx,t - t')\cr
& =  - \frac{N + 8}{24N}{a_4}{u_c}{\Omega _{0q}}\ln(\Lambda t).
\end{align}
Accordingly, by evaluating the four-point function in Eq.~\eqref{eq:four-point-function} at the normalization point, one finds
\begin{align}
 \langle \phi_c \phi_q  \phi_q\phi_q\rangle\biggr|_\text{NP} & = u_c \left[1- u_c \Omega_{0q} \frac{N+8}{24N}a_4 \ln (\Lambda/\mu) \right] \nonumber\\
 & \qquad \qquad  \times \text{finite terms}.
\label{eq:ueff}
\end{align}
In order to render this correlation function finite as $\Lambda \to \infty$, one introduces the dimensionless renormalized coupling constant $g_R$ as
\begin{equation}
\frac{a_4}{16}\Omega_{0q}u_c = \mu^{4-d} Z_g g_R,
\label{eq:def:g}
\end{equation}
where the factor $a_4\Omega_{0q}/16$ (with $a_4 \equiv a_{d=4}$, as defined below Eq.~\eqref{eq:c-coefficients}) is introduced for later convenience.
Note that up to this order in perturbation theory, it is not necessary to introduce a renormalization of the bulk classical and quantum fields.
Accordingly, one concludes that $Z_g$ has to be chosen in such a way as to cancel the logarithmic dependence in Eq.~\eqref{eq:ueff} and therefore
\begin{equation}
Z_g = 1 + u_c \Omega_{0q} \frac{N+8}{24N} a_4 \ln (\Lambda/\mu).
\label{eq:Zg}
\end{equation}

\subsection{Renormalization-group (Callan-Symanzik) equations}
\label{sec:Callan-Symanzik}
Within the renormalization scheme discussed above, one can infer the renormalization-group equations by exploiting the arbitrariness of the scale $\mu$ which was introduced in order to renormalize the critical model. In fact, the original (bare) action is actually independent of the (infrared) scale $\mu$ and therefore the logarithmic derivative with respect to $\mu$ of the corresponding bare coupling constants and correlation functions, once expressed in terms of the renormalized ones, has to vanish when taken with fixed bare coupling constants and cutoff $\Lambda$.
By taking the logarithmic derivative $\mu\, \partial/\partial\mu$ of Eq.~\eqref{eq:def:g} one finds
\begin{equation}
\mu \frac{{\partial {g_R}}}{{\partial \mu }} \equiv \beta (g_R) =  - \epsilon \,{g_R} + \frac{2}{3}\frac{N + 8}{N}\,g_R^2,
\end{equation}
where we used the fact that, from Eq.~\eqref{eq:def:g}, $a_4\Omega_{0q}u_c\mu^{d-4}/16 = g_R + O(g_R^2)$, and we introduced $\epsilon \equiv 4 -d$. Henceforth, the derivative with respect to $\mu$ is always understood as taken with fixed bare parameters.
This $\beta$-function determines the flow of the coupling constant as the critical theory is approached for $\mu \to 0$ and it is characterized by a fixed-point value $g_R^*$ such that $\beta(g_R^*)=0$, with
\begin{equation}
g^*_R = \frac{3}{2}\frac{N}{{N + 8}}\epsilon  + O(\epsilon ^2),
\label{eq:gRst}
\end{equation}
which is infrared (IR) stable for $d<4$, while the Gaussian fixed point $g^*_R=0$ becomes stable for $d>4$.

In order to highlight the consequences of the existence of this fixed point on the correlation functions, we consider  one which involves $N_c$ and $N_q$ classical and quantum fields in the bulk (i.e., with non-vanishing times), respectively, and $N_0$ initial quantum fields, which we schematically indicate as
\begin{equation}
G_{\{N\}} \equiv \langle [\phi_c]^{N_c}[\phi_q]^{N_q} [\phi_{0q}]^{N_0}\rangle  = Z_c^{N_c/2}Z_q^{N_q/2}Z_0^{N_0/2} G^R_{\{N\}}.
\label{eq:GNN}
\end{equation}
Here $\{N\} = ( N_c, N_q, N_{0})$, $G_{\{N\}}^R$ indicates the same quantity $G_{\{N\}}$ expressed in terms of the renormalized fields, while $Z_{c,q}$ are the renormalization constants of the bulk fields with $Z_{c,q}=1$ up to the order in perturbation theory which we are presently interested in. Note that any correlation function
which involves a classical field taken at the boundary $t=0$ vanishes in a deep quench because of the effective initial conditions on the classical field, see Eq.~\eqref{eq:BC}.
By requiring that the logarithmic derivative of the l.h.s.~of Eq.~\eqref{eq:GNN} with respect to the renormalization scale $\mu$ vanishes when taken at fixed bare parameters, and by expressing the r.h.s.~in terms of renormalized quantities, one finds the Callan-Symanzik equation~\cite{ZinnJustinbook}
\begin{equation}
\begin{split}
\left\{\mu\partial_\mu  + \frac{N_c}{2}\gamma_c  + \frac{N_q}{2}\gamma_q + \frac{N_0}{2} \gamma_0  + \beta \partial_g\right\} G^R_{\{N\}} = 0,
\end{split}
\label{eq:RGCS}
\end{equation}
where we introduced the functions
\begin{equation}
\label{eq:gamma-definition}
\gamma_{c,q,0}  \equiv \mu\,\partial_\mu\ln Z_{c,q,0}|_{\rm bare}.
\end{equation}
At the lowest order in perturbation theory one has $\gamma_{c,q}=0$, while
\begin{equation}
\label{eq:gamma0}
\gamma_0 = \frac{N+2}{3N}g_R + O(g_R^2),
\end{equation}
which follows from Eq.~\eqref{eq:Z0}.
Equation~\eqref{eq:RGCS} can be solved in full generality by employing the method of characteristics and, in conjunction with dimensional analysis, renders the scaling behaviour of the renormalized correlation function $G^R_{\{N\}}$. Note that, in principle, $G^R_{\{N\}}$ still depends on the cut-off $\Lambda$. However, dimensional analysis done by using $\mu$ as the reference scale implies that $G^R_{\{N\}}$ actually depends on $\Lambda/\mu$, which diverges as one explores the long-time and large-scale properties of the theory by letting $\mu\to 0$. However, the renormalizations introduced above in Secs.~\ref{sec:RG-Green} and~\ref{sec:RG-coupling} were in fact determined such that this limit (formally $\Lambda\to\infty$) renders finite quantities. 
Accordingly, the leading scaling behaviour of the renormalized quantities can actually be obtained by removing their dependence on $\Lambda$, i.e., by assuming $\Lambda \gg \mu$
For simplicity, we focus here on the scaling behaviour emerging at the fixed point $g^*_R$ (see Eq.~\eqref{eq:gRst}) of the coupling constant. Accordingly,
Eq.~\eqref{eq:RGCS} simplifies as
\begin{equation}
\left\{\mu\partial_\mu + \frac{N_0}{2} \gamma^*_0\right\} G^R_{\{N\}}(\{k,t\};\mu)= 0,
\label{eq:CSfp1}
\end{equation}
where
\begin{equation}
\gamma^*_0 = \frac{\epsilon}{2} \frac{N+2}{N+8} + O(\epsilon^2)
\end{equation}
indicates the value of $\gamma_0$ calculated at the fixed point.
In order to exploit the consequences of dimensional analysis, we define the dimensionless renormalized Green's functions $\hat{G}^R_{\{N\}}$ in the momentum-time representation
{
\begin{equation}
G^R_{\{N\}}(\{k,t\};\mu) = \mu^{d^\kk_{\{N\}}} \hat{G}^R_{\{N\}}\left(\{k/\mu,\mu t\}\right),
\label{eq:RG-dim-an}
\end{equation}
where $\{k,t\}$ indicates the set of times and momenta at which the various fields are calculated, while $d^\kk_{\{N\}}$ takes into account the canonical scaling of the fields in momentum space
and of the delta functions which ensure momentum conservation, i.e.,
\begin{equation}
\label{eq:dN-definition}
d^\kk_{\{N\}} = d+N_c \zeta^\kk_c + N_q \zeta^\kk_q + N_0\zeta^\kk_{q0} + \dots .
\end{equation}
Note that the r.h.s. of Eq.~\eqref{eq:RG-dim-an} is calculated with the dimensionless parameters corresponding to those appearing on the l.h.s.
}
The factors $\zeta^\kk_c, \zeta^\kk_q$ and $\zeta^\kk_{q0}$ denote the canonical scaling of the classical, quantum in the bulk and quantum at the surface fields in momentum space and, from Sec.~\ref{sec:engineering-dimensions}, they read %
\begin{equation}
\label{eq:fields-canonical-scaling}
\zeta^\kk_c = -\frac{d+2}{2}, \qquad \zeta^\kk_q = \zeta^\kk_{q0} = -\frac{d}{2}.
\end{equation}
The dots in Eq.~\eqref{eq:dN-definition} account for the possibility of including derivatives of the fields into the correlation functions~\eqref{eq:GNN}: they scale differently from the fields and therefore they would add new contributions to $d^\kk_{\{N\}}$.
Accordingly, by inserting Eq.~\eqref{eq:RG-dim-an} in  Eq.~\eqref{eq:CSfp1}, one
finds
\begin{equation}
\label{eq:CSfp2}
\left\{\mu\partial_\mu + d^\kk_{\{N\}} + \frac{N_0}{2} \gamma^*_0\right\} \hat{G}^R_{\{N\}}\left(\{k/\mu,\mu t\}\right)= 0.
\end{equation}
Using the method of characteristics it is possible to transform this partial differential equation into an ordinary one in terms of an arbitrary dimensionless parameter $l$:
\begin{equation}
\label{eq:CS-character}
\left\{l \frac{\dd}{\dd l} + d^\kk_{\{N\}} + \frac{N_0}{2} \gamma^*_0\right\} \hat{G}^R_{\{N\}}\left(\left\{k/(\mu l),\mu l\, t\right\}\right) = 0,
\end{equation}
the solution of which is simply given by
\begin{equation}
\label{eq:CS-character-solution}
l^{N_0\gamma^*_0/2 + d^\kk_{\{N\}}} \hat{G}^R_{\{N\}} \left(\left\{\frac{k}{\mu l},\mu l t\right\}\right) = \hat{G}^R_{\{N\}}\left(\left\{\frac{k}{\mu },\mu t\right\}\right),
\end{equation}
which expresses nothing but the fact that the correlation function is a homogeneous (i.e., scale-invariant) function of its argument, as expected at the critical point.
Specializing this general expression to the case of the retarded function $G^R_R$ with two fields in the bulk (i.e., $G^R_{\{N\}}$ with $N_0=0$, $N_c=N_q=1$) one finds, at sufficiently long times:
\begin{equation}
G^R_R(k,t,t') =  l^{-1} G^R_R\left(k/l, l t,lt'\right).
\label{eq:GR-scaling-gen}
\end{equation}
When, instead, one of the fields is at the boundary ($N_q=0$ and $N_c=N_0=1$) one finds
\begin{equation}
\label{eq:GR-scaling}
G^R_R(k,t,0) =  l^{-1 + \gamma_0^*/2} G^R_R\left(k/l, l t, 0\right).
\end{equation}
The scaling of $G_K$ can be thus inferred on the basis of Eq.~\eqref{eq:GK-ren} with the aid of Eq.~\eqref{eq:GR-scaling}, which yield
\begin{equation}
\label{eq:GK-scaling}
G^R_K(k,t,t') =  l^{-2+\gamma_0^*} G^R_K\left(k/l, l t,lt'\right).
\end{equation}
The scaling forms in Eqs.~\eqref{eq:GR-scaling-gen}, \eqref{eq:GR-scaling}, and \eqref{eq:GK-scaling} are consistent  with those inferred from the perturbation theory in Sec.~\ref{sec:Greens-momentum-space}, with the identification
\begin{equation}
\label{eq:theta-gamma}
\theta_N =  \frac{\gamma_0^*}{2}  = \frac{\epsilon}{4} \frac{N+2}{N+8} + O(\epsilon^2).
\end{equation}
For example, by choosing $l \sim (\Lambda t)^{-1} \ll 1$
in Eq.~\eqref{eq:GR-scaling}, one finds $G^R_R(k,t,0) = t^{1-\gamma_0^*/2} G^R_R(kt,1,0) $, which is nothing but the resummed version of the perturbative result reported in Eq.~\eqref{eq:GR-log-div}.

The Callan-Symanzik equation \eqref{eq:RGCS} alone does not provide information on the two-time scaling
suggested in Eq.~\eqref{eq:GR-log}, which actually emerges as a consequence of the fact that the smaller time $t'$ approaches the ``surface'' $t'=0$ at which the anomalous dimension of the field $\vecphi_{0q}$ differs from the one of the quantum field $\vecphi_{q}$ in the bulk.
In order to work out the consequences of this fact, we consider here the equivalent of the standard short-distance expansion of quantum field theory~\cite{ZinnJustinbook,Diehl1997} adapted to $\phi_q(t')$ at short times~\cite{Janssen1988}, which we assume to be valid for $t' \to 0$:
\begin{equation}
\label{eq:OPE}
\phi_q(t') = \sum_{i \ge 0} \sigma_{i,q}(t') O_i,
\end{equation}
where $O_i$ indicates operators located at the surface, ordered such that their scaling dimension increases upon increasing $i$, while $\sigma_{i,q}(t')$ are functions of $t'$. In the limit $t'\to 0$, this sum is dominated by the most relevant term $O_0$, which is $\phi_{0q}$, and therefore we can simply write Eq.~\eqref{eq:OPE} as $\phi_q(t') \simeq \sigma_q(t') \phi_{0q}$, where we dropped the terms of the expansion which are less singular as $t'\to 0$. 
The scaling form of $\sigma_q(t')$ can be derived by inserting this expansion in Eq.~\eqref{eq:CS-character-solution} and by reading out the scaling behavior, which turns out to be
\begin{equation}
\label{eq:sigma-scaling}
\sigma_q(t') = l^{-\gamma_0^*/2} \sigma_q(lt'),
\end{equation}
at the critical point.
Finally, by using Eqs.~\eqref{eq:GR-scaling} and \eqref{eq:sigma-scaling}, one finds:
\begin{equation}
\label{eq:GR-OPE}
\begin{split}
G_R(k,t,t'\to 0) & \simeq \sigma_q(t') G_R(k,t,0) \\
&= t \left(\frac{t'}{t}\right)^{\gamma_0^*/2}G_R(kt,1,0),
\end{split}
\end{equation}
which is the resummed version of Eq.~\eqref{eq:GR-log}, taking into account the relationship between $\gamma_0^*$ and $\theta_N$ in Eq.~\eqref{eq:theta-gamma}.
Equations~\eqref{eq:GR-scaling}, \eqref{eq:GK-scaling} and~\eqref{eq:GR-OPE} demonstrate that the resummation of the leading logarithms which emerge in perturbation theory and which we did somewhat arbitrarily in Sec.~\ref{sec:perturbation-theory}, is fully justified by the existence of an IR-stable fixed point for the flow of the coupling constant.

\subsection{Scaling of magnetization}
\label{sec:magnetization-scaling}

The approach developed above can be conveniently used in order to determine the scaling form of the magnetization $M(t) \equiv {M_1}(t) = \langle {\vecphi _1}(\xx,t)\rangle$ in the presence of a symmetry-breaking term in the pre-quench Hamiltonian $H_0$, which was discussed in Sec.~\ref{sec:magnetization} either within perturbation theory or in the exactly solvable limit $N\to\infty$. In fact, Eq.~\eqref{eq:magnetization-insertion} (which is valid beyond perturbation theory) can be expanded in powers of $M_0$, assuming a homogeneous external field ${\bf h}$ with the components given in Eq.~\eqref{eq:hi-init}; focussing on the components of the fields along direction 1 of the initial field $\mathbf{h}$, this yields
\begin{equation}
\label{eq:M-scaling}
\begin{split}
M(t) & = \sum_{n=1}^{+\infty} \frac{(-i\sqrt{2}M_0)^n}{n!} \int \dd^d x_1 \cdots \dd^d x_n\\
&\quad\quad \times\langle\phi_c(\xx,t)\dot{\phi}_{0q}(\xx_1)\cdots \dot{\phi}_{0q}(\xx_n)\rangle_{h=0},
\end{split}
\end{equation}
(the term with $n=0$ in the previous expansion vanishes due to the choice of the initial condition) from which it follows that the scaling of $M(t)$ is related to that one of correlation functions containing the initial field $\dot{\phi}_{0q} \equiv \dot{\phi}_q(t=0)$. Note that due to spatial translational invariance, the r.h.s. of Eq.~\eqref{eq:M-scaling} is actually independent of $\xx$.
In fact, the logarithmic dependence on $\Lambda$ of the one-loop correction to the magnetization pointed out in Sec.~\ref{sec:magnetization} (see, in particular, Eq.~\eqref{eq:dM})
can be suitably cancelled (according to the approach discussed in this section) with a renormalization of $\dot{\phi}_{0q} = Z^{1/2}_{\dot 0} \dot{\phi}^R_{0q}$, where a simple calculation gives
\begin{equation}
Z_{\dot 0} = 1 + 2 \theta_N \ln (\Lambda/\mu).
\end{equation}
Note that $Z_{\dot 0}$ differs from $Z_0$ in Eq.~\eqref{eq:Z0}, which means that the anomalous dimension (defined as in Eq.~\eqref{eq:gamma-definition}) 
\begin{equation}
\gamma_{\dot 0} \equiv \mu \partial_\mu \ln Z_{\dot 0}|_{\rm{bare}} = - \frac{N+2}{3N}g_R + O(g_R^2)
\end{equation}
of $\dot{\phi}_{0q}$ is different from that one of $\phi_{0q}$, i.e., from $\gamma_0$ in Eq.~\eqref{eq:gamma0}. Accordingly, when $\gamma_{\dot 0}$ is evaluated at the fixed point $g_R^*$ reported in Eq.~\eqref{eq:gRst}, it renders:
\begin{equation}
\gamma^*_{\dot{0}} = -\frac{\epsilon}{2} \frac{N+2}{N+8} + O(\epsilon^2) = -2 \theta_N.
\end{equation}
Following the same line of argument as in Sec.~\ref{sec:Callan-Symanzik}, one can easily derive a
Callan-Symanzik equations for the correlations functions appearing on the r.h.s.~of
Eq.~\eqref{eq:M-scaling}, which are of the form of Eq.~\eqref{eq:GNN} but with the fields $\phi_{0q}$ replaced by $\dot{\phi}_{0q}$ and, correspondingly, $\gamma_0$ by $\gamma_{\dot 0}$ in the equations which follow.  At the critical point, one eventually finds the scaling equation
\begin{equation}
\begin{split}
\label{eq:scaling-phi-dot}
&l^{d^\xx_{\{N\}} - n\theta_N} \langle \phi_c(lt,l\xx) \dot{\phi}_{0q}(l\xx_1)\cdots  \dot{\phi}_{0q}(l\xx_n)\rangle \\
&\quad = \langle \phi_c(t,\xx)  \dot{\phi}_{0q}(\xx_1)\cdots  \dot{\phi}_{0q}(\xx_n) \rangle,
\end{split}
\end{equation}
{
where $d^\xx_{\{N\}}$ accounts for the canonical scaling dimension of the fields in real space and reads:
\begin{equation}
d^\xx_{\{N\}} = N_c\zeta^\xx_c + N_{\dot{0}} \zeta^\xx_{\dot{0}} = [d-2 + n(d+2)]/2,
\end{equation}
for $N_c=1$ and $N_{\dot{0}}= n$, while $\zeta_c^\xx = (d-2)/2$ and $\zeta^\xx_{\dot{0}} = (d+2)/2$ are, respectively, the canonical scaling dimensions of the fields $\phi_c$ and $\dot{\phi}_q$ in real space, which can easily be determined from Sec.~\ref{sec:engineering-dimensions} by noticing that $\dot{\phi}_{0q}$ has the same dimension as $\mu\phi_{0q}$, where $\mu$ is an arbitrary momentum scale, see Eq.~\eqref{eq:GaussianScalingFields}.
}
Substituting Eq.~\eqref{eq:scaling-phi-dot} into Eq.~\eqref{eq:M-scaling} one obtains, after a change of variables in the spatial integrals:
\begin{equation}
\begin{split}
M(t) 		&= l^{d/2-1} \sum_{n=1}^{+\infty} \frac{(-i\sqrt{2})^n}{n!}\left(l^{-d/2+1-\theta_N} M_0\right)^n  \\
& \times
\int \dd^d x_1\cdots\dd x_n  \langle\phi_c(lt,l\xx)\dot{\phi}_{0q}(l\xx_1)\cdots \dot{\phi}_{0q}(l\xx_n)\rangle_{h=0}
\end{split}
\end{equation}
from which, by choosing $l = (\Lambda t)^{-1}$ at long times $t \gg \Lambda^{-1}$, we find a scaling form for the magnetization:
\begin{equation}
M(t) = M_0\, t^{\theta_N}\,\mathcal{M}\left(t^{d/2-1+\theta_N}M_0\right).
\label{eq:scaling-initslip}
\end{equation}
This results entails the existence of a time scale $t_i \sim M_0^{1/(d/2-1+\theta_N)}$ at which the
{increase of the magnetization reported in Eq.~\eqref{eq:scal-M-pt} (supported by both the perturbative and non-perturbative analysis presented in Sec.~\ref{sec:magnetization}) crosses over to a different behaviour.}
For $d \to 4$, this result is in agreement with the $t_i$ estimated in Sec.~\ref{sec:magnetization} for the critical quench of the $O(N)$ model for $N\to \infty$ (see Eq.~\eqref{eq:ti-estimate}).

\section{Conclusions}
\label{sec:conclusions}
In this work we investigated the effects of fluctuations on the dynamical transition beyond mean-field theory which is observed after a quench
of the parameters of an isolated quantum many-body system with $O(N)$ symmetry. In particular, we accounted for fluctuations at the lowest order
in the post-quench coupling constant, extending the analysis of Ref.~\onlinecite{Chiocchetta2015}
to the spatio-temporal structure of response and correlation functions and to the dynamics of the order parameter, while providing the details of the results anticipated therein.
We found that, before thermalization eventually occurs as a consequence of the interaction which is switched on upon quenching,
the system approaches a pre-thermal state within which it undergoes a dynamical transition. We have characterized this transition
within a Wilson's renormalization-group approach, showing that it is associated with a stable fixed point of the RG flow of the relevant
control parameters and that it displays remarkable analogies with the corresponding equilibrium phase transition at finite temperature.
Nevertheless, the breaking of the invariance of the dynamics under time translations caused by the quench induces a
novel algebraic and universal behavior at short times with new features compared to the classical case.
This universal short-time behavior is characterized by a new critical exponent $\theta_N$ calculated here to the leading order
in a dimensional expansion [see Sec.~\ref{sec:RG-Wilson}, Eq.~\eqref{thetaNexpression}]. This result is found in agreement with the expression of the one for a critical quench of the $O(N)$ model in the limit $N\to \infty$ [see Sec.~\ref{sec:magnetization}, Eq.~\eqref{eq:thNinf}], in which the model is exactly solvable.
The correlation functions after the quench are characterized by scaling forms which depend on the exponent $\theta_N$ and which
we determined by using Callan-Symanzik equations [see Sec.~\ref{sec:Callan-Symanzik}, Eq.~\eqref{eq:CS-character}]. Moreover, when an external field breaks the $O(N)$ symmetry of the initial state (restored after the quench),
the dynamics of the order parameter exhibits an algebraic growth, controlled by the initial-slip exponent $\theta_N$, up to a certain macroscopic time $t_i$ [see Sec.~\ref{sec:magnetization-scaling}, Eq.~\eqref{eq:scaling-initslip}].

This slow dynamics, which is referred to as aging, is very similar to the one occurring in both classical dissipative systems and
open quantum system, though it belongs to a different universality class and is characterized by different scaling forms (see Appendix of Ref.~\onlinecite{Maraga2015}).
Among these differences, the most marked ones concern the spatio-temporal dependence of the correlation functions,
since they exhibit here a light cone due {to ballistically propagating excitations in the prethermal state [see Sec.~\ref{sec:light-cone} and Figs.~\ref{fig:lightcone-GR} and \ref{fig:lightcone-GK}].}
Another difference between the open
and closed quantum system is that for the latter, entanglement entropy of the unitarily evolving state
is a good measure of quantum correlations, and a recent study showed how
the aging exponent entering here, also controls the scaling of the entanglement spectrum~\cite{Lemonik2016}.

%
%

As a future perspective, the study of the destabilization of the prethermal regime and the resulting crossover towards the full thermalization
represents an interesting question as well as a technical challenge. In fact, as discussed in Sec.~\ref{sec:dissipative-terms}, the inclusion of the effect of
two-loop diagrams becomes intractable within the usual perturbative techniques because of the appearance of secular terms,
and therefore more elaborate resummation schemes are required.
Moreover, the study of aging in fermionic systems represents an important issue, also because the classical limit of these systems is far from trivial,
and therefore their genuinely quantum nature is expected to bear completely new features.

\begin{acknowledgments}
The authors thank I.~Carusotto, A.~Coser, Y.~Lemonik, A.~Maraga, M.~Marcuzzi, J.~Marino and A.~Silva for invaluable discussions. MT and AM were supported by
National Science Foundation Grant No.~DMR 1303177.
\end{acknowledgments}

\appendix

\section{Functional derivation of Green's functions}
\label{app:functional-Green}
In this Appendix we show how to derive the Green's functions $G_{0K,0R}$ in Eqs.~\eqref{eq:GR-gauss} and~\eqref{eq:GK-gauss} from the Gaussian part of the action~\eqref{eq:SK-s},~\eqref{eq:SK-b} considered in Sec.~\ref{sec:quench-action}. In order to do this, one introduces additional terms in the Keldysh action, in which the various fields are coupled linearly to some external sources $\jj_{c,q}$ and $\jj_{0c,0q}$. Accordingly, any correlation function can be obtained by differentiating the generating function
\begin{align}
\mathcal{Z}_K[\{\jj\} ] = \int D[\vecphi_c,\vecphi_q,\vecphi_{0c},\vecphi_{0q}] \, \ee^{iS}  = \ee^{W[\{\jj\}]}.
\end{align}
where $\{\jj\} \equiv ({\jj_c},{\jj_q},\jj_{0c},\jj_{0q}).$ If one were interested in calculating correlation functions which involve $\dot{\vecphi}_{0c}$ or $\dot{\vecphi}_{0q}$, additional sources should be coupled to these fields.
It turns out that the intermediate results of the actual calculation of $W$ do depend on whether one integrates out first $\vecphi_c$ or $\vecphi_q$, as shown below; however, the final result is the same.\\
First let us integrate out $\vecphi_q$: in order to do this, we integrate by parts the time derivatives in the bulk action in Eq.~\eqref{eq:SK-b}, as
\begin{align}
\label{eq:SK-by-parts}
S_K  & = i\int_\kk  \frac{\omega_{0k}}{2} \left[ \vecphi_{0c}^2 \,\tanh\left(\beta\omega_{0k}/2\right) + \vecphi_{0q}^2 \, \coth\left(\beta \omega_{0k}/2\right) \right]  \nonumber \\
	   & - i\int_\kk \left[  {\jj_{0c}} \, \vecphi_{0q} + {\jj_{0q}} \, \vecphi_{0c} - i\vecphi_{0q}\,\dot{\vecphi}_{0c} \right]\nonumber \\
	   & +  \int_\kk \int_0^\infty \dd t \, \left[-\vecphi_q\left( \ddot{\vecphi}_c + \epsilon \dot{\vecphi}_c +\omega^2_k\vecphi_c - {\jj_c}\,\right) + {\jj_q}\,\vecphi_c \right].
\end{align}
{where we set $\Omega_{0c} = \Omega_{0q} = \Omega_0$ for the sake of simplicity and we defined ${\jj_{0c}}\,{\vecphi _{0q}} = {\jj_{0c}}\, \cdot {\vecphi _{0q}}$}. Note that the integration by parts generates an additional term $-i\vecphi_{0q}\dot{\vecphi}_{0c}$ in the part of the action located at $t=0$ and that we added to $S_K$ an infinitesimal term proportional to $\epsilon > 0$. This term contains the operator $\vecphi_q\dot{\vecphi}_c$, and therefore it can be regarded as an infinitesimal dissipation~\cite{Kamenevbook}. The reason for including it is twofold, as it regularizes the Keldysh functional while ensuring causality. Notice also that such a term breaks the exchange symmetry $\vecphi_c \leftrightarrow \vecphi_q$ of the bulk action, which holds for $\jj_c =\jj_q =0$. The integration of $\vecphi_q$ generates a functional delta function the support of which is on the solution $\bar{\vecphi}_c$ of the equation
\begin{equation}
\left(\partial_t^2 + \epsilon \,\partial_t +\omega^2_k \right)\bar{\vecphi}_c = {\jj_c},
\end{equation}
i.e., on
\begin{align}
\label{eq:SPclassical}
\bar{\vecphi}_c(t)  & =  \vecphi_{0c}\,  \ee^{-\epsilon t}\cos(\omega_k t) + \ee^{-\epsilon t}\dot{\vecphi}_{0c} \, \frac{\sin(\omega_k t)}{\omega_k}  \nonumber \\
						&+ \int _0 ^ t \dd t'\, \frac{\ee^{-\epsilon t}\sin[\omega_k (t-t')]}{\omega_k} {\jj_c}(t'),
\end{align}
where $\vecphi_{0c} \equiv \vecphi_c(t=0)$ and $\dot{\vecphi}_{0c} \equiv \dot{\vecphi}_c(t=0)$.
Note that in Eq.~\eqref{eq:SPclassical} one can safely take the limit $\epsilon \to 0$.
Then, we expand $\vecphi_c$ around $\bar{\vecphi}_c$ and we integrate out the Gaussian fluctuations of $(\vecphi_c -\bar{\vecphi}_c)$. Finally, we integrate out the remaining fields $\vecphi_{0c}$, $\vecphi_{0q}$ and $\dot{\vecphi}_c$, finding
\begin{align}
W[\{{\jj}\}]  &  = \int_\kk  \left\{ \frac{1}{2}\, \jj_{0q}^2\,iG_{0K}(k,0,0)  \right. \nonumber \\
			  & - \int_t\,\left [ \jj_{0q}\, {\jj_q}(t)\, G_{0K}(k,t,0) +  {\jj_{0c}} \, {\jj_q}(t)\, G_{0R}(k,t,0) \right] \nonumber \\
			  & - \int_{t,t'} \, \biggr[   {\jj_q}(t) \,  {\jj_c}(t') \, iG_{0R}(k,t,t')  \nonumber\\
			  & \qquad \left. + \frac{1}{2} \,  {\jj_q}(t) {\jj_q}(t') iG_{0K}(k,t,t') \biggr] \right\},  \label{eq:Wgenerating}
\end{align}
where $\int_t = \int_0^\infty \dd t$ and $G_{0R}, G_{0K}$ are those defined in Eqs. \eqref{eq:GR-gauss} and~\eqref{eq:GK-gauss} and indeed they can be obtained by taking suitable functional derivatives of this $W[\{\jj\}]$, which establish the connection between these functions $G_{0K,0R}$ and the correlation functions of the fields.
Consider now the case in which $\vecphi_c$ and $\vecphi_q$ are integrated in the reversed order: we rewrite $S_K$ as
\begin{align}
S_K  & = i\int_\kk \frac{\omega_{0k}}{2} \left[ \vecphi_{0c}^2 \,\tanh\left(\beta\omega_{0k}/2\right) + \vecphi_{0q}^2 \, \coth\left(\beta \omega_{0k}/2\right) \right] \nonumber \\
& - i\int_\kk \left[ {\jj_{0c}} \, \vecphi_{0q} + {\jj_{0q}} \, \vecphi_{0c} - i\vecphi_{0c}\,\dot{\vecphi}_{0q} \right] \nonumber \\
& +  \int_\kk \int_0^\infty \dd t \, \left[-\vecphi_c \left( \ddot{\vecphi}_q - \epsilon \dot{\vecphi}_q +\omega^2_k\vecphi_q - {\jj_q}\,\right) + {\jj_c}\,\vecphi_q\right],
\end{align}
where the additional term generated in the action located at $t=0$ is different from the one in Eq.~\eqref{eq:SK-by-parts}. Note that, as in Eq.~\eqref{eq:SK-by-parts} we introduced an infinitesimal term with $\epsilon>0$. The integration of $\vecphi_c$ generates a delta function the support of which is on the solution of the equation:
\begin{equation}
(\partial_t^2 - \epsilon \,\partial_t +\omega^2_k)\bar{\vecphi_q} = {\jj_q},
\end{equation}
i.e., on
\begin{align}
\label{eq:vecphiq-bar}
\bar{\vecphi}_q(t) & = \vecphi_{0q}\, \ee^{\epsilon t}\,\cos(\omega_k t) + \dot{\vecphi}_{0q} \,\ee^{\epsilon t}\,\frac{\sin(\omega_k t)}{\omega_k}  \nonumber \\
						& + \int _0 ^ t \dd t'\, \frac{\sin[\omega_k (t-t')]}{\omega_k} {\jj_q}(t'),
\end{align}
Note that, differently from Eq.~\eqref{eq:SPclassical}, the infinitesimal term $\propto \epsilon$ generates an exponential factor which makes $\bar{\vecphi}_q$ grow exponentially in time. In order to avoid this unphysical divergence, we have to impose $\bar{\vecphi}_q(t\to+\infty) =0$, which fixes the value of $\vecphi_{0q}$ and $\dot{\vecphi}_{0q}$ to
\begin{align}
\vecphi_{0q} & = \int_0^{+\infty} \dd t\, \frac{\sin(\omega_k t) }{\omega_k}\, \ee^{-\epsilon t}\, {\jj_q}(t), \label{eq:incond1}\\
\dot{\vecphi}_{0q} & = - \int_0^{+\infty} \dd t\, \cos(\omega_k t) \, \ee^{-\epsilon t}\, {\jj_q}(t),\label{eq:incond2}
\end{align}
from which Eq.~\eqref{eq:vecphiq-bar} can be rewritten as
\begin{equation}
\bar{\vecphi}_q(t)  = - \int_t^{+\infty} \dd t'\, \frac{\sin[\omega_k(t-t')]}{\omega_k}\, \ee^{\epsilon(t-t')}\, {\jj_q}(t').
\end{equation}
Due to the fact that $t' > t$, the limit $\epsilon \to 0$ can now be safely taken in this last expression.
Finally, we expand $\vecphi_q$ around $\bar{\vecphi}_q$ and integrate out the Gaussian fluctuations, we replace the values of $\vecphi_{0q}$ and $\dot{\vecphi}_{0q}$ with Eqs.~\eqref{eq:incond1} and~\eqref{eq:incond2} in the surface action and we eventually integrate out $\vecphi_{0c}$. The final result is the same $W$ as the one given in Eq.~\eqref{eq:Wgenerating}.

\section{Relationship between $G_K$ and $G_R$ in a deep quench}
\label{app:Dyson}
In this Appendix we show that Eq.~\eqref{eq:GKvsGR}, which connect $G_K$ and $G_R$ for a deep quench, is valid to all orders in perturbation theory if the Keldysh self-energy $\Sigma_K$ defined below vanishes. This is the case, e.g., in the limit $N\to\infty$ of the present model, discussed in Ref.~\onlinecite{Maraga2015}.

To show this, we recall that the Dyson equation~\cite{Kamenevbook} for the post-quench propagator $G$ of a theory in the presence of interaction  can be  written symbolically as
\begin{equation}
\label{eq:dys}
\left( G_0^{-1}-\Sigma\right)G=1,
\end{equation}
in terms of the post-quench Gaussian propagator $G_0$ and of the self-energy $\Sigma$ of the system after the quench. In turn, $G$ and $\Sigma$ for a Keldysh action have the structure
\begin{equation}
G=\begin{pmatrix}
G_K & G_R\\
G_A & 0
\end{pmatrix}, \quad
 \Sigma=\begin{pmatrix}
 0 & \Sigma_A \\
 \Sigma_R &\Sigma_K
\end{pmatrix},
\end{equation}
and analogous for the Gaussian propagator. The functions $\Sigma_K$ and $\Sigma_R$ are, respectively, the Keldysh and retarded self-energies, which fully encode the effect of the interaction $u_{c,q}$.
Equation \eqref{eq:dys} is actually a shorthand notation for a set of integral equations of the form (we drop the dependence on the momentum variables $\kk$ since the equations are local in $\kk$)
\begin{align}
\label{eq:GK-dyson}
& G_K(t,t') =	\nonumber \\
& \quad  G_{0K}(t,t') + \int_{t_1,t_2} G_{0R}(t,t_1)\Sigma_K(t_1,t_2)G_R(t',t_2) \nonumber \\
& \quad\quad + \int_{t_1,t_2} G_{0R}(t,t_1)\Sigma_R(t_2,t_1)G_K(t_2,t') \nonumber\\
& \quad\quad + \int_{t_1,t_2} G_{0K}(t,t_1)\Sigma_R(t_1,t_2)G_R(t',t_2),
\end{align}
and
\begin{align}
\label{eq:GR-dyson}
& G_R(t,t') = \nonumber \\
& \quad \,\,\,  G_{0R}(t,t') + \int_{t_1t_2} G_{0R}(t_1,t')\Sigma_R(t_1,t_2)G_R(t,t_2) \nonumber \\
& \quad = G_{0R}(t,t') + \int_{t_1t_2} G_R(t_1,t')\Sigma_R(t_1,t_2)G_{0R}(t,t_2).
\end{align}
In the last equality we have expressed the equation in an equivalent form in which $G_R$ and $G_{0R}$ are interchanged, which corresponds to a different order in the resummation of the one-particle irreducible diagrams contributing to $\Sigma$. In order to simplify the notation, we do not indicate here the two indices of the components of the fields involved in these $G_{R,K}$, $G_{0R,0K}$ and $\Sigma_R$, as they are all diagonal in component space.
In order to prove Eq.~\eqref{eq:GKvsGR}, we assume that $\Sigma_K = 0$ and that, within the Gaussian approximation, $G_{0K}$ and $G_{0R}$ are related via 
\begin{equation}
iG_{0K}(t,t') = \Omega_0G_{0R}(t,0)G_{0R}(t',0),
\label{eq:GK-GRGR}
\end{equation}
which holds for the case we are interested in, see Eqs.~\eqref{eq:GR-gauss} and \eqref{eq:GK-gauss}.
Then, we prove that the function $iG_K^T(t,t') \equiv \Omega_0G_R(t,0)G_R(t',0)$ satisfies the same equation as $G_K$, and therefore they have to be equal, i.e., $iG_K(t,t') = iG^T_K(t,t') = \Omega_0G_R(t,0)G_R(t',0)$, since the Dyson equation is linear.
In fact, by using Eq.~\eqref{eq:GR-dyson}, one can rewrite $G_K^T(t,t')$ as:
\begin{align}
&iG^T_K(t,t') = \Omega_0G_{0R}(t,0)G_{0R}(t',0) \nonumber \\
& \quad + \Omega_0\int_{t_1t_2} G_{0R}(t,0)G_{0R}(t_1,0) \Sigma_R(t_1,t_2)G_R(t',t_2) \nonumber \\	
& \quad + \Omega_0\int_{t_1t_2} G_R(t',0)G_R(t_1,0) \Sigma_R(t_1,t_2) G_{0R}(t,t_2).
\end{align}
The relation \eqref{eq:GK-GRGR} between $G_{0K}$ and $G_{0R}$ implies, after an exchange of the integration variables $t_1$ and $t_2$ in the last integral of the previous equation, that
\begin{align}
&G^T_K(t,t') 	 = G_{0K}(t,t') \nonumber \\
& \qquad + \int_{t_1t_2}G_{0K}(t,t_1)\Sigma_R(t_1,t_2)G_R(t',t_2) \nonumber \\
&  \qquad + \int_{t_1t_2} G_{0R}(t,t_1)\Sigma_R(t_2,t_1)G_K^T(t',t_2),
\end{align}
which is exactly Eq.~\eqref{eq:GK-dyson} satisfied by $G_K$ under the assumption that $\Sigma_K=0$.

An alternative derivation of the same result is the following:
Thinking of $\Omega_0$ as a perturbative parameter, one can now include perturbatively its effects in $\Sigma$. In particular, $\Sigma_K$ then accounts for the Gaussian term $\propto \Omega_0$ while  $G_0$ includes only $G_{0R}$, which is not proportional to $\Omega_0$, while $G_{0K}=0$ --- see Eqs.~\eqref{eq:GR-gauss} and \eqref{eq:GK-gauss}. Accordingly, Eq.~\eqref{eq:dys} implies the integral equation
\begin{equation}
G_K(t,t') = \int_{t_1,t_2} G_R(t,t_1) G_R(t',t_2) \Sigma_K(t_1,t_2),  \label{eq:dy1}
\end{equation}
which holds regardless of whether interactions are present or not.
Note that, in the absence of interactions, Eq.~\eqref{eq:dy1} implies
\begin{equation}
\Sigma_K^0(t,t')=-i\Omega_0\delta(t-t')\delta(t),
\end{equation}
because of Eq.~\eqref{eq:GK-GRGR},
where $\Sigma_K^0$ can be thought of as a Keldysh ``self-energy'' at the zeroth order in the interactions, but which, in fact, simply accounts for the initial distribution after the quench, and it emerges naturally from discretizing the Keldysh action in time and from accounting for the initial density matrix\cite{Kamenevbook}.
Under the assumption that the interaction introduced by $u_c\neq 0$
does not affect $\Sigma_K$ (a condition which we expressed in the derivation above as $\Sigma_K=0$) --- which therefore equals its Gaussian value $\Sigma_K^0$ --- one obtains immediately Eq.~\eqref{eq:GKvsGR}.

\section{Renormalization of $G_R$ in momentum space at initial times}
\label{app:singularities}

In this section we show how a logarithmic dependence on the large cutoff $\Lambda$ appears  in the retarded Green's function $G_R(q,t,t')$ in perturbation theory at one-loop when the earlier time $t'$ is set at the temporal boundary, i.e., with $t'=0$.
Let us focus on Eq.~\eqref{eq:dGR-1}, which expresses the perturbative correction $\delta G_R$ to $G_R$, and assume that the time-independent part $B_0$ of the tadpole $i\mathcal{T}$ (see Eq.~\eqref{eq:Tpole}) has been reabsorbed in a proper redefinition of the parameter $r$, according to Eq.~\eqref{eq:shift-r}.
Due to the causal structure of $G_R$, the domain of integration in Eq.~\eqref{eq:dGR-1} is effectively restricted to the interval $t'\le \tau \le t$  and therefore the integral is finite for $\Lambda \to +\infty$ as long as $t'\neq 0$, because the only possible source of
singularity associated with $B(\tau)\sim \tau^{-2}$ in Eq.~\eqref{eq:B0-crit} is located at $\tau=0$, see Eq.~\eqref{eq:B0-crit3}.
On the other hand, if $t' = 0$, Eq.~\eqref{eq:dGR-1} renders, at the critical point $r=r_c$,
\begin{equation}
\label{eq:GRsing2}
\delta G_R(t,0) = \int_0^{t_2}\!\!\dd t\, G_{0R}(q,t_2,t) B(t) G_{0R}(q,t,0) \equiv I_R(t),
\end{equation}
in which one cannot take $\Lambda \to +\infty$ in $B(\tau)$ from the outset, but this can be done only after a convenient subtraction of the singular term $\sim 1/\tau^2$ has been introduced in the integral $I_R(t)$.
Since $G_{0R}(q,\tau,0) = - \tau + O(\tau^2)$  and $G_{0R}(q,t,\tau) = G_R(q,t,0) + O(\tau)$, a convenient way of writing $I_R(t)$ and of introducing the subtraction is
\begin{align}
I_R(t) &= \int_0^{t} \dd \tau\, \big[G_{0R}(q,t,\tau) G_{0R}(q,\tau,0) \nonumber \\
& + G_{0R}(q,t,0)\,\, \tau \big] B(\tau)  - G_{0R}(q,t,0) \int_0^{t}\dd \tau\, \tau B(\tau),
\label{eq:GRsing1}
\end{align}
where the first integral on the r.h.s.~is finite for $\Lambda \to +\infty$, given that the term within square brackets vanishes $\sim \tau^2$ as $\tau\to0$ and therefore regularizes the singularity that $B(\tau)$ develops as $\Lambda$ grows. The second integral, instead, can be explicitly calculated by using the explicit expression of $B(t)$ in Eq.~\eqref{eq:B0-crit} and it gives
\begin{equation}
\begin{split}
\int_0^{t}\!\!\dd \tau\, \tau B(\tau) &=  - \theta_N  \int_0^{\Lambda t}\!\!\dd\tau\, \tau \frac{1-\tau^2}{(\tau^2 +1 )^2}\\
 & = \theta_N \ln (\Lambda t) + \text{finite terms},
\end{split}
\end{equation}
where $\theta_N$ is given in Eq.~\eqref{thetapert} and the terms which have been omitted are finite as $\Lambda t$ increases. Accordingly, a logarithmic divergence proportional to $G_R(q,t,0)$ occurs in $I_R(t)$ in the limit $\Lambda \to +\infty$.

\section{Renormalization of $G_K$ in momentum space}
\label{app:Keldysh-correction}

In this section we provide some details of the calculation of the one-loop correction $\delta i G_K$ to the Keldysh Green's function $i G_K$ of the model after a deep quench, reported in Eq.~\eqref{eq:dGK-1}.
As in App.~\ref{app:singularities}, the correction involves the tadpole $i\mathcal{T}$, the time-independent part of which can be reabsorbed in a redefinition of the parameter $r$ as in Eq.~\eqref{eq:shift-r}.  As a result, Eq.~\eqref{eq:dGK-1} can be rewritten, at the critical point $r=r_c$ (see after Eq.~\eqref{eq:shift-r}), as
\begin{equation}
\label{eq:GK-correction}
\delta i G_K(q,t,t') = - \frac{\Omega_0}{q^2} [I_K(t) \sin (qt')  + I_K(t') \sin (qt)],
\end{equation}
where the integral $I_K$ is given by
\begin{equation}
\begin{split}
I_K(t)
& = \frac{1}{q}\int_0^t \dd\tau \sin (q(t-\tau)) B(\tau) \sin (q \tau) \\
& = \frac{1}{q}\int_0^{+\infty} \dd\tau \sin (q(t-\tau)) B(\tau) \sin (q \tau)  \\
& \quad - \frac{1}{q}\int_t^{+\infty} \dd\tau \sin (q(t-\tau)) B(\tau) \sin (q \tau) \\
& \equiv I_{K1}(t) + I_{K2}(t),
\end{split}
\label{eq:corr-GR-I12}
\end{equation}
where, for later convenience, we introduced the two integrals $I_{K1,2}$, which we calculate separately.
Given the expression of $B(t)$ in Eq.~\eqref{eq:B0-crit}, $I_{K1}$ can be calculated analytically and it reads:
\begin{equation}
\begin{split}
I_{K1}(t)
& = \theta_N    \left\{   -\frac{\pi}{2} \ee^{-q/\Lambda}  \cos (qt)  - \sin (qt) \times \right. \\
&  \left. \times   \left[\cosh \left(\frac{q}{\Lambda}  \right) \text{Chi}   \left(   \frac{q}{\Lambda}   \right)   - \sinh   \left( \frac{q}{\Lambda}   \right) \text{Shi} \left( \frac{q}{\Lambda} \right) \right] \right\},
\end{split}
\label{eq:GK-I1-app}
\end{equation}
where we introduced the hyperbolic cosine and sine integral, defined as
\begin{align}
\text{Chi}(x) & = \gamma + \ln x + \int_0^x \dd t \,\frac{\cosh t -1}{t}, \\
\text{Shi}(x) & = \int_0^x \dd t \, \frac{\sinh t}{t},
\end{align}
respectively, where $\gamma \simeq 0.577$ is the Euler-Mascheroni constant. The integral $I_{K2}(t)$ in Eq.~\eqref{eq:corr-GR-I12} can be easily calculated for $t \gg \Lambda^{-1}$, in which case the function $B(t)$ is approximated as in
Eq.~\eqref{eq:B0-crit3} and yields
\begin{equation}
I_{K2}(t) = \theta_N\left\{\sin (qt) \,\text{Ci}(2qt) + \cos (qt) \left[ \frac{\pi}{2} - \text{Si}(2qt)\right]\right\},
\label{eq:GK-I2-app}
\end{equation}
where $\text{Ci}$ and $\text{Si}$ were introduced in Eq.~\eqref{eq:CiSi-def}.
Summing Eqs.~\eqref{eq:GK-I1-app} and \eqref{eq:GK-I2-app}  one eventually finds for $I_K(t)$ in
Eq.~\eqref{eq:corr-GR-I12}, at the leading order for $q/\Lambda \ll 1$,
\begin{equation}
\begin{split}
I_K(t) & = - \theta_N\left\{\sin(qt) \left[\ln (q/\Lambda)  - \text{Ci}(2qt)\right]\right. \\
					  &\quad\quad\quad  + \left. \cos (qt)\, \text{Si}(2qt) \right\}.
\end{split}
\label{eq:GK-I1-app-2}
\end{equation}
This expression can be inserted into Eq.~\eqref{eq:GK-correction} which,
taking into account Eq.~\eqref{eq:GR-gauss-dq} for
the Gaussian Keldysh Green's function, gives the explicit form of the correction
\begin{equation}
\begin{split}
\delta iG_K(q,t,t')	& = iG_{0K}(q,t,t') \,\theta_N\,  \left[2\ln(q/\Lambda) -F_K(qt,qt') \right] \nonumber \\
							& \qquad \qquad + \delta g_K(q,t,t'),
\end{split}
\end{equation}
where $F_K(x,y)$ is a scaling function defined as:
\begin{align}
F_K(x,y)	& = \text{Ci}(2x) + \text{Ci}(2y),
\end{align}
while $\delta g_K(q,t,t')$ contains only oscillating and finite corrections:
\begin{align}
\delta g_K(q,t,t')	& = \theta_N \frac{\Omega_0}{q^2} \left[\sin(qt)\cos(qt')\text{Si}(2qt') \right. \nonumber \\
							&	\qquad  \left. + \sin(qt')\cos(qt)\text{Si}(2qt) \right].
\end{align}

\section{Corrections to Green's functions in real space}
\label{app:lc-A}

In this Appendix we analyze the one-loop perturbative corrections to $G_{R,K}$ in real space and time, in order to investigate how their structure is modified by fluctuations.
The starting point is Eq.~\eqref{Gft} which we rewrite for convenience
\begin{equation}
\label{eq:GFT-epsilon}
\begin{split}
G_{R,K}(x,t,t') & =\frac{1}{x^{1-\epsilon/2}\left(2\pi\right)^{2-\epsilon/2}} \int_0^{\infty}\!\!\dd q \, q^{2-\epsilon/2}
\\
& \times J_{1-\epsilon/2}(qx)f\left(q/\Lambda\right) G_{R,K}(q,t,t'),
\end{split}
\end{equation}
in which we introduced the cut-off function $f(x)$, as specified above Eq.~\eqref{eq:GK-lc-computation}.
The first correction $\delta G_{R,K}^{\epsilon}$ to the Gaussian $G_{0R,0K}$ comes from the first-order expansion of this expression in powers of $\epsilon$, which involves an expansion of $J_{1-\epsilon/2}$, given by~\cite{Gradshteyn2007}
\begin{equation}
\label{eq:Bessel-expansion}
\partial_{\nu}J_{\nu}(y)|_{\nu=1}=\frac{\pi}{2}Y_1(y) + \frac{J_0(y)}{y},
\end{equation}
where $Y_1(y)$ is the Bessel function of the second kind. The above expansion does not produce any logarithmic divergence. For the response function,
using the Gaussian retarded Green's function $G_{0R}$ in Eq.~\eqref{eq:GR-gauss-dq}, we may write,
\begin{align}
&\delta G_R^{\epsilon}(x,t,t') = \nonumber \\
& -\frac{1}{x^{1-\epsilon/2}\left(2\pi\right)^{2}}\int_0^{\infty}\!\!\dd q q^{1-\epsilon/2}J_{1}(qx) \sin(q (t-t'))f\left(q/\Lambda\right) \nonumber\\
&+ \frac{1}{x\left(2\pi\right)^{2}}\int_0^{\infty}\!\!\dd q q J_{1}(qx)\sin(q (t-t'))f(q/\Lambda).
\end{align}
Due to the oscillatory nature of the integrands, the two integrals above provide a non-vanishing contribution only if the oscillations from the asymptotic form of the Bessel function $J_1(qx)$ as a function of $q$ are in phase with the oscillations of
$\sin(q(t-t'))$, i.e., only for $x=t-t'$, in which case one finds
\begin{equation}
\begin{split}
&\delta G_R^{\epsilon}(x=t-t') \simeq -\frac{1}{x^{3-\epsilon}\left(2\pi\right)^{2}}\int_0^{\Lambda x}\dd y \frac{y^{1-\epsilon/2}}{\sqrt{y}}\\
&\quad\quad\quad+ \frac{1}{x^{3}\left(2\pi\right)^{2}}\int_0^{\Lambda x} \dd y \frac{y}{\sqrt{y}}\propto \frac{x^{\epsilon/2}-1}{x^{3/2}},
\end{split}
\end{equation}
where we used the fact that $f(y)$ sharply drops to zero for $y\gtrsim 1$.
Expanding this expression in $\epsilon$, we obtain $\delta G_R^{\epsilon}= \left(\epsilon/2\right)(\ln x) G_{0R}$.

Concerning the correction $\delta G_K^{\epsilon}$ to $G_K$, we use again the argument that the term coming from the expansion of
$J_{1-\epsilon/2}$ does not generate any logarithmic divergence,
and therefore we focus on the remaining terms; for simplicity, as in Sec.~\ref{sec:light-cone-G},
we consider on the case of equal times, i.e., of $G_K(x,t,t)$ for which Eqs.~\eqref{eq:GFT-epsilon} and~\eqref{eq:GK-gauss-dq} give
\begin{equation}
\begin{split}
&i\delta G_K^{\epsilon}(x,t,t)= \frac{\Omega_0}{2 x^{1-\epsilon/2}\left(2\pi\right)^{2}}\int_0^{\infty}\!\! \dd q q^{-\epsilon/2}\\
&\qquad \qquad \times J_{1}(qx)
\left[1-\cos(2 q t)\right]f\left(q/\Lambda\right) \\
&- \frac{\Omega_0}{2 x\left(2\pi\right)^{2}}\int_0^{\infty}\!\!\dd q J_{1}(qx)\left[1-\cos(2 q t)\right]
f(q/\Lambda).
\end{split}
\end{equation}
On the light cone $x=2t$, the oscillations of $J_1(qx)$ in the integrand as a function of $q$ are in phase with those of $1-\cos(2 q t)$ and therefore, after the change of variable $y= q x$,
\begin{equation}
\begin{split}
&i\delta G_K^{\epsilon}(x=2t,t,t)\simeq \frac{\Omega_0}{2 x^{2-\epsilon}\left(2\pi\right)^{2}}\int_0^{\Lambda x}\dd y
\frac{y^{-\epsilon/2}}{\sqrt{y}}\\
&\qquad \qquad - \frac{\Omega_0}{2 x^2\left(2\pi\right)^{2}}\int_0^{\Lambda x}\dd y \frac{1}{\sqrt{y}}
\propto \frac{x^{\epsilon/2}-1}{x^{3/2}}.
\end{split}
\end{equation}
Accordingly, on the light cone,
\begin{equation}
\delta G_K^{\epsilon}(x=2t,t,t)= \left(\epsilon/2\right)(\ln x) G_{0K}.
\end{equation}

Inside the light cone, i.e., for $x\ll 2 t$, $\cos(2 q t)$ may be neglected as it is out of phase with the asymptotic oscillations of $J_1$, which
actually provides an UV cutoff to the remaining integral:
\begin{equation}
\begin{split}
&i\delta G_K^{\epsilon}(x\ll 2t,t,t)\simeq \frac{\Omega_0}{2 x^{2-\epsilon}\left(2\pi\right)^{2}}\int_0^{\infty}\!\!\dd y y^{-\epsilon/2}J_1(y)\\
&\quad\quad- \frac{\Omega_0}{2 x^2\left(2\pi\right)^{2}}\int_0^{\infty}\!\!\dd y J_1(y)\propto \frac{x^{\epsilon}-1}{x^2}.
\end{split}
\end{equation}
Accordingly, inside the light cone,
\begin{equation}
\delta G_K^{\epsilon}(x\ll 2t,t,t)= \epsilon( \ln x) G_{0K}.
\end{equation}

Concerning, instead, the interaction corrections $\delta G^u_{R,K}$ at one-loop (see Eq.~\eqref{eq:exp-Grs}), one essentially needs to Fourier transform the corresponding expressions of $\delta G_{R,K}$ obtained in Sec.~\ref{sec:Greens-momentum-space}, which can be calculated directly for $d=4$:
\begin{equation}
\delta G^u_{R,K}(x,t,t')=\frac{1}{4\pi^2 x}\int_0^{\Lambda} \!\!\dd q q^2 J_1(qx)\delta G_{R,K}(q,t,t').
\end{equation}
Using Eq.~\eqref{eq:dGR-2} for $\delta G_R$, we obtain Eq.~(\ref{dGru}) in the main text.
Similarly, using the one-loop interaction correction to $G_K$ given in Eq.~(\ref{eq:dGK-2}), we obtain Eq.~(\ref{dGKU}).
Extracting logarithmic corrections from $\delta G_R^u$ in Eq.~\eqref{dGru} is straightforward and this is discussed in the main text, see Sec.~\ref{sec:light-cone}. Here, instead, we focus on $\delta G_K^u$ and outline how to extract such corrections from the corresponding expression in Eq.~\eqref{dGKU}.
At long times $x \ll 2t$,
\begin{align}
& i\delta G_K^u(x\ll 2 t,t,t)\simeq 2\theta_N\frac{\Omega_0}{8\pi^2 x^2} \int_0^{\infty} \!\!\dd y J_1(y) \ln\biggl(\frac{y}{\Lambda x}\biggr)\nonumber\\
& \simeq -2\theta_N \ln(\Lambda x) G_{0K} + 2\theta_N\frac{\Omega_0}{8\pi^2 x^2}(-\gamma +\ln 2),
\end{align}
where we have used the identity~\cite{Gradshteyn2007} $\int_0^{\infty}\!\dd y J_1(y) \ln(y)=-\gamma +\ln 2 $, where $\gamma \simeq 0.577$ is the Euler-Mascheroni constant.
On the light cone $x=2t$, instead, it is convenient to rewrite Eq.~(\ref{dGKU}) as
\begin{equation}
i\delta G_K^u(x=2t)=-2\theta_N \ln(\Lambda x) G_{0K} + 2\theta_N\frac{\Omega_0}{8\pi^2 x^2}k\left(\Lambda x\right)
\label{dgka2}
\end{equation}
with
\begin{equation}
k(x) = \int_0^x \!\!\dd y J_1(y) \left(1-\cos{y}\right)\left[-{\rm Ci}(y)+ \ln{y}\right],
\label{eq:app-k}
\end{equation}
where we have dropped the $\sin(x){\rm Si}(x)$ term, as it does not provide a logarithmic correction.
Now integrating Eq.~\eqref{eq:app-k} by parts, one finds
\begin{equation}
\begin{split}
&k(x) = \left[-{\rm Ci}(x)+ \ln{x}\right]\int_0^x\!\!\dd y J_1(y)\left(1-\cos{y}\right)\\
&\quad\quad-\int_0^x \!\!\dd y \frac{1-\cos y}{y}\int_0^y\!\!\dd y'J_1(y')\left(1-\cos{y'}\right).
\label{eq:app-k-intp}
\end{split}
\end{equation}
Moreover, since $\int_0^y\dd y' J_1(y')(1-\cos{y'}) \sim \sqrt{y}$ for $y\to\infty$,  the
last term does not give logarithmic corrections.
Accordingly, the logarithmic correction comes only from the first term on the r.h.s.~of Eq.~\eqref{eq:app-k-intp}, which is actually proportional to $G_{0K}$ in real space (see Eq.~\eqref{eq:GK-real-space-d4}) and therefore the logarithmic contribution of the second term $\Omega_0k(\Lambda x\gg 1)/(8\pi^2 x^2) \simeq \ln(\Lambda x)G_{0K}$ on the r.h.s.~of Eq.~\eqref{dgka2} cancels the one coming from the first term. This means that the interaction term, up to this order in perturbation theory, does not affect the leading algebraic behavior of $G_K$ on the light cone.

\section{Wilson's RG: one-loop corrections}
\label{app:Wilson_one_loop}
As discussed in Sec.~\ref{sec:RG-Wilson}, in the Wilson's RG the effective action for the slow fields can be obtained by integrating out the fast fields.
After a rescaling of space and time coordinates and fields, one obtains the recursion relations for the parameters of the (effective) action, as summarized in Eqs.~\eqref{eq:RGrecursion}.
In order to highlight the emergence of a non-trivial fixed point, we will consider the $\epsilon$-expansion around $d_c=4$, with $\epsilon = d_c-d$. The first-order correction to the original (bare) action of the slow fields is
\begin{equation}
\delta S^{(1)} = -u_c\int_{\xx,t} \vecphi_c\cdot\vecphi_q\, I_1(t),
\end{equation}
where, exploiting the spatial translational invariance of the Keldysh Green's function,
\begin{equation}
I_1(t) = \frac{N+2}{12N}iG_{0K}^{>}(\xx=0,t,t).
\end{equation}
The second-order correction, instead, is given by:
\begin{equation}
\delta S^{(2)} = i\frac{1}{2} \langle S_{\mathrm{int}}^2\rangle_{>}^c
\end{equation}
where $S_\text{int}$ is given in Eq.~\eqref{eq:Sint}, while the superscript $c$ indicates that only connected diagrams have to be considered. $\delta S^{(2)}$ contains corrections $\delta S_\text{coh}^{(2)}$ and $\delta S_\text{diss}^{(2)}$ to both the coherent and dissipative vertices, respectively, and is non-local both in time and space.
Let us focus first on the coherent vertices; denoting, for simplicity, $n \equiv (\xx_n,t_n)$, the corrections read:
\begin{widetext}
\begin{align}
\delta S_\text{coh}^{(2)} & =  - u_c^2  \frac{N + 8}{N^2}  \left(  \frac{2}{4!} \right)^2  \sum_{j,l =1}^N \int_{1,2} \, \left\{ \phi _{cl}^2(1)
\phi _{cj}(2)  \phi _{qj}(2)  iG_{0K}(1,2)G_{0R}(2,1) + \phi _{cl}^2(2) \phi _{cj}(1) \phi _{qj}(1) iG_{0K}(1,2) G_{0R}(1,2)\right\}  \nonumber \\
& - u_q u_c \left( \frac{2}{4!} \right)^2\frac{N + 8}{N^2}\sum_{j,l=1}^N \int_{1,2} \, \left\{ \phi _{ql}^2(1)  \phi _{cj}(2)  \phi _{qj}(2)  iG_{0K}(1,2) G_{0R}(2,1) + \phi _{ql}^2(2)\phi _{cj}(1)\phi _{qj}(1)  iG_{0K}(1,2)G_{0R}(1,2) \right\} \nonumber \\
 &=  - u_c 2 \left( \frac{2}{4} \right)^2\frac{N + 8}{N^2}\sum_{j,l=1}^N {\int_{1,2} \, \left\{ u_c\,\phi _{cl}^2(1)\phi _{cj}(2)\phi _{qj}(2) + {u_q}\,\phi _{ql}^2(1)\phi _{cj}(2)\phi _{qj}(2) \right\}iG_{0K}(1,2)G_{0R}(2,1),}
\end{align}
\end{widetext}
where in the last equality we simply relabelled the variables in the second integral. In order to retain only the local part of the resulting interaction, we expand the field $\phi_{c,q}(1)$ around point $2$ as
\begin{equation}
\label{eq:derivative-expansion}
\vecphi_{c,q}(\zz_1) \simeq \vecphi_{c,q}(\zz_2) + (\zz_1-\zz_2)\cdot\frac{\partial \vecphi_{c,q} (\zz_2)}{\partial \zz_2} + \dots
\end{equation}
with $\zz_i \equiv (\xx_i,t_i)$, and we retain only the first term of the expansion $\vecphi_{c,q}(2) \equiv \vecphi_{c,q}(\zz_2)$, as derivatives of higher order are expected to be irrelevant. Accordingly, the correction to the local vertices is
\begin{equation}
\delta S_\text{coh}^{(2)} =  - \frac{2u_c}{4!N}\int_{x,t}\!\!\!\! \left[ u_c\,(\vecphi _q \cdot \vecphi _c)\vecphi_c^2 +u_q\,(\vecphi_c \cdot \vecphi_q)\vecphi_q^2 \right]I_2(t),
\end{equation}
where
\begin{equation}
I_2(t) = \frac{N+8}{6N}\int_{\xx',t'}iG_{0K}(\xx',t,t')G_{0R}(\xx',t-t'),
\end{equation}
where the spatial translational invariance of $G_{R,K}$ has been used.
Finally, assuming for simplicity $T=0$ in the initial state, by using Eqs.~\eqref{eq:GR-gauss} and~\eqref{eq:GK-gauss}, the integrals $I_1$ and $I_2$ read:
\begin{align}
\label{eq:integral1}
&I_1  = \frac{N+2}{12N} \int_>\frac{\dd^d k}{(2\pi)^d} i G_{0K}(k,t,t) \nonumber\\
& \simeq \frac{N+2}{12N} a_d  \frac{\dd\Lambda}{\Lambda}\Lambda^d iG_{0K}(\Lambda,t,t)\nonumber\\
& =  \frac{\dd\Lambda}{\Lambda} \frac{N+2}{12N} a_d \frac{\Lambda^d}{\omega_\Lambda}\left[K_+ + K_-\cos(2 \omega_\Lambda t)\right];
\end{align}
\begin{align}
\label{eq:integral2}
&I_2  =  \frac{N+8}{6N}\int_0^{t} \dd t'  \int_> \frac{\dd^d k}{(2\pi)^d} \,iG_{0K}(k,t,t')G_{0R}(k,t-t') \nonumber \\
& \simeq \frac{N+8}{6N} a_d \frac{\dd\Lambda}{\Lambda} \Lambda^d \, \int_0^{t} \dd t'  \,iG_{0K}(\Lambda,t,t')G_{0R}(\Lambda,t-t') \nonumber \\
& = -\frac{N+8}{24N} a_d \frac{\dd\Lambda}{\Lambda} \frac{\Lambda^d}{\omega_\Lambda^3}\,  \left\{  K_- 2\omega_\Lambda\, t\, \sin(2\omega_\Lambda t) \right. \nonumber \\
& \qquad \qquad \left. + (K_+ - K_-)[1 - \cos(2\omega_\Lambda t)]  \right\},
\end{align}
where $\int_{>} \equiv \int_{\Lambda-\dd\Lambda \leq |\kk|\leq \Lambda }$ and $K_{\pm}$ are given in Eq.~\eqref{eq:Kpm-definition} and their argument is calculated for $k= \Lambda$.
Finally, the generated dissipative vertices $\propto |\vecphi_c|^2 |\vecphi_q|^2$ and $\propto (\vecphi_c\cdot\vecphi_q)^2$ due to the interaction vertices $\propto u_c^2$ and $\propto u_cu_q$ can be calculated analogously. For the latter we find,
\begin{align}
&\delta S_{{\rm{diss}}}^{(2)}\bigg|_{u_cu_q} =  - \,i{\left( {\frac{2}{{4!N}}} \right)^2}\sum_{jklm=1}^N {{u_c}{u_q} \times } \,\,\nonumber\\
&\int_{1,2} {\left[ {2\phi _{qj}^ < (1)\phi _{cj}^ > (1)\phi _{cl}^ > (1)\phi _{cl}^ < (1) + \phi _{qj}^ < (1)\phi _{cj}^ < (1)\phi _{cl}^{ > 2}(1)} \right]} \nonumber\\
&\left[ {2\phi _{qk}^ > (2)\phi _{ck}^ < (2)\phi _{qm}^ > (2)\phi _{qm}^ < (2) + \phi _{qk}^ < (2)\phi _{ck}^ < (2)\phi _{qm}^{ > 2}(2)} \right],
\label{eq:dissipative}
\end{align}
where, for convenience, we reinstated the explicit indication of the fast and the slow components of the involved fields, with the understanding that one has eventually to integrate out the fast components, as discussed in Sec.~\ref{sec:RG-Wilson}.
The previous equation is a sum of four terms, where, for instance, the first one is:
\begin{align}
&  4\left( \frac{2}{4!N} \right)^2  u_c  u_q  \int_{1,2} G_{0R}^2(1,2) \nonumber \\
&  \qquad \qquad \times \biggl\{\left[\vecphi_c^< (2) \cdot \vecphi_q^< (2) \right]^2 + \left[ \vecphi_c^< (2) \right]^2\left[\vecphi_q^< (2) \right]^2 \biggr\}.
\end{align}
All the remaining terms can be analogously evaluated. Summing all terms, including the $u_cu_q$ and $u_c^2$ vertices, one obtains
\begin{align}
& \delta S_\text{diss}^{(2)}= -i\frac{2u_c}{144N^2}\int_{1,2} \biggl[u_qG^2_{0R}(1,2)+u_c
G^2_{0K}(1,2)\biggr] \nonumber \\
& \times \biggl\{(N +6)\biggl({\vecphi_c^<}(1)\cdot{\vecphi_q^<}(1)\biggr)^2+ 2\left|\vecphi_c^<(1)\right|^2
\left|\vecphi_q^<(1)\right|^2
\biggr\}.
\end{align}
Performing the integrations, one obtains,
\begin{align}
&\delta S_\text{diss}^{(2)}  =  - iu_c\frac{\dd\Lambda }{\Lambda } \frac{a_d}{144N^2}\frac{\Lambda ^d}{\omega _\Lambda^2}\left[ u_q - u_c\left( K_+^2 + K_-^2\right) \right] \int_1 t_1 \nonumber \\
&\times  \left\{ (N + 6)\left[ \vecphi _c^ < (1) \cdot \vecphi _q^ < (1) \right]^2  + 2\left[ \vecphi _c^ < (1) \right]^2\left[ \vecphi _q^ < (1) \right]^2 \right\}
\end{align}
where we neglected the oscillating terms coming from the integration of $G_{R,K}$ in time. The coefficients $K_{\pm}$ are given in Eq.~\eqref{eq:Kpm-definition} and their argument is calculated for $k= \Lambda$.

\section{Four-point function at one-loop}
\label{app:second_order_RG}
In this Appendix we calculate explicitly the one-loop contribution to the four-point function introduced in Sec. \ref{sec:RG-CS}. As anticipated in the main text, among the many terms generated by the repeated use of the Wick's theorem, we need to consider only those which renormalize the vertex. In particular, several terms which arise in the perturbative expansion actually renormalize the Green's functions rather then the interaction vertex.
Those we are interested in can be written as
\begin{align}
I_2 & =  - \frac{1}{2}\left( \frac{2u_c}{4!N} \right)^2\int_{1' ,2' } \, \langle \phi_{cj} (1)\phi_{ql} (2)\phi_{qm}(3)\phi_{qn}(4) \nonumber \\
& \times [\vecphi_q(1') \cdot \vecphi_c(1' )]\vecphi^2_c(1' )[\vecphi _q(2')\cdot\vecphi_c(2')\vecphi^2_c(2' )\rangle_0 \nonumber \\
&= \frac{2u_c^2}{4!N}\frac{N + 8}{6}\int_{1' ,2' } \big[ \langle \phi_{cj} (1)\phi_{ql} (2)\phi_{qm}(3)\phi_{qn}(4)\nonumber \\
& \times [\vecphi _c(1' ) \cdot \vecphi_q(1' )]\vecphi^2_c(2' )\rangle_0\,G_{0K}(1' ,2')G_{0R}(1',2')\big ],
\end{align}
where $\langle ... \rangle_0$ denotes the average over the Gaussian part of the action~\eqref{eq:SK-s}, while in the last equality we relabelled the integration variables. In order to select only the local contribution to the vertex, we expand the fields contracted with the external legs around  the point indicated as
$\alpha$ (first term) and we retain only the zeroth order term (cf. Eq.~\eqref{eq:derivative-expansion}). Accordingly, we have
\begin{align}
\label{eq:integral-I2}
&I_2 =\!\frac{2u_c^2}{4!N}\frac{N + 8}{6}\!\int_{1'} \!\! \langle \phi_{cj} (1)\phi_{ql} (2)\phi_{qm}(3)\phi_{qn}(4)(\vecphi_q \cdot \vecphi_c)\vecphi_c^2\rangle_0 \nonumber \\
& \times \int_{2'}  \, G_{0K}(1' ,2')G_{0R}(1' ,2') \nonumber \\
&= \frac{N + 8}{36N}F_{jlmn}\,u_c^2\int_{1'} G_{0R}(1,1' )G_{0R}(1' ,2)G_{0R}(1' ,3)\nonumber \\
& \times G_{0R}(1' ,4) \int_{2'}  \, G_{0K}(1' ,2' )G_{0R}(1' ,2') \nonumber \\
& \equiv\! \frac{- i u_c}{6N}F_{jlmn}\!\!\int_{1'} \!\!\! G_{0R}(1,1')G_{0R}(1',2)G_{0R}(1' ,3)G_{0R}(1',4)J
\end{align}
where $F_{jlmn}$ is defined below Eq.~\eqref{eq:integral-I1-CS}, while the last integral $J$ can be easily calculated:
\begin{align}
\label{eq:J-1}
J & = \frac{N + 8}{6N}u_c\int_\xx \int_0^t \dd t'\, iG_{0K}\left(\xx, t,t'\right)G_{0R}(\xx,t-t')  \nonumber \\
 & = \frac{N + 8}{6N}u_c\int_{\qq} {\int_0^t \dd } t'\,iG_{0K}\left( {q,t,t'} \right)G_{0R}(q,t - t')\,\ee^{ -2q/\Lambda}\nonumber \\
 & = -\frac{N+8}{48N} u_c \Omega_{0q} a_4 \left\{-\frac{(\Lambda t)^2}{1+(\Lambda t)^2} + \ln\left[1 + \left( \Lambda t \right)^2\right] \right\},
\end{align}
where $\ee^{ -2q/\Lambda}$ implements the UV regularization of the integral over momenta $k$, and in the last equality we set $d = 4$. Finally, taking the limit $ \Lambda t \gg 1$, we find the logarithmic dependence on $\Lambda$
\begin{equation}
\label{eq:J-2}
J = - \frac{N+8}{24N} a_4 u_c \Omega_{0q} \ln(\Lambda t),
\end{equation}
reported in Eq.~\eqref{eq:log}.


\bibliography{biblio}

\begin{thebibliography}{123}%
\makeatletter
\providecommand \@ifxundefined [1]{%
 \@ifx{#1\undefined}
}%
\providecommand \@ifnum [1]{%
 \ifnum #1\expandafter \@firstoftwo
 \else \expandafter \@secondoftwo
 \fi
}%
\providecommand \@ifx [1]{%
 \ifx #1\expandafter \@firstoftwo
 \else \expandafter \@secondoftwo
 \fi
}%
\providecommand \natexlab [1]{#1}%
\providecommand \enquote  [1]{``#1''}%
\providecommand \bibnamefont  [1]{#1}%
\providecommand \bibfnamefont [1]{#1}%
\providecommand \citenamefont [1]{#1}%
\providecommand \href@noop [0]{\@secondoftwo}%
\providecommand \href [0]{\begingroup \@sanitize@url \@href}%
\providecommand \@href[1]{\@@startlink{#1}\@@href}%
\providecommand \@@href[1]{\endgroup#1\@@endlink}%
\providecommand \@sanitize@url [0]{\catcode `\\12\catcode `\$12\catcode
  `\&12\catcode `\#12\catcode `\^12\catcode `\_12\catcode `\%12\relax}%
\providecommand \@@startlink[1]{}%
\providecommand \@@endlink[0]{}%
\providecommand \url  [0]{\begingroup\@sanitize@url \@url }%
\providecommand \@url [1]{\endgroup\@href {#1}{\urlprefix }}%
\providecommand \urlprefix  [0]{URL }%
\providecommand \Eprint [0]{\href }%
\providecommand \doibase [0]{http://dx.doi.org/}%
\providecommand \selectlanguage [0]{\@gobble}%
\providecommand \bibinfo  [0]{\@secondoftwo}%
\providecommand \bibfield  [0]{\@secondoftwo}%
\providecommand \translation [1]{[#1]}%
\providecommand \BibitemOpen [0]{}%
\providecommand \bibitemStop [0]{}%
\providecommand \bibitemNoStop [0]{.\EOS\space}%
\providecommand \EOS [0]{\spacefactor3000\relax}%
\providecommand \BibitemShut  [1]{\csname bibitem#1\endcsname}%
\let\auto@bib@innerbib\@empty
\bibitem [{\citenamefont {Lamacraft}\ and\ \citenamefont
  {Moore}(2012)}]{Lamacraft2012}%
  \BibitemOpen
  \bibfield  {author} {\bibinfo {author} {\bibfnamefont {A.}~\bibnamefont
  {Lamacraft}}\ and\ \bibinfo {author} {\bibfnamefont {J.}~\bibnamefont
  {Moore}},\ }in\ \href@noop {} {\emph {\bibinfo {booktitle} {Ultracold Bosonic
  and Fermionic Gases}}},\ \bibinfo {editor} {edited by\ \bibinfo {editor}
  {\bibfnamefont {K.}~\bibnamefont {Levin}}, \bibinfo {editor} {\bibfnamefont
  {A.}~\bibnamefont {Fetter}}, \ and\ \bibinfo {editor} {\bibfnamefont
  {D.}~\bibnamefont {Stamper-Kurn}}}\ (\bibinfo  {publisher} {Elsevier,
  Amsterdam},\ \bibinfo {year} {2012})\ Chap.~\bibinfo {chapter}
  {7}\BibitemShut {NoStop}%
\bibitem [{\citenamefont {Yukalov}(2011)}]{Yukalov2011}%
  \BibitemOpen
  \bibfield  {author} {\bibinfo {author} {\bibfnamefont {V.}~\bibnamefont
  {Yukalov}},\ }\href {\doibase 10.1002/lapl.201110002} {\bibfield  {journal}
  {\bibinfo  {journal} {Laser Phys. Lett.}\ }\textbf {\bibinfo {volume} {8}},\
  \bibinfo {pages} {485} (\bibinfo {year} {2011})}\BibitemShut {NoStop}%
\bibitem [{\citenamefont {Bloch}\ \emph {et~al.}(2008)\citenamefont {Bloch},
  \citenamefont {Dalibard},\ and\ \citenamefont {Zwerger}}]{Bloch2008}%
  \BibitemOpen
  \bibfield  {author} {\bibinfo {author} {\bibfnamefont {I.}~\bibnamefont
  {Bloch}}, \bibinfo {author} {\bibfnamefont {J.}~\bibnamefont {Dalibard}}, \
  and\ \bibinfo {author} {\bibfnamefont {W.}~\bibnamefont {Zwerger}},\ }\href
  {\doibase 10.1103/RevModPhys.80.885} {\bibfield  {journal} {\bibinfo
  {journal} {Rev. Mod. Phys.}\ }\textbf {\bibinfo {volume} {80}},\ \bibinfo
  {pages} {885} (\bibinfo {year} {2008})}\BibitemShut {NoStop}%
\bibitem [{\citenamefont {Greiner}\ \emph {et~al.}(2002)\citenamefont
  {Greiner}, \citenamefont {Mandel}, \citenamefont {Esslinger}, \citenamefont
  {H\"ansch},\ and\ \citenamefont {Bloch}}]{Greiner2002b}%
  \BibitemOpen
  \bibfield  {author} {\bibinfo {author} {\bibfnamefont {M.}~\bibnamefont
  {Greiner}}, \bibinfo {author} {\bibfnamefont {O.}~\bibnamefont {Mandel}},
  \bibinfo {author} {\bibfnamefont {T.}~\bibnamefont {Esslinger}}, \bibinfo
  {author} {\bibfnamefont {T.~W.}\ \bibnamefont {H\"ansch}}, \ and\ \bibinfo
  {author} {\bibfnamefont {I.}~\bibnamefont {Bloch}},\ }\href {\doibase
  10.1038/415039a} {\bibfield  {journal} {\bibinfo  {journal} {Nature}\
  }\textbf {\bibinfo {volume} {415}},\ \bibinfo {pages} {39} (\bibinfo {year}
  {2002})}\BibitemShut {NoStop}%
\bibitem [{\citenamefont {Smallwood}\ \emph {et~al.}(2012)\citenamefont
  {Smallwood}, \citenamefont {Hinton}, \citenamefont {Jozwiak}, \citenamefont
  {Zhang}, \citenamefont {Koralek}, \citenamefont {Eisaki}, \citenamefont
  {Lee}, \citenamefont {Orenstein},\ and\ \citenamefont
  {Lanzara}}]{Smallwood12}%
  \BibitemOpen
  \bibfield  {author} {\bibinfo {author} {\bibfnamefont {C.~L.}\ \bibnamefont
  {Smallwood}}, \bibinfo {author} {\bibfnamefont {J.~P.}\ \bibnamefont
  {Hinton}}, \bibinfo {author} {\bibfnamefont {C.}~\bibnamefont {Jozwiak}},
  \bibinfo {author} {\bibfnamefont {W.}~\bibnamefont {Zhang}}, \bibinfo
  {author} {\bibfnamefont {J.~D.}\ \bibnamefont {Koralek}}, \bibinfo {author}
  {\bibfnamefont {H.}~\bibnamefont {Eisaki}}, \bibinfo {author} {\bibfnamefont
  {D.-H.}\ \bibnamefont {Lee}}, \bibinfo {author} {\bibfnamefont
  {J.}~\bibnamefont {Orenstein}}, \ and\ \bibinfo {author} {\bibfnamefont
  {A.}~\bibnamefont {Lanzara}},\ }\href {\doibase 10.1126/science.1217423}
  {\bibfield  {journal} {\bibinfo  {journal} {Science}\ }\textbf {\bibinfo
  {volume} {336}},\ \bibinfo {pages} {1137} (\bibinfo {year}
  {2012})}\BibitemShut {NoStop}%
\bibitem [{\citenamefont {Mankowsky}\ \emph {et~al.}(2014)\citenamefont
  {Mankowsky}, \citenamefont {Subedi}, \citenamefont {F\"orst}, \citenamefont
  {Mariager}, \citenamefont {Chollet}, \citenamefont {Lemke}, \citenamefont
  {Robinson}, \citenamefont {Glownia}, \citenamefont {Minitti}, \citenamefont
  {Frano}, \citenamefont {Fechner}, \citenamefont {Spaldin}, \citenamefont
  {Loew}, \citenamefont {Keimer}, \citenamefont {Georges},\ and\ \citenamefont
  {Cavalleri}}]{Cavalleri14}%
  \BibitemOpen
  \bibfield  {author} {\bibinfo {author} {\bibfnamefont {R.}~\bibnamefont
  {Mankowsky}}, \bibinfo {author} {\bibfnamefont {A.}~\bibnamefont {Subedi}},
  \bibinfo {author} {\bibfnamefont {M.}~\bibnamefont {F\"orst}}, \bibinfo
  {author} {\bibfnamefont {S.~O.}\ \bibnamefont {Mariager}}, \bibinfo {author}
  {\bibfnamefont {M.}~\bibnamefont {Chollet}}, \bibinfo {author} {\bibfnamefont
  {H.~T.}\ \bibnamefont {Lemke}}, \bibinfo {author} {\bibfnamefont {J.~S.}\
  \bibnamefont {Robinson}}, \bibinfo {author} {\bibfnamefont {J.~M.}\
  \bibnamefont {Glownia}}, \bibinfo {author} {\bibfnamefont {M.~P.}\
  \bibnamefont {Minitti}}, \bibinfo {author} {\bibfnamefont {A.}~\bibnamefont
  {Frano}}, \bibinfo {author} {\bibfnamefont {M.}~\bibnamefont {Fechner}},
  \bibinfo {author} {\bibfnamefont {N.~A.}\ \bibnamefont {Spaldin}}, \bibinfo
  {author} {\bibfnamefont {T.}~\bibnamefont {Loew}}, \bibinfo {author}
  {\bibfnamefont {B.}~\bibnamefont {Keimer}}, \bibinfo {author} {\bibfnamefont
  {A.}~\bibnamefont {Georges}}, \ and\ \bibinfo {author} {\bibfnamefont
  {A.}~\bibnamefont {Cavalleri}},\ }\href@noop {} {\bibfield  {journal}
  {\bibinfo  {journal} {Nature}\ }\textbf {\bibinfo {volume} {516}},\ \bibinfo
  {pages} {71} (\bibinfo {year} {2014})}\BibitemShut {NoStop}%
\bibitem [{\citenamefont {Polkovnikov}\ \emph {et~al.}(2011)\citenamefont
  {Polkovnikov}, \citenamefont {Sengupta}, \citenamefont {Silva},\ and\
  \citenamefont {Vengalattore}}]{PolkovnikovRMP}%
  \BibitemOpen
  \bibfield  {author} {\bibinfo {author} {\bibfnamefont {A.}~\bibnamefont
  {Polkovnikov}}, \bibinfo {author} {\bibfnamefont {K.}~\bibnamefont
  {Sengupta}}, \bibinfo {author} {\bibfnamefont {A.}~\bibnamefont {Silva}}, \
  and\ \bibinfo {author} {\bibfnamefont {M.}~\bibnamefont {Vengalattore}},\
  }\href {\doibase 10.1103/RevModPhys.83.863} {\bibfield  {journal} {\bibinfo
  {journal} {Rev. Mod. Phys.}\ }\textbf {\bibinfo {volume} {83}},\ \bibinfo
  {pages} {863} (\bibinfo {year} {2011})}\BibitemShut {NoStop}%
\bibitem [{\citenamefont {Eisert}\ \emph {et~al.}(2015)\citenamefont {Eisert},
  \citenamefont {Friesdorf},\ and\ \citenamefont {Gogolin}}]{Eisert2015}%
  \BibitemOpen
  \bibfield  {author} {\bibinfo {author} {\bibfnamefont {J.}~\bibnamefont
  {Eisert}}, \bibinfo {author} {\bibfnamefont {M.}~\bibnamefont {Friesdorf}}, \
  and\ \bibinfo {author} {\bibfnamefont {C.}~\bibnamefont {Gogolin}},\
  }\href@noop {} {\bibfield  {journal} {\bibinfo  {journal} {Nat. Phys.}\
  }\textbf {\bibinfo {volume} {11}},\ \bibinfo {pages} {124} (\bibinfo {year}
  {2015})}\BibitemShut {NoStop}%
\bibitem [{\citenamefont {Calabrese}\ and\ \citenamefont
  {Cardy}(2016)}]{Calabrese2016}%
  \BibitemOpen
  \bibfield  {author} {\bibinfo {author} {\bibfnamefont {P.}~\bibnamefont
  {Calabrese}}\ and\ \bibinfo {author} {\bibfnamefont {J.}~\bibnamefont
  {Cardy}},\ }\href {http://stacks.iop.org/1742-5468/2016/i=6/a=064003}
  {\bibfield  {journal} {\bibinfo  {journal} {J. Stat. Mech.}\ }\textbf
  {\bibinfo {volume} {2016}},\ \bibinfo {pages} {064003} (\bibinfo {year}
  {2016})}\BibitemShut {NoStop}%
\bibitem [{\citenamefont {Deutsch}(1991)}]{Deutsch1991}%
  \BibitemOpen
  \bibfield  {author} {\bibinfo {author} {\bibfnamefont {J.~M.}\ \bibnamefont
  {Deutsch}},\ }\href {\doibase 10.1103/PhysRevA.43.2046} {\bibfield  {journal}
  {\bibinfo  {journal} {Phys. Rev. A}\ }\textbf {\bibinfo {volume} {43}},\
  \bibinfo {pages} {2046} (\bibinfo {year} {1991})}\BibitemShut {NoStop}%
\bibitem [{\citenamefont {Srednicki}(1994)}]{Srednicki1994}%
  \BibitemOpen
  \bibfield  {author} {\bibinfo {author} {\bibfnamefont {M.}~\bibnamefont
  {Srednicki}},\ }\href {\doibase 10.1103/PhysRevE.50.888} {\bibfield
  {journal} {\bibinfo  {journal} {Phys. Rev. E}\ }\textbf {\bibinfo {volume}
  {50}},\ \bibinfo {pages} {888} (\bibinfo {year} {1994})}\BibitemShut
  {NoStop}%
\bibitem [{\citenamefont {Rigol}\ \emph {et~al.}(2008)\citenamefont {Rigol},
  \citenamefont {Dunjko},\ and\ \citenamefont {Olshanii}}]{Rigol2008}%
  \BibitemOpen
  \bibfield  {author} {\bibinfo {author} {\bibfnamefont {M.}~\bibnamefont
  {Rigol}}, \bibinfo {author} {\bibfnamefont {V.}~\bibnamefont {Dunjko}}, \
  and\ \bibinfo {author} {\bibfnamefont {M.}~\bibnamefont {Olshanii}},\ }\href
  {\doibase 10.1038/nature06838} {\bibfield  {journal} {\bibinfo  {journal}
  {Nature}\ }\textbf {\bibinfo {volume} {452}},\ \bibinfo {pages} {854}
  (\bibinfo {year} {2008})}\BibitemShut {NoStop}%
\bibitem [{\citenamefont {Biroli}\ \emph {et~al.}(2010)\citenamefont {Biroli},
  \citenamefont {Kollath},\ and\ \citenamefont {L\"auchli}}]{Biroli2010}%
  \BibitemOpen
  \bibfield  {author} {\bibinfo {author} {\bibfnamefont {G.}~\bibnamefont
  {Biroli}}, \bibinfo {author} {\bibfnamefont {C.}~\bibnamefont {Kollath}}, \
  and\ \bibinfo {author} {\bibfnamefont {A.~M.}\ \bibnamefont {L\"auchli}},\
  }\href {\doibase 10.1103/PhysRevLett.105.250401} {\bibfield  {journal}
  {\bibinfo  {journal} {Phys. Rev. Lett.}\ }\textbf {\bibinfo {volume} {105}},\
  \bibinfo {pages} {250401} (\bibinfo {year} {2010})}\BibitemShut {NoStop}%
\bibitem [{\citenamefont {Canovi}\ \emph {et~al.}(2011)\citenamefont {Canovi},
  \citenamefont {Rossini}, \citenamefont {Fazio}, \citenamefont {Santoro},\
  and\ \citenamefont {Silva}}]{Canovi2011}%
  \BibitemOpen
  \bibfield  {author} {\bibinfo {author} {\bibfnamefont {E.}~\bibnamefont
  {Canovi}}, \bibinfo {author} {\bibfnamefont {D.}~\bibnamefont {Rossini}},
  \bibinfo {author} {\bibfnamefont {R.}~\bibnamefont {Fazio}}, \bibinfo
  {author} {\bibfnamefont {G.~E.}\ \bibnamefont {Santoro}}, \ and\ \bibinfo
  {author} {\bibfnamefont {A.}~\bibnamefont {Silva}},\ }\href {\doibase
  10.1103/PhysRevB.83.094431} {\bibfield  {journal} {\bibinfo  {journal} {Phys.
  Rev. B}\ }\textbf {\bibinfo {volume} {83}},\ \bibinfo {pages} {094431}
  (\bibinfo {year} {2011})}\BibitemShut {NoStop}%
\bibitem [{\citenamefont {Tavora}\ and\ \citenamefont
  {Mitra}(2013)}]{Tavora13b}%
  \BibitemOpen
  \bibfield  {author} {\bibinfo {author} {\bibfnamefont {M.}~\bibnamefont
  {Tavora}}\ and\ \bibinfo {author} {\bibfnamefont {A.}~\bibnamefont {Mitra}},\
  }\href {\doibase 10.1103/PhysRevB.88.115144} {\bibfield  {journal} {\bibinfo
  {journal} {Phys. Rev. B}\ }\textbf {\bibinfo {volume} {88}},\ \bibinfo
  {pages} {115144} (\bibinfo {year} {2013})}\BibitemShut {NoStop}%
\bibitem [{\citenamefont {Tavora}\ \emph {et~al.}(2014)\citenamefont {Tavora},
  \citenamefont {Rosch},\ and\ \citenamefont {Mitra}}]{Tavora13}%
  \BibitemOpen
  \bibfield  {author} {\bibinfo {author} {\bibfnamefont {M.}~\bibnamefont
  {Tavora}}, \bibinfo {author} {\bibfnamefont {A.}~\bibnamefont {Rosch}}, \
  and\ \bibinfo {author} {\bibfnamefont {A.}~\bibnamefont {Mitra}},\ }\href
  {\doibase 10.1103/PhysRevLett.113.010601} {\bibfield  {journal} {\bibinfo
  {journal} {Phys. Rev. Lett.}\ }\textbf {\bibinfo {volume} {113}},\ \bibinfo
  {pages} {010601} (\bibinfo {year} {2014})}\BibitemShut {NoStop}%
\bibitem [{\citenamefont {Rigol}(2016)}]{Rigol2016}%
  \BibitemOpen
  \bibfield  {author} {\bibinfo {author} {\bibfnamefont {M.}~\bibnamefont
  {Rigol}},\ }\href {\doibase 10.1103/PhysRevLett.116.100601} {\bibfield
  {journal} {\bibinfo  {journal} {Phys. Rev. Lett.}\ }\textbf {\bibinfo
  {volume} {116}},\ \bibinfo {pages} {100601} (\bibinfo {year}
  {2016})}\BibitemShut {NoStop}%
\bibitem [{\citenamefont {Yin}\ and\ \citenamefont
  {Radzihovsky}(2016)}]{Yin2016}%
  \BibitemOpen
  \bibfield  {author} {\bibinfo {author} {\bibfnamefont {X.}~\bibnamefont
  {Yin}}\ and\ \bibinfo {author} {\bibfnamefont {L.}~\bibnamefont
  {Radzihovsky}},\ }\href {\doibase 10.1103/PhysRevA.93.033653} {\bibfield
  {journal} {\bibinfo  {journal} {Phys. Rev. A}\ }\textbf {\bibinfo {volume}
  {93}},\ \bibinfo {pages} {033653} (\bibinfo {year} {2016})}\BibitemShut
  {NoStop}%
\bibitem [{\citenamefont {Rigol}\ \emph {et~al.}(2007)\citenamefont {Rigol},
  \citenamefont {Dunjko}, \citenamefont {Yurovsky},\ and\ \citenamefont
  {Olshanii}}]{Rigol2007}%
  \BibitemOpen
  \bibfield  {author} {\bibinfo {author} {\bibfnamefont {M.}~\bibnamefont
  {Rigol}}, \bibinfo {author} {\bibfnamefont {V.}~\bibnamefont {Dunjko}},
  \bibinfo {author} {\bibfnamefont {V.}~\bibnamefont {Yurovsky}}, \ and\
  \bibinfo {author} {\bibfnamefont {M.}~\bibnamefont {Olshanii}},\ }\href
  {\doibase 10.1103/PhysRevLett.98.050405} {\bibfield  {journal} {\bibinfo
  {journal} {Phys. Rev. Lett.}\ }\textbf {\bibinfo {volume} {98}},\ \bibinfo
  {pages} {050405} (\bibinfo {year} {2007})}\BibitemShut {NoStop}%
\bibitem [{\citenamefont {Iucci}\ and\ \citenamefont
  {Cazalilla}(2009)}]{Iucci2009}%
  \BibitemOpen
  \bibfield  {author} {\bibinfo {author} {\bibfnamefont {A.}~\bibnamefont
  {Iucci}}\ and\ \bibinfo {author} {\bibfnamefont {M.~A.}\ \bibnamefont
  {Cazalilla}},\ }\href {\doibase 10.1103/PhysRevA.80.063619} {\bibfield
  {journal} {\bibinfo  {journal} {Phys. Rev. A}\ }\textbf {\bibinfo {volume}
  {80}},\ \bibinfo {pages} {063619} (\bibinfo {year} {2009})}\BibitemShut
  {NoStop}%
\bibitem [{\citenamefont {Jaynes}(1957)}]{Jaynes1957}%
  \BibitemOpen
  \bibfield  {author} {\bibinfo {author} {\bibfnamefont {E.~T.}\ \bibnamefont
  {Jaynes}},\ }\href {\doibase 10.1103/PhysRev.106.620} {\bibfield  {journal}
  {\bibinfo  {journal} {Phys. Rev.}\ }\textbf {\bibinfo {volume} {106}},\
  \bibinfo {pages} {620} (\bibinfo {year} {1957})}\BibitemShut {NoStop}%
\bibitem [{\citenamefont {Barthel}\ and\ \citenamefont
  {Schollw{\"o}ck}(2008)}]{Barthel2008}%
  \BibitemOpen
  \bibfield  {author} {\bibinfo {author} {\bibfnamefont {T.}~\bibnamefont
  {Barthel}}\ and\ \bibinfo {author} {\bibfnamefont {U.}~\bibnamefont
  {Schollw{\"o}ck}},\ }\href {\doibase 10.1103/PhysRevLett.100.100601}
  {\bibfield  {journal} {\bibinfo  {journal} {Phys. Rev. Lett.}\ }\textbf
  {\bibinfo {volume} {100}},\ \bibinfo {pages} {100601} (\bibinfo {year}
  {2008})}\BibitemShut {NoStop}%
\bibitem [{\citenamefont {Goldstein}\ and\ \citenamefont
  {Andrei}(2014)}]{Goldstein2014}%
  \BibitemOpen
  \bibfield  {author} {\bibinfo {author} {\bibfnamefont {G.}~\bibnamefont
  {Goldstein}}\ and\ \bibinfo {author} {\bibfnamefont {N.}~\bibnamefont
  {Andrei}},\ }\href {\doibase 10.1103/PhysRevA.90.043625} {\bibfield
  {journal} {\bibinfo  {journal} {Phys. Rev. A}\ }\textbf {\bibinfo {volume}
  {90}},\ \bibinfo {pages} {043625} (\bibinfo {year} {2014})}\BibitemShut
  {NoStop}%
\bibitem [{\citenamefont {Pozsgay}\ \emph {et~al.}(2014)\citenamefont
  {Pozsgay}, \citenamefont {Mesty{\'a}n}, \citenamefont {Werner}, \citenamefont
  {Kormos}, \citenamefont {Zar{\'a}nd},\ and\ \citenamefont
  {Tak{\'a}cs}}]{Pozsgay2014}%
  \BibitemOpen
  \bibfield  {author} {\bibinfo {author} {\bibfnamefont {B.}~\bibnamefont
  {Pozsgay}}, \bibinfo {author} {\bibfnamefont {M.}~\bibnamefont
  {Mesty{\'a}n}}, \bibinfo {author} {\bibfnamefont {M.}~\bibnamefont {Werner}},
  \bibinfo {author} {\bibfnamefont {M.}~\bibnamefont {Kormos}}, \bibinfo
  {author} {\bibfnamefont {G.}~\bibnamefont {Zar{\'a}nd}}, \ and\ \bibinfo
  {author} {\bibfnamefont {G.}~\bibnamefont {Tak{\'a}cs}},\ }\href {\doibase
  10.1103/PhysRevLett.113.117203} {\bibfield  {journal} {\bibinfo  {journal}
  {Phys. Rev. Lett.}\ }\textbf {\bibinfo {volume} {113}},\ \bibinfo {pages}
  {117203} (\bibinfo {year} {2014})}\BibitemShut {NoStop}%
\bibitem [{\citenamefont {Mierzejewski}\ \emph {et~al.}(2014)\citenamefont
  {Mierzejewski}, \citenamefont {Prelov\ifmmode~\check{s}\else \v{s}\fi{}ek},\
  and\ \citenamefont {Prosen}}]{Mierzejewski2014}%
  \BibitemOpen
  \bibfield  {author} {\bibinfo {author} {\bibfnamefont {M.}~\bibnamefont
  {Mierzejewski}}, \bibinfo {author} {\bibfnamefont {P.}~\bibnamefont
  {Prelov\ifmmode~\check{s}\else \v{s}\fi{}ek}}, \ and\ \bibinfo {author}
  {\bibfnamefont {T.}~\bibnamefont {Prosen}},\ }\href {\doibase
  10.1103/PhysRevLett.113.020602} {\bibfield  {journal} {\bibinfo  {journal}
  {Phys. Rev. Lett.}\ }\textbf {\bibinfo {volume} {113}},\ \bibinfo {pages}
  {020602} (\bibinfo {year} {2014})}\BibitemShut {NoStop}%
\bibitem [{\citenamefont {Wouters}\ \emph {et~al.}(2014)\citenamefont
  {Wouters}, \citenamefont {De~Nardis}, \citenamefont {Brockmann},
  \citenamefont {Fioretto}, \citenamefont {Rigol},\ and\ \citenamefont
  {Caux}}]{Wouters2014}%
  \BibitemOpen
  \bibfield  {author} {\bibinfo {author} {\bibfnamefont {B.}~\bibnamefont
  {Wouters}}, \bibinfo {author} {\bibfnamefont {J.}~\bibnamefont {De~Nardis}},
  \bibinfo {author} {\bibfnamefont {M.}~\bibnamefont {Brockmann}}, \bibinfo
  {author} {\bibfnamefont {D.}~\bibnamefont {Fioretto}}, \bibinfo {author}
  {\bibfnamefont {M.}~\bibnamefont {Rigol}}, \ and\ \bibinfo {author}
  {\bibfnamefont {J.-S.}\ \bibnamefont {Caux}},\ }\href {\doibase
  10.1103/PhysRevLett.113.117202} {\bibfield  {journal} {\bibinfo  {journal}
  {Phys. Rev. Lett.}\ }\textbf {\bibinfo {volume} {113}},\ \bibinfo {pages}
  {117202} (\bibinfo {year} {2014})}\BibitemShut {NoStop}%
\bibitem [{\citenamefont {Essler}\ and\ \citenamefont
  {Fagotti}(2016)}]{Essler2016}%
  \BibitemOpen
  \bibfield  {author} {\bibinfo {author} {\bibfnamefont {F.~H.~L.}\
  \bibnamefont {Essler}}\ and\ \bibinfo {author} {\bibfnamefont
  {M.}~\bibnamefont {Fagotti}},\ }\href
  {http://stacks.iop.org/1742-5468/2016/i=6/a=064002} {\bibfield  {journal}
  {\bibinfo  {journal} {J. Stat. Mech.}\ }\textbf {\bibinfo {volume} {2016}},\
  \bibinfo {pages} {064002} (\bibinfo {year} {2016})}\BibitemShut {NoStop}%
\bibitem [{\citenamefont {Berges}\ \emph {et~al.}(2004)\citenamefont {Berges},
  \citenamefont {Bors\'anyi},\ and\ \citenamefont {Wetterich}}]{Berges2004a}%
  \BibitemOpen
  \bibfield  {author} {\bibinfo {author} {\bibfnamefont {J.}~\bibnamefont
  {Berges}}, \bibinfo {author} {\bibfnamefont {S.}~\bibnamefont {Bors\'anyi}},
  \ and\ \bibinfo {author} {\bibfnamefont {C.}~\bibnamefont {Wetterich}},\
  }\href {\doibase 10.1103/PhysRevLett.93.142002} {\bibfield  {journal}
  {\bibinfo  {journal} {Phys. Rev. Lett.}\ }\textbf {\bibinfo {volume} {93}},\
  \bibinfo {pages} {142002} (\bibinfo {year} {2004})}\BibitemShut {NoStop}%
\bibitem [{\citenamefont {Smith}\ \emph {et~al.}(2013)\citenamefont {Smith},
  \citenamefont {Gring}, \citenamefont {Langen}, \citenamefont {Kuhnert},
  \citenamefont {Rauer}, \citenamefont {Geiger}, \citenamefont {Kitagawa},
  \citenamefont {Mazets}, \citenamefont {Demler},\ and\ \citenamefont
  {Schmiedmayer}}]{Langen2013a}%
  \BibitemOpen
  \bibfield  {author} {\bibinfo {author} {\bibfnamefont {D.~A.}\ \bibnamefont
  {Smith}}, \bibinfo {author} {\bibfnamefont {M.}~\bibnamefont {Gring}},
  \bibinfo {author} {\bibfnamefont {T.}~\bibnamefont {Langen}}, \bibinfo
  {author} {\bibfnamefont {M.}~\bibnamefont {Kuhnert}}, \bibinfo {author}
  {\bibfnamefont {B.}~\bibnamefont {Rauer}}, \bibinfo {author} {\bibfnamefont
  {R.}~\bibnamefont {Geiger}}, \bibinfo {author} {\bibfnamefont
  {T.}~\bibnamefont {Kitagawa}}, \bibinfo {author} {\bibfnamefont
  {I.}~\bibnamefont {Mazets}}, \bibinfo {author} {\bibfnamefont
  {E.}~\bibnamefont {Demler}}, \ and\ \bibinfo {author} {\bibfnamefont
  {J.}~\bibnamefont {Schmiedmayer}},\ }\href
  {http://stacks.iop.org/1367-2630/15/i=7/a=075011} {\bibfield  {journal}
  {\bibinfo  {journal} {New J. Phys.}\ }\textbf {\bibinfo {volume} {15}},\
  \bibinfo {pages} {075011} (\bibinfo {year} {2013})}\BibitemShut {NoStop}%
\bibitem [{\citenamefont {Kitagawa}\ \emph {et~al.}(2011)\citenamefont
  {Kitagawa}, \citenamefont {Imambekov}, \citenamefont {Schmiedmayer},\ and\
  \citenamefont {Demler}}]{Kitagawa2011}%
  \BibitemOpen
  \bibfield  {author} {\bibinfo {author} {\bibfnamefont {T.}~\bibnamefont
  {Kitagawa}}, \bibinfo {author} {\bibfnamefont {A.}~\bibnamefont {Imambekov}},
  \bibinfo {author} {\bibfnamefont {J.}~\bibnamefont {Schmiedmayer}}, \ and\
  \bibinfo {author} {\bibfnamefont {E.}~\bibnamefont {Demler}},\ }\href@noop {}
  {\bibfield  {journal} {\bibinfo  {journal} {New J. Phys.}\ }\textbf {\bibinfo
  {volume} {13}},\ \bibinfo {pages} {073018} (\bibinfo {year}
  {2011})}\BibitemShut {NoStop}%
\bibitem [{\citenamefont {Kollar}\ \emph {et~al.}(2011)\citenamefont {Kollar},
  \citenamefont {Wolf},\ and\ \citenamefont {Eckstein}}]{Kollar2011}%
  \BibitemOpen
  \bibfield  {author} {\bibinfo {author} {\bibfnamefont {M.}~\bibnamefont
  {Kollar}}, \bibinfo {author} {\bibfnamefont {F.~A.}\ \bibnamefont {Wolf}}, \
  and\ \bibinfo {author} {\bibfnamefont {M.}~\bibnamefont {Eckstein}},\ }\href
  {\doibase 10.1103/PhysRevB.84.054304} {\bibfield  {journal} {\bibinfo
  {journal} {Phys. Rev. B}\ }\textbf {\bibinfo {volume} {84}},\ \bibinfo
  {pages} {054304} (\bibinfo {year} {2011})}\BibitemShut {NoStop}%
\bibitem [{\citenamefont {Moeckel}\ and\ \citenamefont
  {Kehrein}(2009)}]{Moeckel2009}%
  \BibitemOpen
  \bibfield  {author} {\bibinfo {author} {\bibfnamefont {M.}~\bibnamefont
  {Moeckel}}\ and\ \bibinfo {author} {\bibfnamefont {S.}~\bibnamefont
  {Kehrein}},\ }\href {\doibase http://dx.doi.org/10.1016/j.aop.2009.03.009}
  {\bibfield  {journal} {\bibinfo  {journal} {Ann. Phys.}\ }\textbf {\bibinfo
  {volume} {324}},\ \bibinfo {pages} {2146 } (\bibinfo {year}
  {2009})}\BibitemShut {NoStop}%
\bibitem [{\citenamefont {Moeckel}\ and\ \citenamefont
  {Kehrein}(2010)}]{Moeckel2010}%
  \BibitemOpen
  \bibfield  {author} {\bibinfo {author} {\bibfnamefont {M.}~\bibnamefont
  {Moeckel}}\ and\ \bibinfo {author} {\bibfnamefont {S.}~\bibnamefont
  {Kehrein}},\ }\href@noop {} {\bibfield  {journal} {\bibinfo  {journal} {New
  J. Phys.}\ }\textbf {\bibinfo {volume} {12}},\ \bibinfo {pages} {055016}
  (\bibinfo {year} {2010})}\BibitemShut {NoStop}%
\bibitem [{\citenamefont {Marino}\ and\ \citenamefont
  {Silva}(2012)}]{Marino2012}%
  \BibitemOpen
  \bibfield  {author} {\bibinfo {author} {\bibfnamefont {J.}~\bibnamefont
  {Marino}}\ and\ \bibinfo {author} {\bibfnamefont {A.}~\bibnamefont {Silva}},\
  }\href {\doibase 10.1103/PhysRevB.86.060408} {\bibfield  {journal} {\bibinfo
  {journal} {Phys. Rev. B}\ }\textbf {\bibinfo {volume} {86}},\ \bibinfo
  {pages} {060408} (\bibinfo {year} {2012})}\BibitemShut {NoStop}%
\bibitem [{\citenamefont {Marcuzzi}\ \emph {et~al.}(2013)\citenamefont
  {Marcuzzi}, \citenamefont {Marino}, \citenamefont {Gambassi},\ and\
  \citenamefont {Silva}}]{Marcuzzi2013}%
  \BibitemOpen
  \bibfield  {author} {\bibinfo {author} {\bibfnamefont {M.}~\bibnamefont
  {Marcuzzi}}, \bibinfo {author} {\bibfnamefont {J.}~\bibnamefont {Marino}},
  \bibinfo {author} {\bibfnamefont {A.}~\bibnamefont {Gambassi}}, \ and\
  \bibinfo {author} {\bibfnamefont {A.}~\bibnamefont {Silva}},\ }\href
  {\doibase 10.1103/PhysRevLett.111.197203} {\bibfield  {journal} {\bibinfo
  {journal} {Phys. Rev. Lett.}\ }\textbf {\bibinfo {volume} {111}},\ \bibinfo
  {pages} {197203} (\bibinfo {year} {2013})}\BibitemShut {NoStop}%
\bibitem [{\citenamefont {Mitra}(2013)}]{Mitra2013}%
  \BibitemOpen
  \bibfield  {author} {\bibinfo {author} {\bibfnamefont {A.}~\bibnamefont
  {Mitra}},\ }\href {\doibase 10.1103/PhysRevB.87.205109} {\bibfield  {journal}
  {\bibinfo  {journal} {Phys. Rev. B}\ }\textbf {\bibinfo {volume} {87}},\
  \bibinfo {pages} {205109} (\bibinfo {year} {2013})}\BibitemShut {NoStop}%
\bibitem [{\citenamefont {van~den Worm}\ \emph {et~al.}(2013)\citenamefont
  {van~den Worm}, \citenamefont {Sawyer}, \citenamefont {Bollinger},\ and\
  \citenamefont {Kastner}}]{VanDenWorm2013}%
  \BibitemOpen
  \bibfield  {author} {\bibinfo {author} {\bibfnamefont {M.}~\bibnamefont
  {van~den Worm}}, \bibinfo {author} {\bibfnamefont {B.~C.}\ \bibnamefont
  {Sawyer}}, \bibinfo {author} {\bibfnamefont {J.~J.}\ \bibnamefont
  {Bollinger}}, \ and\ \bibinfo {author} {\bibfnamefont {M.}~\bibnamefont
  {Kastner}},\ }\href@noop {} {\bibfield  {journal} {\bibinfo  {journal} {New
  J. Phys.}\ }\textbf {\bibinfo {volume} {15}},\ \bibinfo {pages} {083007}
  (\bibinfo {year} {2013})}\BibitemShut {NoStop}%
\bibitem [{\citenamefont {Bertini}\ \emph {et~al.}(2015)\citenamefont
  {Bertini}, \citenamefont {Essler}, \citenamefont {Groha},\ and\ \citenamefont
  {Robinson}}]{Bertini2015}%
  \BibitemOpen
  \bibfield  {author} {\bibinfo {author} {\bibfnamefont {B.}~\bibnamefont
  {Bertini}}, \bibinfo {author} {\bibfnamefont {F.~H.~L.}\ \bibnamefont
  {Essler}}, \bibinfo {author} {\bibfnamefont {S.}~\bibnamefont {Groha}}, \
  and\ \bibinfo {author} {\bibfnamefont {N.~J.}\ \bibnamefont {Robinson}},\
  }\href {\doibase 10.1103/PhysRevLett.115.180601} {\bibfield  {journal}
  {\bibinfo  {journal} {Phys. Rev. Lett.}\ }\textbf {\bibinfo {volume} {115}},\
  \bibinfo {pages} {180601} (\bibinfo {year} {2015})}\BibitemShut {NoStop}%
\bibitem [{\citenamefont {Berges}\ \emph {et~al.}(2008)\citenamefont {Berges},
  \citenamefont {Rothkopf},\ and\ \citenamefont {Schmidt}}]{Berges2008}%
  \BibitemOpen
  \bibfield  {author} {\bibinfo {author} {\bibfnamefont {J.}~\bibnamefont
  {Berges}}, \bibinfo {author} {\bibfnamefont {A.}~\bibnamefont {Rothkopf}}, \
  and\ \bibinfo {author} {\bibfnamefont {J.}~\bibnamefont {Schmidt}},\ }\href
  {\doibase 10.1103/PhysRevLett.101.041603} {\bibfield  {journal} {\bibinfo
  {journal} {Phys. Rev. Lett.}\ }\textbf {\bibinfo {volume} {101}},\ \bibinfo
  {pages} {041603} (\bibinfo {year} {2008})}\BibitemShut {NoStop}%
\bibitem [{\citenamefont {Berges}\ and\ \citenamefont
  {Hoffmeister}(2009)}]{Berges2009}%
  \BibitemOpen
  \bibfield  {author} {\bibinfo {author} {\bibfnamefont {J.}~\bibnamefont
  {Berges}}\ and\ \bibinfo {author} {\bibfnamefont {G.}~\bibnamefont
  {Hoffmeister}},\ }\href {\doibase
  http://dx.doi.org/10.1016/j.nuclphysb.2008.12.017} {\bibfield  {journal}
  {\bibinfo  {journal} {Nucl. Phys. B}\ }\textbf {\bibinfo {volume} {813}},\
  \bibinfo {pages} {383 } (\bibinfo {year} {2009})}\BibitemShut {NoStop}%
\bibitem [{\citenamefont {Langen}\ \emph {et~al.}(2016)\citenamefont {Langen},
  \citenamefont {Gasenzer},\ and\ \citenamefont {Schmiedmayer}}]{Langen2016}%
  \BibitemOpen
  \bibfield  {author} {\bibinfo {author} {\bibfnamefont {T.}~\bibnamefont
  {Langen}}, \bibinfo {author} {\bibfnamefont {T.}~\bibnamefont {Gasenzer}}, \
  and\ \bibinfo {author} {\bibfnamefont {J.}~\bibnamefont {Schmiedmayer}},\
  }\href {http://stacks.iop.org/1742-5468/2016/i=6/a=064009} {\bibfield
  {journal} {\bibinfo  {journal} {J. Stat. Mech.}\ }\textbf {\bibinfo {volume}
  {2016}},\ \bibinfo {pages} {064009} (\bibinfo {year} {2016})}\BibitemShut
  {NoStop}%
\bibitem [{\citenamefont {Berges}\ and\ \citenamefont
  {Sexty}(2011)}]{Berges2011}%
  \BibitemOpen
  \bibfield  {author} {\bibinfo {author} {\bibfnamefont {J.}~\bibnamefont
  {Berges}}\ and\ \bibinfo {author} {\bibfnamefont {D.}~\bibnamefont {Sexty}},\
  }\href {\doibase 10.1103/PhysRevD.83.085004} {\bibfield  {journal} {\bibinfo
  {journal} {Phys. Rev. D}\ }\textbf {\bibinfo {volume} {83}},\ \bibinfo
  {pages} {085004} (\bibinfo {year} {2011})}\BibitemShut {NoStop}%
\bibitem [{\citenamefont {Berges}\ and\ \citenamefont
  {Sexty}(2012)}]{Berges2012}%
  \BibitemOpen
  \bibfield  {author} {\bibinfo {author} {\bibfnamefont {J.}~\bibnamefont
  {Berges}}\ and\ \bibinfo {author} {\bibfnamefont {D.}~\bibnamefont {Sexty}},\
  }\href {\doibase 10.1103/PhysRevLett.108.161601} {\bibfield  {journal}
  {\bibinfo  {journal} {Phys. Rev. Lett.}\ }\textbf {\bibinfo {volume} {108}},\
  \bibinfo {pages} {161601} (\bibinfo {year} {2012})}\BibitemShut {NoStop}%
\bibitem [{\citenamefont {Pi\~neiro Orioli}\ \emph {et~al.}(2015)\citenamefont
  {Pi\~neiro Orioli}, \citenamefont {Boguslavski},\ and\ \citenamefont
  {Berges}}]{PineiroOrioli2015}%
  \BibitemOpen
  \bibfield  {author} {\bibinfo {author} {\bibfnamefont {A.}~\bibnamefont
  {Pi\~neiro Orioli}}, \bibinfo {author} {\bibfnamefont {K.}~\bibnamefont
  {Boguslavski}}, \ and\ \bibinfo {author} {\bibfnamefont {J.}~\bibnamefont
  {Berges}},\ }\href {\doibase 10.1103/PhysRevD.92.025041} {\bibfield
  {journal} {\bibinfo  {journal} {Phys. Rev. D}\ }\textbf {\bibinfo {volume}
  {92}},\ \bibinfo {pages} {025041} (\bibinfo {year} {2015})}\BibitemShut
  {NoStop}%
\bibitem [{\citenamefont {Nowak}\ \emph {et~al.}(2012)\citenamefont {Nowak},
  \citenamefont {Schole}, \citenamefont {Sexty},\ and\ \citenamefont
  {Gasenzer}}]{Nowak2012}%
  \BibitemOpen
  \bibfield  {author} {\bibinfo {author} {\bibfnamefont {B.}~\bibnamefont
  {Nowak}}, \bibinfo {author} {\bibfnamefont {J.}~\bibnamefont {Schole}},
  \bibinfo {author} {\bibfnamefont {D.}~\bibnamefont {Sexty}}, \ and\ \bibinfo
  {author} {\bibfnamefont {T.}~\bibnamefont {Gasenzer}},\ }\href {\doibase
  10.1103/PhysRevA.85.043627} {\bibfield  {journal} {\bibinfo  {journal} {Phys.
  Rev. A}\ }\textbf {\bibinfo {volume} {85}},\ \bibinfo {pages} {043627}
  (\bibinfo {year} {2012})}\BibitemShut {NoStop}%
\bibitem [{\citenamefont {Karl}\ \emph {et~al.}(2013)\citenamefont {Karl},
  \citenamefont {Nowak},\ and\ \citenamefont {Gasenzer}}]{Karl2013}%
  \BibitemOpen
  \bibfield  {author} {\bibinfo {author} {\bibfnamefont {M.}~\bibnamefont
  {Karl}}, \bibinfo {author} {\bibfnamefont {B.}~\bibnamefont {Nowak}}, \ and\
  \bibinfo {author} {\bibfnamefont {T.}~\bibnamefont {Gasenzer}},\ }\href
  {\doibase 10.1103/PhysRevA.88.063615} {\bibfield  {journal} {\bibinfo
  {journal} {Phys. Rev. A}\ }\textbf {\bibinfo {volume} {88}},\ \bibinfo
  {pages} {063615} (\bibinfo {year} {2013})}\BibitemShut {NoStop}%
\bibitem [{\citenamefont {Schole}\ \emph {et~al.}(2012)\citenamefont {Schole},
  \citenamefont {Nowak},\ and\ \citenamefont {Gasenzer}}]{Schole2012}%
  \BibitemOpen
  \bibfield  {author} {\bibinfo {author} {\bibfnamefont {J.}~\bibnamefont
  {Schole}}, \bibinfo {author} {\bibfnamefont {B.}~\bibnamefont {Nowak}}, \
  and\ \bibinfo {author} {\bibfnamefont {T.}~\bibnamefont {Gasenzer}},\ }\href
  {\doibase 10.1103/PhysRevA.86.013624} {\bibfield  {journal} {\bibinfo
  {journal} {Phys. Rev. A}\ }\textbf {\bibinfo {volume} {86}},\ \bibinfo
  {pages} {013624} (\bibinfo {year} {2012})}\BibitemShut {NoStop}%
\bibitem [{\citenamefont {Lux}\ \emph {et~al.}(2014)\citenamefont {Lux},
  \citenamefont {M\"uller}, \citenamefont {Mitra},\ and\ \citenamefont
  {Rosch}}]{Lux14}%
  \BibitemOpen
  \bibfield  {author} {\bibinfo {author} {\bibfnamefont {J.}~\bibnamefont
  {Lux}}, \bibinfo {author} {\bibfnamefont {J.}~\bibnamefont {M\"uller}},
  \bibinfo {author} {\bibfnamefont {A.}~\bibnamefont {Mitra}}, \ and\ \bibinfo
  {author} {\bibfnamefont {A.}~\bibnamefont {Rosch}},\ }\href {\doibase
  10.1103/PhysRevA.89.053608} {\bibfield  {journal} {\bibinfo  {journal} {Phys.
  Rev. A}\ }\textbf {\bibinfo {volume} {89}},\ \bibinfo {pages} {053608}
  (\bibinfo {year} {2014})}\BibitemShut {NoStop}%
\bibitem [{\citenamefont {Eckstein}\ \emph {et~al.}(2009)\citenamefont
  {Eckstein}, \citenamefont {Kollar},\ and\ \citenamefont
  {Werner}}]{Eckstein2009}%
  \BibitemOpen
  \bibfield  {author} {\bibinfo {author} {\bibfnamefont {M.}~\bibnamefont
  {Eckstein}}, \bibinfo {author} {\bibfnamefont {M.}~\bibnamefont {Kollar}}, \
  and\ \bibinfo {author} {\bibfnamefont {P.}~\bibnamefont {Werner}},\ }\href
  {\doibase 10.1103/PhysRevLett.103.056403} {\bibfield  {journal} {\bibinfo
  {journal} {Phys. Rev. Lett.}\ }\textbf {\bibinfo {volume} {103}},\ \bibinfo
  {pages} {056403} (\bibinfo {year} {2009})}\BibitemShut {NoStop}%
\bibitem [{\citenamefont {Sciolla}\ and\ \citenamefont
  {Biroli}(2010)}]{Sciolla2010}%
  \BibitemOpen
  \bibfield  {author} {\bibinfo {author} {\bibfnamefont {B.}~\bibnamefont
  {Sciolla}}\ and\ \bibinfo {author} {\bibfnamefont {G.}~\bibnamefont
  {Biroli}},\ }\href {\doibase 10.1103/PhysRevLett.105.220401} {\bibfield
  {journal} {\bibinfo  {journal} {Phys. Rev. Lett.}\ }\textbf {\bibinfo
  {volume} {105}},\ \bibinfo {pages} {220401} (\bibinfo {year}
  {2010})}\BibitemShut {NoStop}%
\bibitem [{\citenamefont {Gambassi}\ and\ \citenamefont
  {Calabrese}(2011)}]{Gambassi2011}%
  \BibitemOpen
  \bibfield  {author} {\bibinfo {author} {\bibfnamefont {A.}~\bibnamefont
  {Gambassi}}\ and\ \bibinfo {author} {\bibfnamefont {P.}~\bibnamefont
  {Calabrese}},\ }\href {http://stacks.iop.org/0295-5075/95/i=6/a=66007}
  {\bibfield  {journal} {\bibinfo  {journal} {Europhys. Lett.}\ }\textbf
  {\bibinfo {volume} {95}},\ \bibinfo {pages} {66007} (\bibinfo {year}
  {2011})}\BibitemShut {NoStop}%
\bibitem [{\citenamefont {Schir\'o}\ and\ \citenamefont
  {Fabrizio}(2010)}]{Schiro2010}%
  \BibitemOpen
  \bibfield  {author} {\bibinfo {author} {\bibfnamefont {M.}~\bibnamefont
  {Schir\'o}}\ and\ \bibinfo {author} {\bibfnamefont {M.}~\bibnamefont
  {Fabrizio}},\ }\href {\doibase 10.1103/PhysRevLett.105.076401} {\bibfield
  {journal} {\bibinfo  {journal} {Phys. Rev. Lett.}\ }\textbf {\bibinfo
  {volume} {105}},\ \bibinfo {pages} {076401} (\bibinfo {year}
  {2010})}\BibitemShut {NoStop}%
\bibitem [{\citenamefont {Schir\'o}\ and\ \citenamefont
  {Fabrizio}(2011)}]{Schiro2011}%
  \BibitemOpen
  \bibfield  {author} {\bibinfo {author} {\bibfnamefont {M.}~\bibnamefont
  {Schir\'o}}\ and\ \bibinfo {author} {\bibfnamefont {M.}~\bibnamefont
  {Fabrizio}},\ }\href {\doibase 10.1103/PhysRevB.83.165105} {\bibfield
  {journal} {\bibinfo  {journal} {Phys. Rev. B}\ }\textbf {\bibinfo {volume}
  {83}},\ \bibinfo {pages} {165105} (\bibinfo {year} {2011})}\BibitemShut
  {NoStop}%
\bibitem [{\citenamefont {Sciolla}\ and\ \citenamefont
  {Biroli}(2011)}]{Sciolla2011}%
  \BibitemOpen
  \bibfield  {author} {\bibinfo {author} {\bibfnamefont {B.}~\bibnamefont
  {Sciolla}}\ and\ \bibinfo {author} {\bibfnamefont {G.}~\bibnamefont
  {Biroli}},\ }\href@noop {} {\bibfield  {journal} {\bibinfo  {journal} {J.
  Stat. Mech.}\ }\textbf {\bibinfo {volume} {2011}},\ \bibinfo {pages} {P11003}
  (\bibinfo {year} {2011})}\BibitemShut {NoStop}%
\bibitem [{\citenamefont {Sciolla}\ and\ \citenamefont
  {Biroli}(2013)}]{Sciolla2013}%
  \BibitemOpen
  \bibfield  {author} {\bibinfo {author} {\bibfnamefont {B.}~\bibnamefont
  {Sciolla}}\ and\ \bibinfo {author} {\bibfnamefont {G.}~\bibnamefont
  {Biroli}},\ }\href {\doibase 10.1103/PhysRevB.88.201110} {\bibfield
  {journal} {\bibinfo  {journal} {Phys. Rev. B}\ }\textbf {\bibinfo {volume}
  {88}},\ \bibinfo {pages} {201110} (\bibinfo {year} {2013})}\BibitemShut
  {NoStop}%
\bibitem [{\citenamefont {Chandran}\ \emph {et~al.}(2013)\citenamefont
  {Chandran}, \citenamefont {Nanduri}, \citenamefont {Gubser},\ and\
  \citenamefont {Sondhi}}]{Chandran2013}%
  \BibitemOpen
  \bibfield  {author} {\bibinfo {author} {\bibfnamefont {A.}~\bibnamefont
  {Chandran}}, \bibinfo {author} {\bibfnamefont {A.}~\bibnamefont {Nanduri}},
  \bibinfo {author} {\bibfnamefont {S.~S.}\ \bibnamefont {Gubser}}, \ and\
  \bibinfo {author} {\bibfnamefont {S.~L.}\ \bibnamefont {Sondhi}},\ }\href
  {\doibase 10.1103/PhysRevB.88.024306} {\bibfield  {journal} {\bibinfo
  {journal} {Phys. Rev. B}\ }\textbf {\bibinfo {volume} {88}},\ \bibinfo
  {pages} {024306} (\bibinfo {year} {2013})}\BibitemShut {NoStop}%
\bibitem [{\citenamefont {Smacchia}\ \emph {et~al.}(2015)\citenamefont
  {Smacchia}, \citenamefont {Knap}, \citenamefont {Demler},\ and\ \citenamefont
  {Silva}}]{Smacchia2015}%
  \BibitemOpen
  \bibfield  {author} {\bibinfo {author} {\bibfnamefont {P.}~\bibnamefont
  {Smacchia}}, \bibinfo {author} {\bibfnamefont {M.}~\bibnamefont {Knap}},
  \bibinfo {author} {\bibfnamefont {E.}~\bibnamefont {Demler}}, \ and\ \bibinfo
  {author} {\bibfnamefont {A.}~\bibnamefont {Silva}},\ }\href {\doibase
  10.1103/PhysRevB.91.205136} {\bibfield  {journal} {\bibinfo  {journal} {Phys.
  Rev. B}\ }\textbf {\bibinfo {volume} {91}},\ \bibinfo {pages} {205136}
  (\bibinfo {year} {2015})}\BibitemShut {NoStop}%
\bibitem [{\citenamefont {Chiocchetta}\ \emph {et~al.}(2015)\citenamefont
  {Chiocchetta}, \citenamefont {Tavora}, \citenamefont {Gambassi},\ and\
  \citenamefont {Mitra}}]{Chiocchetta2015}%
  \BibitemOpen
  \bibfield  {author} {\bibinfo {author} {\bibfnamefont {A.}~\bibnamefont
  {Chiocchetta}}, \bibinfo {author} {\bibfnamefont {M.}~\bibnamefont {Tavora}},
  \bibinfo {author} {\bibfnamefont {A.}~\bibnamefont {Gambassi}}, \ and\
  \bibinfo {author} {\bibfnamefont {A.}~\bibnamefont {Mitra}},\ }\href
  {\doibase 10.1103/PhysRevB.91.220302} {\bibfield  {journal} {\bibinfo
  {journal} {Phys. Rev. B}\ }\textbf {\bibinfo {volume} {91}},\ \bibinfo
  {pages} {220302} (\bibinfo {year} {2015})}\BibitemShut {NoStop}%
\bibitem [{\citenamefont {{Maraga}}\ \emph {et~al.}(2015)\citenamefont
  {{Maraga}}, \citenamefont {{Chiocchetta}}, \citenamefont {{Mitra}},\ and\
  \citenamefont {{Gambassi}}}]{Maraga2015}%
  \BibitemOpen
  \bibfield  {author} {\bibinfo {author} {\bibfnamefont {A.}~\bibnamefont
  {{Maraga}}}, \bibinfo {author} {\bibfnamefont {A.}~\bibnamefont
  {{Chiocchetta}}}, \bibinfo {author} {\bibfnamefont {A.}~\bibnamefont
  {{Mitra}}}, \ and\ \bibinfo {author} {\bibfnamefont {A.}~\bibnamefont
  {{Gambassi}}},\ }\href {\doibase 10.1103/PhysRevE.92.042151} {\bibfield
  {journal} {\bibinfo  {journal} {Phys. Rev. E}\ }\textbf {\bibinfo {volume}
  {92}},\ \bibinfo {pages} {042151} (\bibinfo {year} {2015})}\BibitemShut
  {NoStop}%
\bibitem [{\citenamefont {{Chichinadze}}\ \emph {et~al.}(2016)\citenamefont
  {{Chichinadze}}, \citenamefont {{Ribeiro}}, \citenamefont {{Shchadilova}},\
  and\ \citenamefont {{Rubtsov}}}]{Chichinadze2016}%
  \BibitemOpen
  \bibfield  {author} {\bibinfo {author} {\bibfnamefont {D.~V.}\ \bibnamefont
  {{Chichinadze}}}, \bibinfo {author} {\bibfnamefont {P.}~\bibnamefont
  {{Ribeiro}}}, \bibinfo {author} {\bibfnamefont {Y.~E.}\ \bibnamefont
  {{Shchadilova}}}, \ and\ \bibinfo {author} {\bibfnamefont {A.~N.}\
  \bibnamefont {{Rubtsov}}},\ }\href@noop {} {\bibfield  {journal} {\bibinfo
  {journal} {arXiv:1604.01995}\ } (\bibinfo {year} {2016})}\BibitemShut
  {NoStop}%
\bibitem [{\citenamefont {Barankov}\ and\ \citenamefont
  {Levitov}(2006)}]{Barankov2006}%
  \BibitemOpen
  \bibfield  {author} {\bibinfo {author} {\bibfnamefont {R.~A.}\ \bibnamefont
  {Barankov}}\ and\ \bibinfo {author} {\bibfnamefont {L.~S.}\ \bibnamefont
  {Levitov}},\ }\href {\doibase 10.1103/PhysRevLett.96.230403} {\bibfield
  {journal} {\bibinfo  {journal} {Phys. Rev. Lett.}\ }\textbf {\bibinfo
  {volume} {96}},\ \bibinfo {pages} {230403} (\bibinfo {year}
  {2006})}\BibitemShut {NoStop}%
\bibitem [{\citenamefont {Yuzbashyan}\ \emph {et~al.}(2006)\citenamefont
  {Yuzbashyan}, \citenamefont {Tsyplyatyev},\ and\ \citenamefont
  {Altshuler}}]{Yuzbashyan2006b}%
  \BibitemOpen
  \bibfield  {author} {\bibinfo {author} {\bibfnamefont {E.~A.}\ \bibnamefont
  {Yuzbashyan}}, \bibinfo {author} {\bibfnamefont {O.}~\bibnamefont
  {Tsyplyatyev}}, \ and\ \bibinfo {author} {\bibfnamefont {B.~L.}\ \bibnamefont
  {Altshuler}},\ }\href {\doibase 10.1103/PhysRevLett.96.097005} {\bibfield
  {journal} {\bibinfo  {journal} {Phys. Rev. Lett.}\ }\textbf {\bibinfo
  {volume} {96}},\ \bibinfo {pages} {097005} (\bibinfo {year}
  {2006})}\BibitemShut {NoStop}%
\bibitem [{\citenamefont {{Maraga}}\ \emph {et~al.}(2016)\citenamefont
  {{Maraga}}, \citenamefont {{Smacchia}},\ and\ \citenamefont
  {{Silva}}}]{Maraga2016}%
  \BibitemOpen
  \bibfield  {author} {\bibinfo {author} {\bibfnamefont {A.}~\bibnamefont
  {{Maraga}}}, \bibinfo {author} {\bibfnamefont {P.}~\bibnamefont
  {{Smacchia}}}, \ and\ \bibinfo {author} {\bibfnamefont {A.}~\bibnamefont
  {{Silva}}},\ }\href@noop {} {\bibfield  {journal} {\bibinfo  {journal}
  {arXiv:1602.01763}\ } (\bibinfo {year} {2016})}\BibitemShut {NoStop}%
\bibitem [{\citenamefont {Calabrese}\ and\ \citenamefont
  {Gambassi}(2005)}]{Gambassi2005}%
  \BibitemOpen
  \bibfield  {author} {\bibinfo {author} {\bibfnamefont {P.}~\bibnamefont
  {Calabrese}}\ and\ \bibinfo {author} {\bibfnamefont {A.}~\bibnamefont
  {Gambassi}},\ }\href {http://stacks.iop.org/0305-4470/38/i=18/a=R01}
  {\bibfield  {journal} {\bibinfo  {journal} {J. Phys. A: Math. Gen.}\ }\textbf
  {\bibinfo {volume} {38}},\ \bibinfo {pages} {R133} (\bibinfo {year}
  {2005})}\BibitemShut {NoStop}%
\bibitem [{\citenamefont {{Biroli}}(2015)}]{Biroli2015}%
  \BibitemOpen
  \bibfield  {author} {\bibinfo {author} {\bibfnamefont {G.}~\bibnamefont
  {{Biroli}}},\ }\href@noop {} {\bibfield  {journal} {\bibinfo  {journal}
  {arXiv:1507.05858}\ } (\bibinfo {year} {2015})}\BibitemShut {NoStop}%
\bibitem [{\citenamefont {Baumann}\ and\ \citenamefont
  {Gambassi}(2007)}]{Baumann2007}%
  \BibitemOpen
  \bibfield  {author} {\bibinfo {author} {\bibfnamefont {F.}~\bibnamefont
  {Baumann}}\ and\ \bibinfo {author} {\bibfnamefont {A.}~\bibnamefont
  {Gambassi}},\ }\href {http://stacks.iop.org/1742-5468/2007/i=01/a=P01002}
  {\bibfield  {journal} {\bibinfo  {journal} {J. Stat. Mech.}\ }\textbf
  {\bibinfo {volume} {2007}},\ \bibinfo {pages} {P01002} (\bibinfo {year}
  {2007})}\BibitemShut {NoStop}%
\bibitem [{\citenamefont {Janssen}\ \emph {et~al.}(1989)\citenamefont
  {Janssen}, \citenamefont {Schaub},\ and\ \citenamefont
  {Schmittmann}}]{Janssen1988}%
  \BibitemOpen
  \bibfield  {author} {\bibinfo {author} {\bibfnamefont {H.}~\bibnamefont
  {Janssen}}, \bibinfo {author} {\bibfnamefont {B.}~\bibnamefont {Schaub}}, \
  and\ \bibinfo {author} {\bibfnamefont {B.}~\bibnamefont {Schmittmann}},\
  }\href {\doibase 10.1007/BF01319383} {\bibfield  {journal} {\bibinfo
  {journal} {Z. Phys. B}\ }\textbf {\bibinfo {volume} {73}},\ \bibinfo {pages}
  {539} (\bibinfo {year} {1989})}\BibitemShut {NoStop}%
\bibitem [{\citenamefont {Buchhold}\ and\ \citenamefont
  {Diehl}(2015)}]{Buchold2014}%
  \BibitemOpen
  \bibfield  {author} {\bibinfo {author} {\bibfnamefont {M.}~\bibnamefont
  {Buchhold}}\ and\ \bibinfo {author} {\bibfnamefont {S.}~\bibnamefont
  {Diehl}},\ }\href {\doibase 10.1103/PhysRevA.92.013603} {\bibfield  {journal}
  {\bibinfo  {journal} {Phys. Rev. A}\ }\textbf {\bibinfo {volume} {92}},\
  \bibinfo {pages} {013603} (\bibinfo {year} {2015})}\BibitemShut {NoStop}%
\bibitem [{\citenamefont {Gagel}\ \emph {et~al.}(2014)\citenamefont {Gagel},
  \citenamefont {Orth},\ and\ \citenamefont {Schmalian}}]{Gagel2014}%
  \BibitemOpen
  \bibfield  {author} {\bibinfo {author} {\bibfnamefont {P.}~\bibnamefont
  {Gagel}}, \bibinfo {author} {\bibfnamefont {P.~P.}\ \bibnamefont {Orth}}, \
  and\ \bibinfo {author} {\bibfnamefont {J.}~\bibnamefont {Schmalian}},\ }\href
  {\doibase 10.1103/PhysRevLett.113.220401} {\bibfield  {journal} {\bibinfo
  {journal} {Phys. Rev. Lett.}\ }\textbf {\bibinfo {volume} {113}},\ \bibinfo
  {pages} {220401} (\bibinfo {year} {2014})}\BibitemShut {NoStop}%
\bibitem [{\citenamefont {Gagel}\ \emph {et~al.}(2015)\citenamefont {Gagel},
  \citenamefont {Orth},\ and\ \citenamefont {Schmalian}}]{Gagel2015}%
  \BibitemOpen
  \bibfield  {author} {\bibinfo {author} {\bibfnamefont {P.}~\bibnamefont
  {Gagel}}, \bibinfo {author} {\bibfnamefont {P.~P.}\ \bibnamefont {Orth}}, \
  and\ \bibinfo {author} {\bibfnamefont {J.}~\bibnamefont {Schmalian}},\ }\href
  {\doibase 10.1103/PhysRevB.92.115121} {\bibfield  {journal} {\bibinfo
  {journal} {Phys. Rev. B}\ }\textbf {\bibinfo {volume} {92}},\ \bibinfo
  {pages} {115121} (\bibinfo {year} {2015})}\BibitemShut {NoStop}%
\bibitem [{\citenamefont {{Lang}}\ and\ \citenamefont
  {{Piazza}}(2016)}]{Lang2016}%
  \BibitemOpen
  \bibfield  {author} {\bibinfo {author} {\bibfnamefont {J.}~\bibnamefont
  {{Lang}}}\ and\ \bibinfo {author} {\bibfnamefont {F.}~\bibnamefont
  {{Piazza}}},\ }\href@noop {} {\bibfield  {journal} {\bibinfo  {journal}
  {arXiv:1602.05102}\ } (\bibinfo {year} {2016})}\BibitemShut {NoStop}%
\bibitem [{\citenamefont {Yin}\ \emph {et~al.}(2014)\citenamefont {Yin},
  \citenamefont {Mai},\ and\ \citenamefont {Zhong}}]{Yin2014}%
  \BibitemOpen
  \bibfield  {author} {\bibinfo {author} {\bibfnamefont {S.}~\bibnamefont
  {Yin}}, \bibinfo {author} {\bibfnamefont {P.}~\bibnamefont {Mai}}, \ and\
  \bibinfo {author} {\bibfnamefont {F.}~\bibnamefont {Zhong}},\ }\href
  {\doibase 10.1103/PhysRevB.89.144115} {\bibfield  {journal} {\bibinfo
  {journal} {Phys. Rev. B}\ }\textbf {\bibinfo {volume} {89}},\ \bibinfo
  {pages} {144115} (\bibinfo {year} {2014})}\BibitemShut {NoStop}%
\bibitem [{\citenamefont {Calabrese}\ and\ \citenamefont
  {Cardy}(2006)}]{Calabrese2006}%
  \BibitemOpen
  \bibfield  {author} {\bibinfo {author} {\bibfnamefont {P.}~\bibnamefont
  {Calabrese}}\ and\ \bibinfo {author} {\bibfnamefont {J.}~\bibnamefont
  {Cardy}},\ }\href {\doibase 10.1103/PhysRevLett.96.136801} {\bibfield
  {journal} {\bibinfo  {journal} {Phys. Rev. Lett.}\ }\textbf {\bibinfo
  {volume} {96}},\ \bibinfo {pages} {136801} (\bibinfo {year}
  {2006})}\BibitemShut {NoStop}%
\bibitem [{\citenamefont {Calabrese}\ and\ \citenamefont
  {Cardy}(2007)}]{Calabrese2007}%
  \BibitemOpen
  \bibfield  {author} {\bibinfo {author} {\bibfnamefont {P.}~\bibnamefont
  {Calabrese}}\ and\ \bibinfo {author} {\bibfnamefont {J.}~\bibnamefont
  {Cardy}},\ }\href {http://stacks.iop.org/1742-5468/2007/i=06/a=P06008}
  {\bibfield  {journal} {\bibinfo  {journal} {J. Stat. Mech.}\ }\textbf
  {\bibinfo {volume} {2007}},\ \bibinfo {pages} {P06008} (\bibinfo {year}
  {2007})}\BibitemShut {NoStop}%
\bibitem [{\citenamefont {Foini}\ \emph {et~al.}(2011)\citenamefont {Foini},
  \citenamefont {Cugliandolo},\ and\ \citenamefont {Gambassi}}]{Foini2011}%
  \BibitemOpen
  \bibfield  {author} {\bibinfo {author} {\bibfnamefont {L.}~\bibnamefont
  {Foini}}, \bibinfo {author} {\bibfnamefont {L.}~\bibnamefont {Cugliandolo}},
  \ and\ \bibinfo {author} {\bibfnamefont {A.}~\bibnamefont {Gambassi}},\
  }\href@noop {} {\bibfield  {journal} {\bibinfo  {journal} {Phys. Rev. B}\
  }\textbf {\bibinfo {volume} {84}},\ \bibinfo {pages} {212404} (\bibinfo
  {year} {2011})}\BibitemShut {NoStop}%
\bibitem [{\citenamefont {Foini}\ \emph {et~al.}(2012)\citenamefont {Foini},
  \citenamefont {Cugliandolo},\ and\ \citenamefont {Gambassi}}]{Foini2012}%
  \BibitemOpen
  \bibfield  {author} {\bibinfo {author} {\bibfnamefont {L.}~\bibnamefont
  {Foini}}, \bibinfo {author} {\bibfnamefont {L.}~\bibnamefont {Cugliandolo}},
  \ and\ \bibinfo {author} {\bibfnamefont {A.}~\bibnamefont {Gambassi}},\
  }\href@noop {} {\bibfield  {journal} {\bibinfo  {journal} {J. Stat. Mech.}\
  }\textbf {\bibinfo {volume} {2012}},\ \bibinfo {pages} {P09011} (\bibinfo
  {year} {2012})}\BibitemShut {NoStop}%
\bibitem [{\citenamefont {Marcuzzi}\ and\ \citenamefont
  {Gambassi}(2014)}]{Marcuzzi2014}%
  \BibitemOpen
  \bibfield  {author} {\bibinfo {author} {\bibfnamefont {M.}~\bibnamefont
  {Marcuzzi}}\ and\ \bibinfo {author} {\bibfnamefont {A.}~\bibnamefont
  {Gambassi}},\ }\href {\doibase 10.1103/PhysRevB.89.134307} {\bibfield
  {journal} {\bibinfo  {journal} {Phys. Rev. B}\ }\textbf {\bibinfo {volume}
  {89}},\ \bibinfo {pages} {134307} (\bibinfo {year} {2014})}\BibitemShut
  {NoStop}%
\bibitem [{\citenamefont {Sotiriadis}\ \emph {et~al.}(2009)\citenamefont
  {Sotiriadis}, \citenamefont {Calabrese},\ and\ \citenamefont
  {Cardy}}]{Sotiriadis2009}%
  \BibitemOpen
  \bibfield  {author} {\bibinfo {author} {\bibfnamefont {S.}~\bibnamefont
  {Sotiriadis}}, \bibinfo {author} {\bibfnamefont {P.}~\bibnamefont
  {Calabrese}}, \ and\ \bibinfo {author} {\bibfnamefont {J.}~\bibnamefont
  {Cardy}},\ }\href {http://stacks.iop.org/0295-5075/87/i=2/a=20002} {\bibfield
   {journal} {\bibinfo  {journal} {Europhys. Lett.}\ }\textbf {\bibinfo
  {volume} {87}},\ \bibinfo {pages} {20002} (\bibinfo {year}
  {2009})}\BibitemShut {NoStop}%
\bibitem [{\citenamefont {Sotiriadis}\ and\ \citenamefont
  {Cardy}(2010)}]{Sotiriadis2010}%
  \BibitemOpen
  \bibfield  {author} {\bibinfo {author} {\bibfnamefont {S.}~\bibnamefont
  {Sotiriadis}}\ and\ \bibinfo {author} {\bibfnamefont {J.}~\bibnamefont
  {Cardy}},\ }\href {\doibase 10.1103/PhysRevB.81.134305} {\bibfield  {journal}
  {\bibinfo  {journal} {Phys. Rev. B}\ }\textbf {\bibinfo {volume} {81}},\
  \bibinfo {pages} {134305} (\bibinfo {year} {2010})}\BibitemShut {NoStop}%
\bibitem [{\citenamefont {Fisher}\ and\ \citenamefont
  {Hohenberg}(1988)}]{Fisher1988}%
  \BibitemOpen
  \bibfield  {author} {\bibinfo {author} {\bibfnamefont {D.~S.}\ \bibnamefont
  {Fisher}}\ and\ \bibinfo {author} {\bibfnamefont {P.~C.}\ \bibnamefont
  {Hohenberg}},\ }\href {\doibase 10.1103/PhysRevB.37.4936} {\bibfield
  {journal} {\bibinfo  {journal} {Phys. Rev. B}\ }\textbf {\bibinfo {volume}
  {37}},\ \bibinfo {pages} {4936} (\bibinfo {year} {1988})}\BibitemShut
  {NoStop}%
\bibitem [{\citenamefont {Sondhi}\ \emph {et~al.}(1997)\citenamefont {Sondhi},
  \citenamefont {Girvin}, \citenamefont {Carini},\ and\ \citenamefont
  {Shahar}}]{Sondhi1997}%
  \BibitemOpen
  \bibfield  {author} {\bibinfo {author} {\bibfnamefont {S.~L.}\ \bibnamefont
  {Sondhi}}, \bibinfo {author} {\bibfnamefont {S.~M.}\ \bibnamefont {Girvin}},
  \bibinfo {author} {\bibfnamefont {J.~P.}\ \bibnamefont {Carini}}, \ and\
  \bibinfo {author} {\bibfnamefont {D.}~\bibnamefont {Shahar}},\ }\href
  {\doibase 10.1103/RevModPhys.69.315} {\bibfield  {journal} {\bibinfo
  {journal} {Rev. Mod. Phys.}\ }\textbf {\bibinfo {volume} {69}},\ \bibinfo
  {pages} {315} (\bibinfo {year} {1997})}\BibitemShut {NoStop}%
\bibitem [{\citenamefont {Sachdev}(2011)}]{Sachdevbook}%
  \BibitemOpen
  \bibfield  {author} {\bibinfo {author} {\bibfnamefont {S.}~\bibnamefont
  {Sachdev}},\ }\href@noop {} {\emph {\bibinfo {title} {Quantum Phase
  Transitions}}}\ (\bibinfo  {publisher} {Cambridge University Press},\
  \bibinfo {year} {2011})\BibitemShut {NoStop}%
\bibitem [{\citenamefont {Zinn-Justin}(1989)}]{ZinnJustinbook}%
  \BibitemOpen
  \bibfield  {author} {\bibinfo {author} {\bibfnamefont {J.}~\bibnamefont
  {Zinn-Justin}},\ }\href@noop {} {\emph {\bibinfo {title} {Quantum Field
  Theory and Critical Phenomena}}}\ (\bibinfo  {publisher} {Oxford Clarendon
  Press},\ \bibinfo {year} {1989})\BibitemShut {NoStop}%
\bibitem [{\citenamefont {Diehl}\ and\ \citenamefont
  {Dietrich}(1981)}]{Diehl1981}%
  \BibitemOpen
  \bibfield  {author} {\bibinfo {author} {\bibfnamefont {H.~W.}\ \bibnamefont
  {Diehl}}\ and\ \bibinfo {author} {\bibfnamefont {S.}~\bibnamefont
  {Dietrich}},\ }\href {\doibase 10.1007/BF01298293} {\bibfield  {journal}
  {\bibinfo  {journal} {Z. Phys. B}\ }\textbf {\bibinfo {volume} {42}},\
  \bibinfo {pages} {65} (\bibinfo {year} {1981})}\BibitemShut {NoStop}%
\bibitem [{\citenamefont {Diehl}(1997)}]{Diehl1997}%
  \BibitemOpen
  \bibfield  {author} {\bibinfo {author} {\bibfnamefont {H.~W.}\ \bibnamefont
  {Diehl}},\ }\href {\doibase 10.1142/S0217979297001751} {\bibfield  {journal}
  {\bibinfo  {journal} {Int. J. Mod. Phys. B}\ }\textbf {\bibinfo {volume}
  {11}},\ \bibinfo {pages} {3503} (\bibinfo {year} {1997})}\BibitemShut
  {NoStop}%
\bibitem [{\citenamefont {Diehl}(1986)}]{Diehl1986}%
  \BibitemOpen
  \bibfield  {author} {\bibinfo {author} {\bibfnamefont {H.~W.}\ \bibnamefont
  {Diehl}},\ }in\ \href@noop {} {\emph {\bibinfo {booktitle} {Phase Transition
  and Critical Phenomena}}},\ Vol.~\bibinfo {volume} {10},\ \bibinfo {editor}
  {edited by\ \bibinfo {editor} {\bibfnamefont {C.}~\bibnamefont {Domb}}\ and\
  \bibinfo {editor} {\bibfnamefont {J.~L.}\ \bibnamefont {Lebowitz}}}\
  (\bibinfo  {publisher} {Academic, London},\ \bibinfo {year}
  {1986})\BibitemShut {NoStop}%
\bibitem [{\citenamefont {Sachdev}(1997)}]{Sachdev1997}%
  \BibitemOpen
  \bibfield  {author} {\bibinfo {author} {\bibfnamefont {S.}~\bibnamefont
  {Sachdev}},\ }\href {\doibase 10.1103/PhysRevB.55.142} {\bibfield  {journal}
  {\bibinfo  {journal} {Phys. Rev. B}\ }\textbf {\bibinfo {volume} {55}},\
  \bibinfo {pages} {142} (\bibinfo {year} {1997})}\BibitemShut {NoStop}%
\bibitem [{\citenamefont {Cardy}(1988)}]{Cardy1988}%
  \BibitemOpen
  \bibinfo {editor} {\bibfnamefont {J.}~\bibnamefont {Cardy}},\ ed.,\
  \href@noop {} {\emph {\bibinfo {title} {Finite-size scaling}}}\ (\bibinfo
  {publisher} {North-Holland, Amsterdam},\ \bibinfo {year} {1988})\BibitemShut
  {NoStop}%
\bibitem [{\citenamefont {Brankov}\ \emph {et~al.}(2000)\citenamefont
  {Brankov}, \citenamefont {Dantchev},\ and\ \citenamefont
  {Tonchev}}]{Brankov2000}%
  \BibitemOpen
  \bibfield  {author} {\bibinfo {author} {\bibfnamefont {J.~G.}\ \bibnamefont
  {Brankov}}, \bibinfo {author} {\bibfnamefont {D.~M.}\ \bibnamefont
  {Dantchev}}, \ and\ \bibinfo {author} {\bibfnamefont {N.~S.}\ \bibnamefont
  {Tonchev}},\ }\href@noop {} {\emph {\bibinfo {title} {The Theory of Critical
  Phenomena in Finite-Size Systems --- Scaling and Quantum Effects}}}\
  (\bibinfo  {publisher} {World Scientific, Singapore},\ \bibinfo {year}
  {2000})\BibitemShut {NoStop}%
\bibitem [{\citenamefont {Ma}(2000)}]{Mabook}%
  \BibitemOpen
  \bibfield  {author} {\bibinfo {author} {\bibfnamefont {S.~K.}\ \bibnamefont
  {Ma}},\ }\href {https://books.google.de/books?id=DZvXOwntNZ4C} {\emph
  {\bibinfo {title} {Modern Theory of Critical Phenomena}}}\ (\bibinfo
  {publisher} {Perseus},\ \bibinfo {year} {2000})\BibitemShut {NoStop}%
\bibitem [{\citenamefont {Goldenfeld}(1992)}]{Goldenfeldbook}%
  \BibitemOpen
  \bibfield  {author} {\bibinfo {author} {\bibfnamefont {N.}~\bibnamefont
  {Goldenfeld}},\ }\href@noop {} {\emph {\bibinfo {title} {Lectures on Phase
  Transitions and the Renormalization Group}}}\ (\bibinfo  {publisher}
  {Addison-Wesley, Reading},\ \bibinfo {year} {1992})\BibitemShut {NoStop}%
\bibitem [{\citenamefont {Kubo}(1966)}]{Kubo1966}%
  \BibitemOpen
  \bibfield  {author} {\bibinfo {author} {\bibfnamefont {R.}~\bibnamefont
  {Kubo}},\ }\href@noop {} {\bibfield  {journal} {\bibinfo  {journal} {Rep.
  Prog. Phys.}\ }\textbf {\bibinfo {volume} {29}},\ \bibinfo {pages} {255}
  (\bibinfo {year} {1966})}\BibitemShut {NoStop}%
\bibitem [{\citenamefont {Cugliandolo}(2011)}]{Cugliandolo2011}%
  \BibitemOpen
  \bibfield  {author} {\bibinfo {author} {\bibfnamefont {L.}~\bibnamefont
  {Cugliandolo}},\ }\href@noop {} {\bibfield  {journal} {\bibinfo  {journal}
  {J. Phys. A: Math. Theor.}\ }\textbf {\bibinfo {volume} {44}},\ \bibinfo
  {pages} {483001} (\bibinfo {year} {2011})}\BibitemShut {NoStop}%
\bibitem [{\citenamefont {Mitra}\ and\ \citenamefont
  {Giamarchi}(2012)}]{Mitra2012}%
  \BibitemOpen
  \bibfield  {author} {\bibinfo {author} {\bibfnamefont {A.}~\bibnamefont
  {Mitra}}\ and\ \bibinfo {author} {\bibfnamefont {T.}~\bibnamefont
  {Giamarchi}},\ }\href {\doibase 10.1103/PhysRevB.85.075117} {\bibfield
  {journal} {\bibinfo  {journal} {Phys. Rev. B}\ }\textbf {\bibinfo {volume}
  {85}},\ \bibinfo {pages} {075117} (\bibinfo {year} {2012})}\BibitemShut
  {NoStop}%
\bibitem [{\citenamefont {Cheneau}\ \emph {et~al.}(2012)\citenamefont
  {Cheneau}, \citenamefont {Barmettler}, \citenamefont {Poletti}, \citenamefont
  {Endres}, \citenamefont {Schausz}, \citenamefont {Fukuhara}, \citenamefont
  {Gross}, \citenamefont {Bloch}, \citenamefont {Kollath},\ and\ \citenamefont
  {Kuhr}}]{Cheneau2012}%
  \BibitemOpen
  \bibfield  {author} {\bibinfo {author} {\bibfnamefont {M.}~\bibnamefont
  {Cheneau}}, \bibinfo {author} {\bibfnamefont {P.}~\bibnamefont {Barmettler}},
  \bibinfo {author} {\bibfnamefont {D.}~\bibnamefont {Poletti}}, \bibinfo
  {author} {\bibfnamefont {M.}~\bibnamefont {Endres}}, \bibinfo {author}
  {\bibfnamefont {P.}~\bibnamefont {Schausz}}, \bibinfo {author} {\bibfnamefont
  {T.}~\bibnamefont {Fukuhara}}, \bibinfo {author} {\bibfnamefont
  {C.}~\bibnamefont {Gross}}, \bibinfo {author} {\bibfnamefont
  {I.}~\bibnamefont {Bloch}}, \bibinfo {author} {\bibfnamefont
  {C.}~\bibnamefont {Kollath}}, \ and\ \bibinfo {author} {\bibfnamefont
  {S.}~\bibnamefont {Kuhr}},\ }\href@noop {} {\bibfield  {journal} {\bibinfo
  {journal} {Nature}\ }\textbf {\bibinfo {volume} {481}},\ \bibinfo {pages}
  {484} (\bibinfo {year} {2012})}\BibitemShut {NoStop}%
\bibitem [{\citenamefont {Langen}\ \emph {et~al.}(2013)\citenamefont {Langen},
  \citenamefont {Geiger}, \citenamefont {Kuhnert}, \citenamefont {Rauer},\ and\
  \citenamefont {Schmiedmayer}}]{Langen2013b}%
  \BibitemOpen
  \bibfield  {author} {\bibinfo {author} {\bibfnamefont {T.}~\bibnamefont
  {Langen}}, \bibinfo {author} {\bibfnamefont {R.}~\bibnamefont {Geiger}},
  \bibinfo {author} {\bibfnamefont {M.}~\bibnamefont {Kuhnert}}, \bibinfo
  {author} {\bibfnamefont {B.}~\bibnamefont {Rauer}}, \ and\ \bibinfo {author}
  {\bibfnamefont {J.}~\bibnamefont {Schmiedmayer}},\ }\href@noop {} {\bibfield
  {journal} {\bibinfo  {journal} {Nat. Phys.}\ }\textbf {\bibinfo {volume}
  {9}},\ \bibinfo {pages} {640} (\bibinfo {year} {2013})}\BibitemShut {NoStop}%
\bibitem [{\citenamefont {Stein}\ and\ \citenamefont
  {Weiss}(1971)}]{Stein1971}%
  \BibitemOpen
  \bibfield  {author} {\bibinfo {author} {\bibfnamefont {E.~M.}\ \bibnamefont
  {Stein}}\ and\ \bibinfo {author} {\bibfnamefont {G.~L.}\ \bibnamefont
  {Weiss}},\ }\href@noop {} {\emph {\bibinfo {title} {Introduction to Fourier
  Analysis on Euclidean Spaces}}},\ Vol.~\bibinfo {volume} {1}\ (\bibinfo
  {publisher} {Princeton University Press},\ \bibinfo {year}
  {1971})\BibitemShut {NoStop}%
\bibitem [{\citenamefont {Abramowitz}\ and\ \citenamefont
  {Stegun}(1965)}]{Abramowitzbook}%
  \BibitemOpen
  \bibfield  {author} {\bibinfo {author} {\bibfnamefont {M.}~\bibnamefont
  {Abramowitz}}\ and\ \bibinfo {author} {\bibfnamefont {I.}~\bibnamefont
  {Stegun}},\ }\href@noop {} {\emph {\bibinfo {title} {Handbook of Mathematical
  Functions}}}\ (\bibinfo  {publisher} {Dover Publications},\ \bibinfo {year}
  {1965})\BibitemShut {NoStop}%
\bibitem [{\citenamefont {Altland}\ and\ \citenamefont
  {Simons}(2010)}]{Altlandbook2010}%
  \BibitemOpen
  \bibfield  {author} {\bibinfo {author} {\bibfnamefont {A.}~\bibnamefont
  {Altland}}\ and\ \bibinfo {author} {\bibfnamefont {B.~D.}\ \bibnamefont
  {Simons}},\ }\href@noop {} {\emph {\bibinfo {title} {Condensed Matter Field
  Theory}}}\ (\bibinfo  {publisher} {Cambridge University Press},\ \bibinfo
  {year} {2010})\BibitemShut {NoStop}%
\bibitem [{\citenamefont {Kamenev}(2011)}]{Kamenevbook}%
  \BibitemOpen
  \bibfield  {author} {\bibinfo {author} {\bibfnamefont {A.}~\bibnamefont
  {Kamenev}},\ }\href@noop {} {\emph {\bibinfo {title} {Field Theory of
  Non-Equilibrium Systems}}}\ (\bibinfo  {publisher} {Cambridge University
  Press},\ \bibinfo {year} {2011})\BibitemShut {NoStop}%
\bibitem [{\citenamefont {Calzetta}\ and\ \citenamefont
  {Hu}(1988)}]{Calzetta1988}%
  \BibitemOpen
  \bibfield  {author} {\bibinfo {author} {\bibfnamefont {E.}~\bibnamefont
  {Calzetta}}\ and\ \bibinfo {author} {\bibfnamefont {B.~L.}\ \bibnamefont
  {Hu}},\ }\href {\doibase 10.1103/PhysRevD.37.2878} {\bibfield  {journal}
  {\bibinfo  {journal} {Phys. Rev. D}\ }\textbf {\bibinfo {volume} {37}},\
  \bibinfo {pages} {2878} (\bibinfo {year} {1988})}\BibitemShut {NoStop}%
\bibitem [{\citenamefont {Lubensky}\ and\ \citenamefont
  {Rubin}(1975)}]{Lubensky1975}%
  \BibitemOpen
  \bibfield  {author} {\bibinfo {author} {\bibfnamefont {T.~C.}\ \bibnamefont
  {Lubensky}}\ and\ \bibinfo {author} {\bibfnamefont {M.~H.}\ \bibnamefont
  {Rubin}},\ }\href {\doibase 10.1103/PhysRevB.11.4533} {\bibfield  {journal}
  {\bibinfo  {journal} {Phys. Rev. B}\ }\textbf {\bibinfo {volume} {11}},\
  \bibinfo {pages} {4533} (\bibinfo {year} {1975})}\BibitemShut {NoStop}%
\bibitem [{\citenamefont {Bray}\ and\ \citenamefont
  {Moore}(1977{\natexlab{a}})}]{Bray1977a}%
  \BibitemOpen
  \bibfield  {author} {\bibinfo {author} {\bibfnamefont {A.~J.}\ \bibnamefont
  {Bray}}\ and\ \bibinfo {author} {\bibfnamefont {M.~A.}\ \bibnamefont
  {Moore}},\ }\href {http://stacks.iop.org/0305-4470/10/i=11/a=021} {\bibfield
  {journal} {\bibinfo  {journal} {J. Phys. A: Math. Gen.}\ }\textbf {\bibinfo
  {volume} {10}},\ \bibinfo {pages} {1927} (\bibinfo {year}
  {1977}{\natexlab{a}})}\BibitemShut {NoStop}%
\bibitem [{\citenamefont {Bray}\ and\ \citenamefont
  {Moore}(1977{\natexlab{b}})}]{Bray1977b}%
  \BibitemOpen
  \bibfield  {author} {\bibinfo {author} {\bibfnamefont {A.~J.}\ \bibnamefont
  {Bray}}\ and\ \bibinfo {author} {\bibfnamefont {M.~A.}\ \bibnamefont
  {Moore}},\ }\href {\doibase 10.1103/PhysRevLett.38.735} {\bibfield  {journal}
  {\bibinfo  {journal} {Phys. Rev. Lett.}\ }\textbf {\bibinfo {volume} {38}},\
  \bibinfo {pages} {735} (\bibinfo {year} {1977}{\natexlab{b}})}\BibitemShut
  {NoStop}%
\bibitem [{\citenamefont {Sieberer}\ \emph {et~al.}(2015)\citenamefont
  {Sieberer}, \citenamefont {Chiocchetta}, \citenamefont {Gambassi},
  \citenamefont {T\"auber},\ and\ \citenamefont {Diehl}}]{Sieberer2015}%
  \BibitemOpen
  \bibfield  {author} {\bibinfo {author} {\bibfnamefont {L.~M.}\ \bibnamefont
  {Sieberer}}, \bibinfo {author} {\bibfnamefont {A.}~\bibnamefont
  {Chiocchetta}}, \bibinfo {author} {\bibfnamefont {A.}~\bibnamefont
  {Gambassi}}, \bibinfo {author} {\bibfnamefont {U.~C.}\ \bibnamefont
  {T\"auber}}, \ and\ \bibinfo {author} {\bibfnamefont {S.}~\bibnamefont
  {Diehl}},\ }\href {\doibase 10.1103/PhysRevB.92.134307} {\bibfield  {journal}
  {\bibinfo  {journal} {Phys. Rev. B}\ }\textbf {\bibinfo {volume} {92}},\
  \bibinfo {pages} {134307} (\bibinfo {year} {2015})}\BibitemShut {NoStop}%
\bibitem [{\citenamefont {Diehl}\ \emph {et~al.}(2006)\citenamefont {Diehl},
  \citenamefont {Grüneberg},\ and\ \citenamefont {Shpot}}]{Diehl2006}%
  \BibitemOpen
  \bibfield  {author} {\bibinfo {author} {\bibfnamefont {H.~W.}\ \bibnamefont
  {Diehl}}, \bibinfo {author} {\bibfnamefont {D.}~\bibnamefont {Grüneberg}}, \
  and\ \bibinfo {author} {\bibfnamefont {M.~A.}\ \bibnamefont {Shpot}},\ }\href
  {http://stacks.iop.org/0295-5075/75/i=2/a=241} {\bibfield  {journal}
  {\bibinfo  {journal} {Europhys. Lett.}\ }\textbf {\bibinfo {volume} {75}},\
  \bibinfo {pages} {241} (\bibinfo {year} {2006})}\BibitemShut {NoStop}%
\bibitem [{\citenamefont {Diehl}\ and\ \citenamefont
  {Schmidt}(2011)}]{Diehl2011}%
  \BibitemOpen
  \bibfield  {author} {\bibinfo {author} {\bibfnamefont {H.~W.}\ \bibnamefont
  {Diehl}}\ and\ \bibinfo {author} {\bibfnamefont {F.~M.}\ \bibnamefont
  {Schmidt}},\ }\href {http://stacks.iop.org/1367-2630/13/i=12/a=123025}
  {\bibfield  {journal} {\bibinfo  {journal} {New J. Phys.}\ }\textbf {\bibinfo
  {volume} {13}},\ \bibinfo {pages} {123025} (\bibinfo {year}
  {2011})}\BibitemShut {NoStop}%
\bibitem [{\citenamefont {Mitra}\ and\ \citenamefont
  {Giamarchi}(2011)}]{Mitra2011}%
  \BibitemOpen
  \bibfield  {author} {\bibinfo {author} {\bibfnamefont {A.}~\bibnamefont
  {Mitra}}\ and\ \bibinfo {author} {\bibfnamefont {T.}~\bibnamefont
  {Giamarchi}},\ }\href {\doibase 10.1103/PhysRevLett.107.150602} {\bibfield
  {journal} {\bibinfo  {journal} {Phys. Rev. Lett.}\ }\textbf {\bibinfo
  {volume} {107}},\ \bibinfo {pages} {150602} (\bibinfo {year}
  {2011})}\BibitemShut {NoStop}%
\bibitem [{\citenamefont {Mitra}(2012)}]{Mitra12b}%
  \BibitemOpen
  \bibfield  {author} {\bibinfo {author} {\bibfnamefont {A.}~\bibnamefont
  {Mitra}},\ }\href {\doibase 10.1103/PhysRevLett.109.260601} {\bibfield
  {journal} {\bibinfo  {journal} {Phys. Rev. Lett.}\ }\textbf {\bibinfo
  {volume} {109}},\ \bibinfo {pages} {260601} (\bibinfo {year}
  {2012})}\BibitemShut {NoStop}%
\bibitem [{\citenamefont {Dantchev}\ \emph {et~al.}(2003)\citenamefont
  {Dantchev}, \citenamefont {Krech},\ and\ \citenamefont
  {Dietrich}}]{Dantchev2003}%
  \BibitemOpen
  \bibfield  {author} {\bibinfo {author} {\bibfnamefont {D.}~\bibnamefont
  {Dantchev}}, \bibinfo {author} {\bibfnamefont {M.}~\bibnamefont {Krech}}, \
  and\ \bibinfo {author} {\bibfnamefont {S.}~\bibnamefont {Dietrich}},\ }\href
  {\doibase 10.1103/PhysRevE.67.066120} {\bibfield  {journal} {\bibinfo
  {journal} {Phys. Rev. E}\ }\textbf {\bibinfo {volume} {67}},\ \bibinfo
  {pages} {066120} (\bibinfo {year} {2003})}\BibitemShut {NoStop}%
\bibitem [{\citenamefont {T{\"a}uber}(2014)}]{TauberBook2014}%
  \BibitemOpen
  \bibfield  {author} {\bibinfo {author} {\bibfnamefont {U.~C.}\ \bibnamefont
  {T{\"a}uber}},\ }\href@noop {} {\emph {\bibinfo {title} {Critical Dynamics: a
  Field Theory Approach to Equilibrium and Non-Equilibrium Scaling Behavior}}}\
  (\bibinfo  {publisher} {Cambridge University Press},\ \bibinfo {year}
  {2014})\BibitemShut {NoStop}%
\bibitem [{\citenamefont {Wilson}\ and\ \citenamefont
  {Kogut}(1974)}]{Wilson1974}%
  \BibitemOpen
  \bibfield  {author} {\bibinfo {author} {\bibfnamefont {K.~G.}\ \bibnamefont
  {Wilson}}\ and\ \bibinfo {author} {\bibfnamefont {J.}~\bibnamefont {Kogut}},\
  }\href {\doibase http://dx.doi.org/10.1016/0370-1573(74)90023-4} {\bibfield
  {journal} {\bibinfo  {journal} {Phys. Rep.}\ }\textbf {\bibinfo {volume}
  {12}},\ \bibinfo {pages} {75 } (\bibinfo {year} {1974})}\BibitemShut
  {NoStop}%
\bibitem [{\citenamefont {Giraud}\ and\ \citenamefont
  {Serreau}(2010)}]{Giraud2010}%
  \BibitemOpen
  \bibfield  {author} {\bibinfo {author} {\bibfnamefont {A.}~\bibnamefont
  {Giraud}}\ and\ \bibinfo {author} {\bibfnamefont {J.}~\bibnamefont
  {Serreau}},\ }\href {\doibase 10.1103/PhysRevLett.104.230405} {\bibfield
  {journal} {\bibinfo  {journal} {Phys. Rev. Lett.}\ }\textbf {\bibinfo
  {volume} {104}},\ \bibinfo {pages} {230405} (\bibinfo {year}
  {2010})}\BibitemShut {NoStop}%
\bibitem [{\citenamefont {Calabrese}\ \emph {et~al.}(2011)\citenamefont
  {Calabrese}, \citenamefont {Essler},\ and\ \citenamefont
  {Fagotti}}]{Fagotti2011}%
  \BibitemOpen
  \bibfield  {author} {\bibinfo {author} {\bibfnamefont {P.}~\bibnamefont
  {Calabrese}}, \bibinfo {author} {\bibfnamefont {F.~H.~L.}\ \bibnamefont
  {Essler}}, \ and\ \bibinfo {author} {\bibfnamefont {M.}~\bibnamefont
  {Fagotti}},\ }\href {\doibase 10.1103/PhysRevLett.106.227203} {\bibfield
  {journal} {\bibinfo  {journal} {Phys. Rev. Lett.}\ }\textbf {\bibinfo
  {volume} {106}},\ \bibinfo {pages} {227203} (\bibinfo {year}
  {2011})}\BibitemShut {NoStop}%
\bibitem [{\citenamefont {Calabrese}\ \emph
  {et~al.}(2012{\natexlab{a}})\citenamefont {Calabrese}, \citenamefont
  {Essler},\ and\ \citenamefont {Fagotti}}]{Fagotti2012a}%
  \BibitemOpen
  \bibfield  {author} {\bibinfo {author} {\bibfnamefont {P.}~\bibnamefont
  {Calabrese}}, \bibinfo {author} {\bibfnamefont {F.~H.~L.}\ \bibnamefont
  {Essler}}, \ and\ \bibinfo {author} {\bibfnamefont {M.}~\bibnamefont
  {Fagotti}},\ }\href {http://stacks.iop.org/1742-5468/2012/i=07/a=P07016}
  {\bibfield  {journal} {\bibinfo  {journal} {J. Stat. Mech.}\ }\textbf
  {\bibinfo {volume} {2012}},\ \bibinfo {pages} {P07016} (\bibinfo {year}
  {2012}{\natexlab{a}})}\BibitemShut {NoStop}%
\bibitem [{\citenamefont {Calabrese}\ \emph
  {et~al.}(2012{\natexlab{b}})\citenamefont {Calabrese}, \citenamefont
  {Essler},\ and\ \citenamefont {Fagotti}}]{Fagotti2012b}%
  \BibitemOpen
  \bibfield  {author} {\bibinfo {author} {\bibfnamefont {P.}~\bibnamefont
  {Calabrese}}, \bibinfo {author} {\bibfnamefont {F.~H.~L.}\ \bibnamefont
  {Essler}}, \ and\ \bibinfo {author} {\bibfnamefont {M.}~\bibnamefont
  {Fagotti}},\ }\href {http://stacks.iop.org/1742-5468/2012/i=07/a=P07022}
  {\bibfield  {journal} {\bibinfo  {journal} {J. Stat. Mech.}\ }\textbf
  {\bibinfo {volume} {2012}},\ \bibinfo {pages} {P07022} (\bibinfo {year}
  {2012}{\natexlab{b}})}\BibitemShut {NoStop}%
\bibitem [{\citenamefont {Nicklas}\ \emph {et~al.}(2015)\citenamefont
  {Nicklas}, \citenamefont {Karl}, \citenamefont {H\"ofer}, \citenamefont
  {Johnson}, \citenamefont {Muessel}, \citenamefont {Strobel}, \citenamefont
  {Tomkovi\ifmmode~\check{c}\else \v{c}\fi{}}, \citenamefont {Gasenzer},\ and\
  \citenamefont {Oberthaler}}]{Nicklas2015}%
  \BibitemOpen
  \bibfield  {author} {\bibinfo {author} {\bibfnamefont {E.}~\bibnamefont
  {Nicklas}}, \bibinfo {author} {\bibfnamefont {M.}~\bibnamefont {Karl}},
  \bibinfo {author} {\bibfnamefont {M.}~\bibnamefont {H\"ofer}}, \bibinfo
  {author} {\bibfnamefont {A.}~\bibnamefont {Johnson}}, \bibinfo {author}
  {\bibfnamefont {W.}~\bibnamefont {Muessel}}, \bibinfo {author} {\bibfnamefont
  {H.}~\bibnamefont {Strobel}}, \bibinfo {author} {\bibfnamefont
  {J.}~\bibnamefont {Tomkovi\ifmmode~\check{c}\else \v{c}\fi{}}}, \bibinfo
  {author} {\bibfnamefont {T.}~\bibnamefont {Gasenzer}}, \ and\ \bibinfo
  {author} {\bibfnamefont {M.~K.}\ \bibnamefont {Oberthaler}},\ }\href
  {\doibase 10.1103/PhysRevLett.115.245301} {\bibfield  {journal} {\bibinfo
  {journal} {Phys. Rev. Lett.}\ }\textbf {\bibinfo {volume} {115}},\ \bibinfo
  {pages} {245301} (\bibinfo {year} {2015})}\BibitemShut {NoStop}%
\bibitem [{\citenamefont {Berges}(2004)}]{Berges2004b}%
  \BibitemOpen
  \bibfield  {author} {\bibinfo {author} {\bibfnamefont {J.}~\bibnamefont
  {Berges}},\ }\href {\doibase http://dx.doi.org/10.1063/1.1843591} {\bibfield
  {journal} {\bibinfo  {journal} {AIP Conf. Proc.}\ }\textbf {\bibinfo {volume}
  {739}},\ \bibinfo {pages} {3} (\bibinfo {year} {2004})}\BibitemShut {NoStop}%
\bibitem [{\citenamefont {{Berges}}(2015)}]{Berges2015}%
  \BibitemOpen
  \bibfield  {author} {\bibinfo {author} {\bibfnamefont {J.}~\bibnamefont
  {{Berges}}},\ }\href@noop {} {\bibfield  {journal} {\bibinfo  {journal}
  {arXiv:1503.02907}\ } (\bibinfo {year} {2015})}\BibitemShut {NoStop}%
\bibitem [{\citenamefont {Moeckel}\ and\ \citenamefont
  {Kehrein}(2008)}]{Moeckel2008}%
  \BibitemOpen
  \bibfield  {author} {\bibinfo {author} {\bibfnamefont {M.}~\bibnamefont
  {Moeckel}}\ and\ \bibinfo {author} {\bibfnamefont {S.}~\bibnamefont
  {Kehrein}},\ }\href {\doibase 10.1103/PhysRevLett.100.175702} {\bibfield
  {journal} {\bibinfo  {journal} {Phys. Rev. Lett.}\ }\textbf {\bibinfo
  {volume} {100}},\ \bibinfo {pages} {175702} (\bibinfo {year}
  {2008})}\BibitemShut {NoStop}%
\bibitem [{\citenamefont {Wegner}\ and\ \citenamefont
  {Houghton}(1973)}]{Wegner1973}%
  \BibitemOpen
  \bibfield  {author} {\bibinfo {author} {\bibfnamefont {F.~J.}\ \bibnamefont
  {Wegner}}\ and\ \bibinfo {author} {\bibfnamefont {A.}~\bibnamefont
  {Houghton}},\ }\href {\doibase 10.1103/PhysRevA.8.401} {\bibfield  {journal}
  {\bibinfo  {journal} {Phys. Rev. A}\ }\textbf {\bibinfo {volume} {8}},\
  \bibinfo {pages} {401} (\bibinfo {year} {1973})}\BibitemShut {NoStop}%
\bibitem [{\citenamefont {Lemonik}\ and\ \citenamefont
  {Mitra}(2016)}]{Lemonik2016}%
  \BibitemOpen
  \bibfield  {author} {\bibinfo {author} {\bibfnamefont {Y.}~\bibnamefont
  {Lemonik}}\ and\ \bibinfo {author} {\bibfnamefont {A.}~\bibnamefont
  {Mitra}},\ }\href {\doibase 10.1103/PhysRevB.94.024306} {\bibfield  {journal}
  {\bibinfo  {journal} {Phys. Rev. B}\ }\textbf {\bibinfo {volume} {94}},\
  \bibinfo {pages} {024306} (\bibinfo {year} {2016})}\BibitemShut {NoStop}%
\bibitem [{\citenamefont {Gradshteyn}\ and\ \citenamefont
  {Ryzhik}(2007)}]{Gradshteyn2007}%
  \BibitemOpen
  \bibfield  {author} {\bibinfo {author} {\bibfnamefont {I.~S.}\ \bibnamefont
  {Gradshteyn}}\ and\ \bibinfo {author} {\bibfnamefont {I.~M.}\ \bibnamefont
  {Ryzhik}},\ }\href@noop {} {\emph {\bibinfo {title} {Table of Integrals,
  Series, and Products}}},\ \bibinfo {edition} {seventh}\ ed.\ (\bibinfo
  {publisher} {Elsevier/Academic Press, Amsterdam},\ \bibinfo {year}
  {2007})\BibitemShut {NoStop}%
\end{thebibliography}%

\end{document}